\documentclass[11pt,a4paper]{article}
\pdfoutput=1

\usepackage{jheppub}

\usepackage{amsmath, amssymb} 
\usepackage{mathtools}
\usepackage{bm}
\usepackage{dsfont}
\usepackage{braket}
\usepackage{eqparbox}

\numberwithin{equation}{section}

\usepackage[dvipsnames]{xcolor}
\usepackage{tikz}
\usepackage{tikz-cd}
\usepackage{pgfplots}
\usetikzlibrary{intersections,backgrounds,calc}
\usepgfplotslibrary{fillbetween}
\pgfplotsset{compat=newest}
\usepackage{subcaption}

\usepackage[numbers, sort&compress]{natbib}
\usepackage{hyperref}
\hypersetup{colorlinks=true, linktoc=page, linkcolor=purple, citecolor=blue}

\newcommand{\cA}{\mathcal{A}}

\newcommand{\cC}{\mathcal{C}}

\newcommand{\cL}{\mathcal{L}}

\newcommand{\cN}{\mathcal{N}}
\newcommand{\cO}{\mathcal{O}}

\newcommand{\cS}{\mathcal{S}}

\newcommand{\mfs}{\mathfrak{s}}

\newcommand{\bbZ}{\mathbb{Z}}

\newcommand{\e}{\varepsilon}

\renewcommand{\[}{\left[}

\newcommand{\p}{\partial}

\DeclareMathOperator{\Tr}{Tr}

\renewcommand{\Re}{\operatorname{Re}}

\newcommand{\floor}[1]{\cramped{\left\lfloor #1 \right\rfloor}}

\newcommand{\half}{\tfrac{1}{2}}

\newcommand\tvdots{\raisebox{3pt}{$\scalebox{.75}{$\vdots$}$}}
\newcommand\tddots{\raisebox{3pt}{$\scalebox{.75}{$\ddots$}$}}
    
\newcommand{\no}{\nonumber}

\begin{document}

\title{The Baker-Coon-Romans $N$-point amplitude and an exact field theory limit of the Coon amplitude}

\date{\today}

\author[a]{Nicholas Geiser}

\affiliation[a]{
    Leinweber Center for Theoretical Physics, Randall Labratory of Physics\\
    University of Michigan, Ann Arbor\\
    450 Church St, Ann Arbor, MI 48109-1040, USA}

\emailAdd{ngeiser@umich.edu}

\preprint{LCTP-23-16}


\abstract{We study the $N$-point Coon amplitude discovered first by Baker and Coon in the 1970s and then again independently by Romans in the 1980s. This Baker-Coon-Romans (BCR) amplitude retains several properties of tree-level string amplitudes, namely duality and factorization, with a $q$-deformed version of the string spectrum. Although the formula for the $N$-point BCR  amplitude is only valid for ${q > 1}$, the four-point case admits a straightforward extension to all ${q \geq 0}$ which reproduces the usual expression for the four-point Coon amplitude. At five points, there are inconsistencies with factorization when pushing ${q < 1}$. Despite these issues, we find a new relation between the five-point BCR amplitude and Cheung and Remmen's four-point basic hypergeometric amplitude, placing the latter within the broader family of Coon amplitudes. Finally, we compute the $q \to \infty$ limit of the $N$-point BCR amplitudes and discover an exact correspondence between these amplitudes and the field theory amplitudes of a scalar transforming in the adjoint representation of a global symmetry group with an infinite set of non-derivative single-trace interaction terms. This correspondence at $q = \infty$ is the first definitive realization of the Coon amplitude (in any limit) from a field theory described by an explicit Lagrangian.}


\maketitle


\section{Introduction}
\label{sec:intro}

The four-point Coon amplitude, discovered in 1969 by D.~D.~Coon~\cite{Coon:1969yw, Coon:1972qz, Baker:1976en}, is a deformation of string theory's famous Veneziano amplitude~\cite{Veneziano:1968yb} with a non-linear Regge trajectory and a free deformation parameter ${q \geq 0}$. At $q=1$, Coon amplitudes become tree-level open string amplitudes. The early studies of Coon amplitudes in the 1970s were phenomenologically motivated, in parallel with the height of dual-resonance models for the strong interaction and the birth of string theory~\cite{Cappelli:2012cto}. During this period, Coon and his collaborator M.~Baker discovered a putative $N$-point generalization of the four-point Coon amplitude, valid for ${q > 1}$~\cite{Baker:1970vxk}. Despite this initial flurry of research, interest in Coon amplitudes died off by the mid 1970s. In the late 1980s, L.~J.~Romans independently rediscovered the original four-point Coon amplitude along with its putative $N$-point generalization~\cite{Romans:1988qs, Romans:1989di}.\footnote{Romans' first paper~\cite{Romans:1988qs} on Coon amplitudes was never published, but the preprint can be found \href{https://lib-extopc.kek.jp/preprints/PDF/1989/8904/8904530.pdf}{here}.} Throughout the following two decades, Coon amplitudes were again forgotten. Then in 2016, Coon amplitudes reappeared in the context of the modern S-matrix bootstrap program~\cite{Caron-Huot:2016icg}.

In the last two years, Coon amplitudes have received a burst of renewed interest, beginning with F.~Figueroa and P.~Tourkine's 2022 paper~\cite{Figueroa:2022onw}. Several groups have studied the low-energy expansion of Coon amplitudes~\cite{Figueroa:2022onw, Geiser:2022icl}, the unitarity properties of Coon amplitudes~\cite{Figueroa:2022onw, Chakravarty:2022vrp, Bhardwaj:2022lbz, Geiser:2022icl, Jepsen:2023sia}, various extensions and generalizations of Coon amplitudes~\cite{Geiser:2022exp, Cheung:2022mkw, Cheung:2023adk, Cheung:2023uwn, Duhr:2023his, Rigatos:2023asb}, and possible physical models realizing the Coon spectrum~\cite{Maldacena:2022ckr, Li:2023hce, Bhardwaj:2023eus}.

It is now a well-established result that the four-point Coon amplitude with ${q > 1}$ is non-unitary~\cite{Figueroa:2022onw, Geiser:2022icl, Bhardwaj:2022lbz, Chakravarty:2022vrp}. In contrast, the four-point Coon amplitude with ${q < 1}$ seems to satisfy the constraints of unitarity on its poles (below a $q$-dependent critical dimension)~\cite{Figueroa:2022onw, Geiser:2022icl, Bhardwaj:2022lbz, Chakravarty:2022vrp}. There is, however, an open question regarding the unitarity properties of the branch cut in the four-point Coon amplitude with ${q < 1}$. In~\cite{Jepsen:2023sia}, it was shown that the imaginary discontinuity along the branch cut does not decompose into a positively weighted sum of partial waves. The author of~\cite{Jepsen:2023sia} discusses possible interpretations of this fact and presents several strategies for excising the negativity from the partial wave coefficients. In any case, even though Coon amplitudes with ${q \neq 1}$ may violate unitarity, the ${q \to 1}$ limit is necessarily unitary by the no-ghost theorem of string theory~\cite{Brower:1972wj, Goddard:1972iy, Thorn:1983cz}. For now, we set aside these issues of unitarity and set our sights on the general $N$-point Coon amplitude in an arbitrary number ${d \geq 3}$ of spacetime dimension.

Despite all the recent work, there is still no definitive formulation of the $N$-point Coon amplitude valid for all values of the deformation parameter ${q \geq 0}$. The $N$-point amplitude discovered by Baker, Coon, and Romans is only valid for $q > 1$. For $q < 1$, there is a single paper from 1975 which proposes a worldsheet-integral-esque formula for the $N$-point Coon amplitude~\cite{GonzalezMestres:1975ord}, but it is not clear whether this formula is consistent.

Moreover, there is no definitive field theory or string theory realization of the Coon amplitude (besides, of course, string theory at $q=1$). Accumulation point spectra like those exhibited by Coon amplitudes with $q < 1$ were recently found in a setup involving open strings ending on a D\nobreakdash-brane~\cite{Maldacena:2022ckr}. Accumulation point spectra have also appeared in various contexts in the modern S-matrix bootstrap program~\cite{Caron-Huot:2020cmc}, so it is imperative to better understand the Coon amplitude’s physical origins.

In this paper, we study the $N$-point Coon amplitude discovered first by Baker and Coon in the 1970s~\cite{Baker:1970vxk} and then again independently by Romans in the 1980s~\cite{Romans:1988qs, Romans:1989di}. The formula for this $N$-point ``Baker-Coon-Romans" amplitude is remarkably compact. Moreover, the Baker-Coon-Romans amplitude retains several important physical properties of string amplitudes, namely duality and factorization, with $q$-deformed versions of the string spectrum and Regge trajectory. In several ways, the $N$-point Baker-Coon-Romans formula is simpler than its ${q \to 1}$ limit, the corresponding $N$-point open string disk integral. In principle, we may thus perform a computation at ${q \neq 1}$ and then take the $q \to 1$ limit to indirectly study string theory. For this reason alone, Coon amplitudes are worthy of study. 

Unfortunately, the naive formula for the $N$-point Baker-Coon-Romans amplitude only converges for ${q > 1}$. At four points, the amplitude admits a straightforward extension to all ${q \geq 0}$ which reproduces the usual expression for the four-point Coon amplitude. At five points, however, there are inconsistencies with factorization when pushing ${q < 1}$.

Despite these issues, the five-point Baker-Coon-Romans amplitude is still interesting. In this paper, we find a new relation between the five-point Baker-Coon-Romans amplitude and C.~Cheung and G.~Remmen's recently discovered four-point ``basic hypergeometric" (or ``hypergeometric Coon") amplitude~\cite{Cheung:2023adk, Rigatos:2023asb}, firmly placing the latter within the broader family of Coon amplitudes.

In this paper, we also compute the $q \to \infty$ limit of the $N$-point Baker-Coon-Romans amplitudes and discover an exact correspondence between these amplitudes and the tree-level amplitudes of a particular field theory. The field theory is that of a scalar field transforming in the adjoint representation of a global symmetry group with an infinite set of non-derivative single-trace interaction terms and no higher-derivative interactions. This correspondence at $q = \infty$ is the first definitive realization of the Coon amplitude (in any limit other than $q = 1$) from a field theory described by an explicit Lagrangian.

It is also known that Coon amplitudes at $q=0$ are described by a field theory~\cite{Figueroa:2022onw, Geiser:2022icl}, but the exact form of this field theory in unknown. Using our exact results at ${q = \infty}$ and the four-point Coon amplitude at ${q=0}$, we conjecture that the field theory at ${q=0}$ is given by a similar adjoint scalar theory with a slightly different set of interactions. Our results take a major step towards realizing the Coon amplitude for general $q$ in terms of a definitive field theory or worldsheet model.

In the remainder of this section, we recall some basic facts about Coon amplitudes, and we describe the field theory whose amplitudes reproduce the Coon amplitudes at $q=\infty$.

\subsection{Preliminaries}

Coon amplitudes describe an infinite spectrum of particles with mass-squared $m_n^2$ and spins~$j_n$ with ${0 \leq j_n \leq n}$ for integer ${n \geq 0}$ (as well as particles in more general representations of the Lorentz group like the anti-symmetric tensor $B_{\mu\nu}$ of string theory). The Coon spectrum has three adjustable parameters $\mu^2$, $\delta$, and $q$ related to the masses by
\begin{align}
\label{eq:spectrum}
    m_n^2
    &=
    \mu^2
    \bigg(
    \frac{1-q^n}{1-q^{\phantom{n}}}
    + \delta
    \bigg)
    =
    \mu^2 [n]_q + m_0^2
    \, ,
\end{align}
where $m_0^2 = \mu^2 \delta$ is the lowest mass-squared and $[n]_q$ are the $q$-deformed integers,
\begin{align}
    [n]_q 
    =
    \frac{1-q^n}{1-q^{\phantom{n}}}
    =
    1 + q + \cdots + q^{n-1}
    \, ,
\end{align}
which become the usual integers in the limit $q \to 1$. The dimensionful parameter $\mu^2$ sets the mass scale of the theory and can be arbitrarily adjusted through a choice of units. The dimensionless parameter $\delta$ (or equivalently the dimensionful $m_0^2$) is physical and sets the Regge intercept of the spectrum. Unitarity constrains ${-1 \leq \delta \leq \frac{1}{3}}$~\cite{Figueroa:2022onw,Cheung:2022mkw}. The dimensionless deformation parameter $q$ is taken to be non-negative. When ${q < 1}$, the spectrum has a finite accumulation point ${m_\infty^2 = (1-q)^{-1} \mu^2 + m_0^2}$. When ${q \geq 1}$, the spectrum is unbounded. Unitarity constrains ${q \leq 1}$~\cite{Figueroa:2022onw, Geiser:2022icl, Bhardwaj:2022lbz, Chakravarty:2022vrp}. At $q = 1$, the Coon spectrum reproduces the spectrum ${m_n^2 = \mu^2 (n+\delta)}$ of string theory. The deformation away from the string spectrum may be understood using the mathematical theory of $q$-deformations or $q$-analogs~\cite{gasper_rahman_2004}. Along these lines, the four-point Coon amplitude can be understood as a simple $q$-deformation of the Veneziano amplitude~\cite{Geiser:2022icl}.

There are some inconsistencies in the literature with the deformation parameter~$q$. The~$q$ defined here (and in all of the modern literature, e.g.~\cite{Figueroa:2022onw, Geiser:2022exp, Geiser:2022icl}) is actually $q^{-1}$ in much of the early literature (e.g.~\cite{Coon:1969yw, Baker:1970vxk, Romans:1988qs, Romans:1989di}). We have defined $q$ so that the Coon spectrum is given by~\eqref{eq:spectrum} for all values of $q \geq 0$. In Coon's original work, there are two distinct four-point Coon amplitudes, each defined for $0 \leq q \leq 1$, which match at $q=1$~\cite{Baker:1976en}. In~\cite{Geiser:2022icl}, it was shown that these two Coon amplitudes can be written as a single function defined for all ${q \geq 0}$ by taking one of the two amplitudes and replacing ${q \mapsto q^{-1}}$.

In this paper, we only consider the scattering of scalars in the lowest mass-level $m_0^2$ of the Coon spectrum in $d \geq 3$ spacetime dimensions. In general, these scalars may be massless, massive, or tachyonic. We further assume that these scalars transform in the adjoint representation of some global flavor symmetry group $G$ so that the full $N$-point Coon amplitudes are given by sums of ordered $N$-point partial amplitudes multiplied by single-trace Chan-Paton factors, just as in tree-level open string \mbox{theory~\cite{Green:1987sp, Polchinski:1998rq}}. In string theory, multi-trace Chan-Paton factors only appear at loop level. Since Coon amplitudes become tree-level open string amplitudes in the limit ${q \to 1}$, we can only have single traces, regardless of the group $G$. This decomposition into partial amplitudes and single-trace structures is also shared by tree-level gluon amplitudes in Yang-Mills theory~\cite{Dixon:1996wi, Dixon:2013uaa, Cheung:2017pzi, Elvang:2013cua}.

Hence, the Coon amplitude for $N$ scalars of mass-squared $m_0^2$ transforming in the adjoint representation of the group $G$ is given by
\begin{align}
    \cA^{(N)}_{q; \, G}
    =
    \sum_{\sigma \in S_N / \bbZ_N }
    \Tr
    (
    T^{a_{\sigma(1)}} T^{a_{\sigma(2)}} \cdots T^{a_{\sigma(N)}}
    )
    \,
    \cA^{(N)}_{q}
    (
    \sigma(1), \sigma(2), \hdots, \sigma(N)
    )
    \, ,
\end{align}
where the sum is over elements $\sigma$ of the permutation group $S_N$ modulo the group of cyclic permutations $\bbZ_N$ acting on the labels $(1, 2, \dots, N)$ (or equivalently over elements of the permutation group $S_{N-1}$ acting on the labels $(2, \dots, N)$ with the first label fixed). The matrices $T^a$ are the generators of the group $G$ in the defining representation normalized by $\Tr(T^a T^b) = \delta^{ab}$, and $a, b, \dots$ are adjoint indices. The partial amplitudes $\cA^{(N)}_{q}$ depend on a cyclic ordering of the labels $(1, 2, \dots, N)$. 

For $q > 1$, these $N$-point partial amplitudes with the canonical ordering $(1,2,\hdots,N)$ are given by the Baker-Coon-Romans formula, which we introduce in~\autoref{sec:BCR}. For $q=1$, the $N$-point partial amplitudes may be computed using worldsheet perturbative string theory. For $q<1$, there is no general formula or method to calculate the $N$-point partial Coon amplitudes. We discuss this subtlety in~\autoref{sec:q<1}.

\subsection{The \texorpdfstring{$q = \infty$}{q = infinity} field theory} 

The limits $q \to \infty$ and $q \to 0$ can be roughly interpreted as integrating out the infinite tower of excited states in the Coon spectrum, leaving only the lightest scalar with mass-squared~$m_0^2$. The resulting theories should then be field theories described by a Lagrangian.

For the unitary group $G = \mathrm{U}(N_F)$ (with arbitrary $N_F \geq 2$), the field theory at $q = \infty$ is given by the following Lagrangian,
\begin{align}
\label{eq:LagAS1}
    \cL_{\text{AS}}
    =
    - \frac{1}{2} \Tr \p_\mu \phi \, \p^\mu \phi
    - \frac{1}{2} \, m_0^2 \Tr \phi^2
    - \sum_{n \geq 3}
    \frac{1}{n} \,
    \tilde{g}^{n-2} \,
    \Lambda^{n+d-\frac{1}{2}nd} 
    \Tr \phi^n
    \, ,
\end{align}
where the scalar field $\phi = T^a \phi^a$ transforms in the adjoint representation of the global flavor symmetry group $\mathrm{U}(N_F)$. The coupling constant $\tilde{g}$ is dimensionless, and $\Lambda$ is a mass scale. This Lagrangian can be resummed as follows,
\begin{align}
    \cL_{\text{AS}}
    =
    - \frac{1}{2} \Tr \p_\mu \phi \, \p^\mu \phi
    - \frac{1}{2} (m_0^2 - \Lambda^2) \Tr \phi^2
    &
    + \frac{\Lambda^d}{\tilde{g}^{\phantom{2}}}
        \Tr \phi / \Lambda^{\frac{d-2}{2}}
\no \\
    &
    + \frac{\Lambda^d}{\tilde{g}^2}
        \Tr \ln (1 - \tilde{g} \, \phi / \Lambda^{\frac{d-2}{2}} )
    \, ,
\end{align}
and is amenable to a semiclassical analysis. In \autoref{sec:app2} we demonstrate that the field theory defined by this Lagrangian has a stable vacuum (at least classically) and is thus a viable theory in its own right.

The $N$-point tree-level amplitudes of this adjoint scalar theory are exactly equal to the $N$-point Baker-Coon-Romans amplitudes in the limit $q \to \infty$. That is,
\begin{align}
\label{eq:AqAAS1}
    \lim_{q \to \infty} 
    \cA^{(N)}_{q; \, \mathrm{U}(N_F)}
    &=
    \cA^{(N) \, \text{tree}}
       _{\text{AS}; \, \mathrm{U}(N_F)}
    \, .
\end{align}
For the special unitary group $G = \mathrm{SU}(N_F)$, a similar equality holds at large $N_F$,
\begin{align}
\label{eq:AqAAS2}
    \lim_{q \to \infty} 
    \cA^{(N)}_{q; \, \mathrm{SU}(N_F)}
    &=
    \cA^{(N) \, \text{tree}}
       _{\text{AS}; \, \mathrm{SU}(N_F)}
    \Big(
    1 + \cO \big( N_F^{-1} \big)
    \Big)
    \, .
\end{align}
We prove these equalities in~\autoref{sec:AS}. Then in~\autoref{sec:disc}, we provide evidence for a conjecture about a similar field theory at $q = 0$.

\subsection{Outline}

This paper is organized as follows. In~\autoref{sec:N}, we review the basic details of $N$-point scattering amplitudes. (Some technical details are included in~\autoref{sec:app}.) In~\autoref{sec:BCR}, we introduce the $N$-point Baker-Coon-Romans partial amplitudes, review their properties, and compute the ${q \to \infty}$ limit. In~\autoref{sec:q<1}, we discuss the subtleties involved with extending the Baker-Coon-Romans formula to ${q < 1}$. In~\autoref{sec:AS}, we compute the amplitudes of the adjoint scalar theory defined in~\eqref{eq:LagAS1}, and we prove the equalities~\eqref{eq:AqAAS1} and~\eqref{eq:AqAAS2}. Finally, in~\autoref{sec:disc}, we discuss some open problems and present a conjecture for the field theory at ${q=0}$.  In \autoref{sec:app2} we perform a brief classical analysis of the adjoint scalar theory defined in~\eqref{eq:LagAS1}. 


\section{\texorpdfstring{$N$}{N}-point scattering}
\label{sec:N}

In this section, we review the basic details of $N$-point scattering amplitudes relevant for our subsequent discussion of the $N$-point Baker-Coon-Romans amplitude. Some useful reviews of scattering amplitudes with far more detail include~\cite{Dixon:1996wi, Dixon:2013uaa, Cheung:2017pzi, Elvang:2013cua}.

Our conventions are as follows. We work in arbitrary spacetime dimension $d \geq 3$. We use the mostly-plus signature $\eta_{\mu \nu} = \operatorname{diag}(-1,1,\dots,1)$ so that the on-shell relation is ${p_i^2 = -m_i^2}$. We use all incoming momenta so that the statement of momentum conservation for $N$-point scattering is simply ${p_1 + p_2 + \cdots + p_N = 0}$.

\subsection{Kinematics}

The kinematic invariants or \textit{Mandelstam variables} for $N$-point scattering are defined by
\begin{align}
    s_{i_1 i_2 \cdots i_n}
    =
    - ( p_{i_1} + p_{i_2} + \cdots + p_{i_n} )^2
    \, ,
\end{align}
with $1 \leq i_\ell \leq N$ and each $i_\ell$ unique. These variables may be interrelated using momentum conservation and on-shell relations. For $N$-point scattering, there are $\half N(N-3)$ independent Mandelstam variables.

For the scattering of four identical scalars with mass-squared $m_0^2$, the
Mandelstam variables are written as
\begin{alignat}{3}
    s
    &=
    s_{12}
    =
    s_{34}
    =
    \phantom{-} 4 E^2
    &&
    {} \geq 4 m_0^2
    \, ,
\no \\
    t
    &=
    s_{23}
    =
    s_{14}
    =
    -2 (E^2 - m_0^2) (1-\cos\theta)
    &&
    {} \leq 0
    \, ,
\no \\
    u
    &=
    s_{13}
    =
    s_{24}
    =
    -2 (E^2 - m_0^2) (1+\cos\theta)
    &&
    {} \leq 0
    \, .
\end{alignat}
These three variables satisfy the on-shell relation~${s+t+u=4m_0^2}$, leaving two independent variables. Here~$E$ and~$\theta$ are the center-of-mass energy and scattering angle, respectively. The inequalities refer to the physical scattering regime with real~$s_{ij}$.

\subsection{Planar channels}

Each Mandelstam variable corresponds to a \textit{scattering channel}. A channel is a partition of the external particles into two disjoint sets of two or more particles each. We are primarily concerned with so-called \textit{planar channels} formed from sets of consecutive particles. We denote the planar channel formed from the set of consecutive particles ${\{i,i+1,\dots,j-1,j\}}$ with ${1 \leq i < j \leq N}$ by the ordered pair $(i,j)$. Because of momentum conservation, a channel cannot be distinguished from its complement. Hence,
\begin{align}
    (i,j)
    =
    \{ i, \dots, j \}
    =
    \{ 1, \dots, i-1, j+1, \dots, N\}
    \, ,
\end{align}
and we can restrict $j \leq N-1$ without loss of generality. The set $\cC^{(N)}$ of independent planar channels for $N$-point scattering is given by
\begin{align}
    \cC^{(N)}
    =
    \{
    (i,j)
    \, : \,
    1 \leq i < j \leq N-1
    \, , \,
    (i,j) \neq (1,N-1)
    \}
    \, ,
\end{align}
with $(1,N-1)$ excluded since this does not represent a partition of the particles into two sets of two or more particles each. The set of planar channels has order
\begin{align}
    |\cC^{(N)}|
    =
    \eqmakebox[00][c]{$\displaystyle\sum_{i=1   \mathstrut}^{N-2} \,$}
    \eqmakebox[00][c]{$\displaystyle\sum_{j=i+1 \mathstrut}^{N-1} \,$}
    ( 1 - \delta_{(i,j),(1,N-1)} )
    =
    \frac{1}{2} N(N-3)
    \, ,
\end{align}
which is equal to the number of independent Mandelstam variables. The Mandelstam variable corresponding to the planar channel $(i,j)$ will be denoted by
\begin{align}
\label{eq:CN}
    s_{(i,j)}
    =
    s_{i \cdots j}
    =
    - ( p_{i} + \cdots + p_{j} )^2
    \, .
\end{align}
The corresponding set of independent Mandelstam variables is given by
\begin{align}
    \cS^{(N)}
    =
    \{ s_{(i,j)} : (i,j) \in \cC^{(N)} \}
    \, .
\end{align}
For ${N = 3,4,5,6}$ (the cases which we explicitly consider in this paper), we have
\begin{align}
    \cS^{(3)}
    &=
    \emptyset
    \, ,
    \vphantom{\{s_{(}}
\no \\[1.5ex]
    \cS^{(4)}
    &=
    \{
    s,
    t
    \}
    \vphantom{\{s_{(}}
\no \\
    &=
    \{
    s_{(1,2)},
    s_{(2,3)}
    \}
    \, ,
\no \\[1.5ex]
    \cS^{(5)}
    &=
    \{
    s_{12},
    s_{23},
    s_{34},
    s_{45},
    s_{51}
    \}
    \vphantom{\{s_{(}}
\no \\
    &=
    \{
    s_{(1,2)},
    s_{(1,3)},
    s_{(2,3)},
    s_{(2,4)},
    s_{(3,4)}
    \}
    \, ,
\no \\[1.5ex]
    \cS^{(6)}
    &=
    \{
    s_{12},
    s_{23},
    s_{34},
    s_{45},
    s_{56},
    s_{61},
    s_{123},
    s_{234},
    s_{345}
    \}
    \vphantom{\{s_{(}}
\no \\
    &=
    \{
    s_{(1,2)},
    s_{(1,3)},
    s_{(1,4)},
    s_{(2,3)},
    s_{(2,4)},
    s_{(2,5)},
    s_{(3,4)},
    s_{(3,5)},
    s_{(4,5)}
    \}
    \, ,
\end{align}
where we have presented each set in both the ``$i \cdots j$" notation and the ``$(i,j)$" notation. The  $N=3$ set is empty because there is no kinematic freedom at three points.

\subsection{Overlapping and non-overlapping channels}

Any pair of channels are either \textit{overlapping} or \textit{non-overlapping}. Two channels are non-overlapping if one channel contains or is disjoint from the other. Otherwise they are overlapping. At five points for example, the planar channel $(1,2) = \{1,2\} = \{3,4,5\}$ is overlapping with
\begin{align}
    (2,3) &= \{2,3\} = \{1,4,5\}
    \quad \text{and} \quad
    (2,4) = \{2,3,4\} = \{1,5\}
\intertext{and non-overlapping with}
    (1,3) &= \{1,2,3\} = \{4,5\}
    \quad \text{and} \quad
    (3,4) = \{3,4\} = \{1,2,5\}
    \, .
\end{align}
In this example, we have written each channel using both the ordered pair notation and the set notation to be as clear as possible.

Given these definitions, we may define the following sets whose elements are subsets of the full set of planar channels $\cC^{(N)}$:
\begin{itemize}
\item \makebox[\widthof{$\cN^{(N)}_{[n]}$}][r]{$\cO^{(N)}$}, the set of pairs of overlapping planar channels
\item $\cN^{(N)}_{[n]}$, the set of $n$-tuples (with $n \geq 2$) of mutually non-overlapping planar channels
\end{itemize}
For later convenience, we also define
\begin{align}
    \cN^{(N)}_{[1]}
    =
    \big\{
    \{ (i,j) \}
    \, : \,
    (i,j) \in  \cC^{(N)}
    \big\}
    \,.
\end{align}
Although the preceding sentences implicitly define these sets, it will be useful to have explicit expressions which list their elements without double-counting. We derive these expressions in~\autoref{sec:app}. The reader, however, should not be concerned with the expressions themselves. In~\autoref{sec:app}, we also derive two useful results. First, the set of overlapping pairs of planar channels has order
\begin{align}
    |\cO^{(N)}|
    =
    \frac{1}{24}N(N-1)(N-2)(N-3)
    \, .
\end{align}
Second, the sets \smash{$\cN^{(N)}_{[n]}$} with $n \geq N-2$ are empty. In other words, there are at most $N-3$ mutually non-overlapping planar channels in $N$-point scattering.


\section{The Baker-Coon-Romans formula}
\label{sec:BCR}

In this section, we review the general properties of the Baker-Coon-Romans $N$-point partial amplitude~\cite{Baker:1970vxk, Romans:1988qs} and compute its ${q \to \infty}$ limit. In particular, we review the amplitude's convergence properties, its spectrum, its duality properties, and its factorization properties.

\subsection{\texorpdfstring{$N$}{N}-point Baker-Coon-Romans partial amplitudes}

Despite fitting on just two lines, the Baker-Coon-Romans formula contains a lot of physics. The $N$-point Baker-Coon-Romans partial amplitude for the canonical ordering $(1,2,\dots,N)$ is given by
\begin{align}
\label{eq:BCRN}
    \cA^{(N)}_q
    (1, 2, \dots, N)
    &=
    - g^{N-2}
    \mu^{N+d-\frac{1}{2}Nd}
    \big[
    (1-q) (q^{-1};q^{-1})_\infty
    \big]^{N-3}
\no \\
    & \quad
    \times
    \sum_{n_I \geq 0 \, , \, I \in \cC^{(N)} \mathstrut}
    \bigg\{
    \prod_{J \in \cC^{(N)} \mathstrut}
    \frac{ a_J^{n_J} }{ (q^{-1};q^{-1})_{n_J} }
    \prod_{\{K,L\} \in \cO^{(N)}}
    q^{-n_K n_L}
    \bigg\}
    \, .
\end{align}
Here $g$ is a dimensionless coupling constant, $\mu$ is the mass scale which appears in the Coon spectrum~\eqref{eq:spectrum}, and ${(x;q)_n = \prod_{\ell=0}^{n-1} (1-x q^\ell)}$ is the $q$-Pochhammer symbol. Our normalization, i.e.\ the first line of~\eqref{eq:BCRN}, differs from the previous literature~\cite{Baker:1970vxk, Romans:1988qs} but is chosen to facilitate various limits. The uppercase Latin letters ${I,J,K,L \in \cC^{(N)}}$ denote planar scattering channels, and $\cO^{(N)}$ is the set of overlapping pairs of planar channels defined in the previous section. Finally, $a_I$ is a dimensionless affine transformation of the Mandelstam variable $s_I$ defined by
\begin{align}
\label{eq:aI}
    a_I(s_I)
    =
    1 + (q-1) \frac{s_I-m_0^2}{\mu^2}
    \, ,
\end{align}
which satisfies $a_I(m_n^2) = q^n$. At $N$ points, there are ${|\cC^{(N)}| = \frac{1}{2} N(N-3)}$ independent summation variables $n_I$, and the two products within the curly brackets on the second line respectively contribute ${|\cC^{(N)}| = \frac{1}{2} N(N-3)}$ and ${|\cO^{(N)}| = \frac{1}{24}N(N-1)(N-2)(N-3)}$ factors to the summand. We now examine~\eqref{eq:BCRN} for the first few values of $N$.

\subsubsection{$N=3$}

The only consistent three-point amplitude is a constant~\cite{Dixon:1996wi, Dixon:2013uaa, Cheung:2017pzi, Elvang:2013cua}. At three points, there are no independent Mandelstam variables, and ${\cC^{(3)} = \cO^{(3)} = \emptyset}$. Thus, only the prefactor on the first line of~\eqref{eq:BCRN} contributes, and the three-point Baker-Coon-Romans partial amplitude is simply given by a constant,
\begin{align}
\label{eq:BCR3}
    \cA^{(3)}_q
    (1,2,3)
    &=
    - g
    \mu^{3-d/2}
    \, ,
\end{align}
as expected. This amplitude is independent of $q$ and exactly matches the three-point partial amplitude computed from an adjoint scalar field theory with a cubic interaction term $\cL_{\text{AS}} \supset -\frac{1}{3} g \mu^{3-d/2} \Tr \phi^3$.

\subsubsection{$N=4$}

At four points, there are two independent planar channels $(1,2)$ and $(2,3)$ and one pair of overlapping planar channels $\{(1,2),(2,3)\}$. Thus, the second line of~\eqref{eq:BCRN} becomes a double sum with ${2+1=3}$ total factors in its summand. The four-point Baker-Coon-Romans partial amplitude is then given by
\begin{align}
\label{eq:BCR4}
    \cA^{(4)}_q
    (1,2,3,4)
    &=
    - g^2
    \mu^{4-d}
    (1-q) (q^{-1};q^{-1})_\infty
\no \\
    & \quad
    \times
    \sum_{n_{(1,2)} \geq 0} \,\,
    \sum_{n_{(2,3)} \geq 0} \,\,
    \frac{ a_{(1,2)}^{n_{(1,2)}} }{ (q^{-1};q^{-1})_{n_{(1,2)}} }
    \frac{ a_{(2,3)}^{n_{(2,3)}} }{ (q^{-1};q^{-1})_{n_{(2,3)}} }
    q^{-n_{(1,2)} n_{(2,3)}}
    \, ,
\end{align}
with $a_I$ defined in~\eqref{eq:aI}. We analyze this expression in more detail in~\autoref{sec:q<1}.

\subsubsection{$N=5$}

At five points, there are five planar channels $\{(1,2),(1,3),(2,3),(2,4),(2,5)\}$ and five pairs of overlapping planar channels. Thus, the second line of~\eqref{eq:BCRN} contributes a sum over five variables with ${5+5=10}$ total factors in the summand. The five-point Baker-Coon-Romans partial amplitude is then given by
\begin{align}
\label{eq:BCR5}
    \cA^{(5)}_q
    (1,2,3,4,5)
    &=
    - g^3
    \mu^{5-3d/2}
    \big[
    (1-q) (q^{-1};q^{-1})_\infty
    \big]^2
    \smash[t]{
        \vphantom{
        \frac{ a_{(1,2)}^{n_{(1,2)}} }{ (q^{-1};q^{-1})_{n_{(1,2)}} }}}
\no \\
    & \quad
    \times
    \sum_{n_{(1,2)} \geq 0} \,\,
    \sum_{n_{(1,3)} \geq 0} \,\,
    \sum_{n_{(2,3)} \geq 0} \,\,
    \sum_{n_{(2,4)} \geq 0} \,\,
    \sum_{n_{(3,4)} \geq 0} 
\no \\
    & \quad
    \times
    \frac{ a_{(1,2)}^{n_{(1,2)}} }{ (q^{-1};q^{-1})_{n_{(1,2)}} }
    \frac{ a_{(1,3)}^{n_{(1,2)}} }{ (q^{-1};q^{-1})_{n_{(1,3)}} }
    \frac{ a_{(2,3)}^{n_{(1,2)}} }{ (q^{-1};q^{-1})_{n_{(2,3)}} }
\no \\
    & \quad
    \times
    \frac{ a_{(2,4)}^{n_{(1,2)}} }{ (q^{-1};q^{-1})_{n_{(2,4)}} }
    \frac{ a_{(3,4)}^{n_{(1,2)}} }{ (q^{-1};q^{-1})_{n_{(3,4)}} }
\no \\
    & \quad
    \smash[b]{
        \vphantom{
        \frac{ a_{(1,2)}^{n_{(1,2)}} }{ (q^{-1};q^{-1})_{n_{(1,2)}} }}}
    \times
    q^{ - n_{(1,2)} n_{(2,3)}
        - n_{(2,3)} n_{(3,4)}
        - n_{(3,4)} n_{(1,3)} 
        - n_{(1,3)} n_{(2,4)} 
        - n_{(2,4)} n_{(1,2)} }
    \, ,
\end{align}
with $a_I$ defined in~\eqref{eq:aI}. We analyze this expression in more detail in~\autoref{sec:q<1}.

\subsubsection{$N=6$}

At six points, there are nine planar channels and fifteen pairs of overlapping planar channels. Thus, the second line of~\eqref{eq:BCRN} contributes a sum over nine variables with ${9+15=24}$ total factors in the summand. The explicit expression for the six-point Baker-Coon-Romans partial amplitude is quite large, so we omit it and now return to the general case. 

\subsection{Convergence}

The convergence properties of the Baker-Coon-Romans formula were first discussed in the original works~\cite{Baker:1970vxk, Romans:1988qs}. Here we review the convergence properties of the general $N$-point Baker-Coon-Romans formula~\eqref{eq:BCRN} using the ratio test. (It is also worth noting that the $q$-Pochhammer symbols on the first line of~\eqref{eq:BCRN} only converge for $q \geq 1$.) The summation over each variable $n_I$ converges if ${q > 1}$ and ${|a_I| < 1}$. For fixed ${q > 1}$, the latter condition is satisfied within an open set on the complex $s_I$ plane. Outside of this region, the Baker-Coon-Romans partial amplitude~\eqref{eq:BCRN} is defined by an analytic continuation in the Mandelstam variables $s_I$ (excluding of course the infinite sequence of simple poles in the complex $s_I$ plane at ${s_I = m_n^2}$). This process of analytic continuation in the kinematic variables is familiar to string theory amplitudes~\cite{Witten:2013pra}. For example, the four-point tree-level open string amplitude (i.e.\ the Veneziano amplitude) is most properly defined from a worldsheet disk integral with imaginary values of $s$ and $t$. The integral evaluates to Euler's beta function and can then be analytically continued to physical real-valued $s$ and $t$. Similarly, the four-point and five-point Baker-Coon-Romans amplitudes can be written in terms of known special functions, and these special functions provide the analytic continuation to physical values of the Mandelstam variables~\cite{Romans:1988qs}. We discuss these special functions in more detail in~\autoref{sec:q<1}. At higher points we assume that there is no obstruction to a similar analytic continuation in the Mandelstam variables~\cite{Romans:1988qs}. Continuation to $q < 1$ is a different story which we discuss in~\autoref{sec:q<1}.

\subsection{Spectrum}

The spectrum of the Baker-Coon-Romans formula was discussed in~\cite{Baker:1970vxk, Romans:1988qs, Coon:1972te}. Here we briefly review the fact that~\eqref{eq:BCRN} correctly describes the Coon spectrum~\eqref{eq:spectrum}. To do so, we use the $q$-binomial theorem~\cite{gasper_rahman_2004}, which relates infinite series involving $q$-Pochhammer symbols as follows,
\begin{align}
\label{eq:qbinom}
    \frac{ (ab;q)_\infty }
         { (a;q)_\infty (b;q)_\infty }
    =
    \sum_{n = 0}^\infty
    \frac{ a^n }
         { (q;q)_n \, (b q^n;q)_\infty }
    =
    \sum_{n = 0}^\infty
    \frac{ b^n }
         { (q;q)_n \, (a q^n;q)_\infty }
    \, .
\end{align}
The series in~\eqref{eq:qbinom} converge when ${|q|, |a|, |b| < 1}$.

Now, the Coon spectrum should appear as a sequence of simple poles at each $m_n^2$ in any planar scattering channel. We may isolate the contribution to~\eqref{eq:BCRN} from any particular planar channel $I$ as follows,
\begin{align}
    \cA^{(N)}_q
    (1, 2, \dots, N)
    &\propto
    \sum_{n_I \geq 0 }
    \frac{ a_I^{n_I} }{ (q^{-1};q^{-1})_{n_I} }
    \prod_{ J \in \cO^{(N)}_I }
    q^{-n_I n_J}
    =
     \sum_{n_I \geq 0 }
    \frac{ \big( a_I q^{-\tilde{n}_I} \big)^{n_I} }
         { (q^{-1};q^{-1})_{n_I} }
    \, .
\end{align}
Here \smash{$\cO^{(N)}_I$} is the set of planar channels overlapping with $I$, and $\tilde{n}_I$ is a non-negative integer given by the sum of summation variables for the planar channels which overlap with $I$,
\begin{align}
    \tilde{n}_I
    =
    \sum_{J \in \cO^{(N)}_I } n_J
    \geq
    0
    \, .
\end{align}
We can use the $q$-binomial theorem~\eqref{eq:qbinom} with $a = a_I q^{-\tilde{n}_I}$ and $b=0$ to perform the sum over $n_I$ and find
\begin{align}
    \cA^{(N)}_q
    (1, 2, \dots, N)
    &\propto
    \frac{1}{ (a_I q^{-\tilde{n}_I} ; q^{-1})_\infty }
    \propto
    \frac{1}{ \prod_{n=0}^\infty (s_I - m_{n+\tilde{n}_I}^2) } 
    \, ,
\end{align}
which has simple poles at $s_I = m_n^2$ for each integer $n \geq 0$ (since $\tilde{n}_I \geq 0$). The residues of these poles are polynomials in the overlapping Mandelstam variables,
\begin{align}
    \cA^{(N)}_q
    (1, 2, \dots, N)
    \sim
    \frac{1}{s_I - m_n^2}
    \Big(
    \text{degree-}n
    \text{ polynomial in }
    s_J
    \text{ with }
    J \in \cO^{(N)}_I
    \Big)
    \, ,
\end{align}
which indicates the exchange of particles of mass-squared $m_n^2$ and spin $j_n$ with $0 \leq j_n \leq n$ as claimed in~\eqref{eq:spectrum}. When the residue is a polynomial in more than one of the overlapping Mandelstam variables $s_J$, the decomposition in terms of irreducible representations of the Lorentz group also includes representations corresponding to more general Young tableaux. This feature is evident in the spectrum of string theory which has an excitation corresponding to the anti-symmetric tensor $B_{\mu\nu}$ whose representation is not labeled by a spin in the canonical sense.\footnote{We thank L.~Lindwasser for clarifying this subtlety.} These details are discussed in~\cite{Coon:1972te} for the Coon amplitude with $q > 1$ and in~\cite{Fubini:1969qb} for the dual resonance model (i.e.\ string theory) at $q=1$.

\subsection{Duality}

The duality properties of the Baker-Coon-Romans formula were first discussed in Baker, Coon, and Romans' original works~\cite{Baker:1970vxk, Romans:1988qs}. Here we briefly review the fact that $N$-point Baker-Coon-Romans partial amplitude is duality invariant in the sense of dual resonance models~\cite{Cappelli:2012cto}. A more detailed proof is given in~\cite{Romans:1988qs}. Duality invariance refers to invariance under combined cyclic and anti-cyclic permutations of the external particles. For $N$-point scattering, duality transformations are generated by the cyclic shift $\phi$ and the reflection $\theta$,
\begin{align}
    \phi : \,
    &
    (1,2,\dots,N)
    \mapsto
    (2,\dots,N,1)
    \, ,
\no \\
    \theta : \,
    &
    (1,2,\dots,N)
    \mapsto
    (N,\dots,2,1)
    \, ,
\end{align}
which satisfy $\phi^N = \theta^2 = (\phi \, \circ \, \theta)^2 = \mathds{1}$. The Baker-Coon-Romans formula~\eqref{eq:BCRN} only depends on the external particle labels through the sets of planar channels $\cC^{(N)}$ and overlapping planar channels $\cO^{(N)}$. Both $\cC^{(N)}$ and $\cO^{(N)}$ are invariant under duality transformations. One can explicitly verify this fact using the formulae for these sets in~\autoref{sec:app}. Thus, we conclude that the Baker-Coon-Romans partial amplitude~\eqref{eq:BCRN} is invariant under any duality transformation.

\subsection{Factorization}

The factorization properties of the Baker-Coon-Romans formula were discussed in detail in~\cite{Baker:1970vxk, Romans:1988qs, Coon:1972te, Yu:1972fz, Baker:1972zv}. Here we briefly review the high-level results. Factorization refers to the property that tree-level amplitudes factorize into products of lower-point amplitudes on their poles. We can schematically demonstrate the factorization of the Baker-Coon-Romans $N$-point partial amplitude~\eqref{eq:BCRN} on any given planar channel using the $q$-binomial theorem. Because~\eqref{eq:BCRN} is invariant under duality transformations, it is sufficient to consider only the planar channels $(1,M)$ with ${M=2,3,\dots,\floor{N/2}}$. Schematically, the Baker-Coon-Romans partial amplitudes should obey,
\begin{align}
    \cA^{(N)}_q
    (1,2,\dots,N)
    &\sim
    \sum_{n, \ell}
    \tilde{\cA}^{(M+1)}_q
    (1,2,\dots,M,X_{n,\ell})
\no \\
    & \quad \phantom{\sum}
    \times
    \frac{ \cN_{n,\ell} }{s_{(1,M)}-m_n^2}
    \times
    \tilde{\cA}^{(N-M+1)}_q
    (-X_{n,\ell},M,M+1,\dots,N)
    \, ,
\end{align}
where the sum is over states (i.e.\ over mass-squared $m_n^2$ and spins $\ell$) in the Coon spectrum~\eqref{eq:spectrum} exchanged in the $(1,M)$ channel. Here $\cN_{n,\ell}$ is the numerator of the propagator for a particle of mass-squared $m_n^2$ and spin $\ell$. The $(M+1)$-point and $(N-M+1)$-point Coon amplitudes $\tilde{\cA}_q$ respectively describe the scattering of $M$ and $N-M$ massless scalars with one massive spinning state (labeled by $X_{n,\ell}$). Using the $q$-binomial theorem~\eqref{eq:qbinom} to perform the sum over $n_{(1,M)}$ in~\eqref{eq:BCRN} produces exactly this factorization. In fact, the Baker-Coon-Romans formula itself exhibits the full factorization of an $N$-point tree-level amplitude into a product of three-point amplitudes. For more details, we refer the reader to~\cite{Coon:1972te, Yu:1972fz, Baker:1972zv, Romans:1988qs}. Of particular interest is~\cite{Yu:1972fz, Baker:1972zv}, where thrice-iterated factorization was used to derive the most general three-point Coon amplitude for any three massive spinning states.

\subsection{\texorpdfstring{$q \to \infty$}{q to infinity}}

Although the $q \to \infty$ limit has been computed for some lower point Coon amplitudes~\cite{Romans:1988qs}, the general result is yet unknown. For the first time, we compute the $q \to \infty$ limit of the $N$-point Baker-Coon-Romans partial amplitude~\eqref{eq:BCRN}.

This limit only makes sense if we simultaneously take $\mu \to \infty$ with both $m_0^2$ and a new scale $\Lambda^2 = q^{-1} \mu^2$ fixed so that the dimensionless quantities $a_I$ defined in~\eqref{eq:aI} remain finite. This combined limit of $q \to \infty$ and $\mu \to \infty$ sends $m_n^2 \to \infty$ for all $n \geq 1$ which effectively integrates out all the particles in the Coon spectrum except for the lowest-lying scalars with mass-squared $m_0^2$. From this perspective, we expect the resulting amplitudes to correspond to a field theory describing the self-interactions of these scalars. We also define a new dimensionless coupling constant $\tilde{g} = q^{(6-d)/4} g$ which we keep fixed in the limit $q \to \infty$ so that the combination ${\tilde{g} \Lambda^{3-d/2} = g \mu^{3-d/2}}$ (and thus the three-point Coon amplitude~\eqref{eq:BCR3}) remains finite. For $d < 6$, this limit corresponds to a weak coupling limit in $g$. For $d > 6$, this limit corresponds to a strong coupling limit in~$g$.

In summary, we are considering the limit
\begin{align}
    q \to \infty
    \text{ with fixed }
    m_0^2,
    \text{ }
    \Lambda^2 = q^{-1} \mu^2,
    \text{ and }
    \tilde{g} = q^{(6-d)/4} g
    \, .
\end{align}
We can take this limit directly from the Baker-Coon-Romans formula~\eqref{eq:BCRN}. We find
\begin{align}
\label{eq:Aqlim}
    \smash[b]{
    \lim_{q \to \infty}
    \cA^{(N)}_q
    (1, 2, \dots, N)
    }
    &=
    (-)^N \tilde{g}^{N-2}
    \Lambda^{N+d-\frac{1}{2}Nd}
\no \\
    & \quad
    \times
    \sum_{n_I \geq 0 \, , \, I \in \cC^{(N)} \mathstrut}
    \bigg\{
    \prod_{J \in \cC^{(N)} \mathstrut}
    \tilde{a}_J^{n_J}
    \prod_{\{K,L\} \in \cO^{(N)}}
    \delta_{n_K n_L , 0}
    \bigg\}
    \, ,
\end{align}
where we have defined ${\tilde{a}_I = \lim_{q \to \infty} a_I = 1+(s_I-m_0^2)/\Lambda^2}$. We recall that uppercase Latin letters ${I,J,K,L \in \cC^{(N)}}$ denote planar scattering channels, and $\cO^{(N)}$ is the set of overlapping pairs of planar channels defined in the previous section. We have replaced the factors of $q^{-n_K n_L}$ with the Kronecker deltas $\delta_{n_K n_L , 0}$ since ${\lim_{q \to \infty} q^{-n_K n_L} = 0}$ vanishes unless ${n_K n_L = 0}$. We can then replace these deltas with Kronecker deltas of a single summation variable using ${\delta_{n_K n_L , 0} = \delta_{n_K,0} + \delta_{n_L,0} - \delta_{n_K,0} \delta_{n_L,0}}$. If we ignore the Kronecker deltas, the remaining sums are all geometric series,
\begin{align}
    \sum_{n_I \geq 0}
    \tilde{a}_I^{n_I}
    =
    \frac{1}{1-\tilde{a}_I}
    =
    -\frac{\Lambda^2}{s_I-m_0^2}
    =
    -\frac{1}{\mfs_I}
    \, ,
\end{align}
where we have defined the dimensionless Mandelstam variable $\mfs_{I} = (s_I - m_0^2)/\Lambda^2$. The Kronecker deltas, however, complicate things. Before tackling the general case, we explicitly examine the first few values of $N$.

\subsubsection{$N=4$}

For $N=4$, the expression~\eqref{eq:Aqlim} becomes
\begin{align}
    \lim_{q \to \infty}
    \cA^{(4)}_q
    (1, 2, 3, 4)
    =
    \tilde{g}^{2}
    \Lambda^{4-d}
    \sum_{n_{(1,2)} \geq 0} \,\,
    \sum_{n_{(2,3)} \geq 0} \,\,
    &
    \tilde{a}_{(1,2)}^{n_{(1,2)}} \,\,
    \tilde{a}_{(2,3)}^{n_{(2,3)}}
\\ \no
    &
    \times
    \big(
      \delta_{n_{(1,2)},0}
    + \delta_{n_{(2,3)},0}
    - \delta_{n_{(1,2)},0} \delta_{n_{(2,3)},0}
    \big)
    \, .
\end{align}
Performing the sums is straightforward and yields
\begin{align}
\label{eq:Aqinf4}
    \lim_{q \to \infty}
    \cA^{(4)}_q
    (1, 2, 3, 4)
    &=
    - \tilde{g}^2
    \Lambda^{4-d}
    \bigg(
      \frac{\Lambda^2}{s-m_0^2}
    + \frac{\Lambda^2}{t-m_0^2}
    + 1
    \bigg)
    \, ,
\end{align}
which is precisely the four-point partial amplitude computed from an adjoint scalar theory with both cubic and quarter interaction terms. We revisit this observation in~\autoref{sec:AS}.

\subsubsection{$N=5$}

For $N=5$, the calculation is more involved. In this case, the expression~\eqref{eq:Aqlim} becomes
\begin{align}
    \lim_{q \to \infty}
    \cA^{(5)}_q
    (1, 2, 3, 4, 5)
    =
    - \tilde{g}^3
    \Lambda^{5-3d/2}
    &
    \sum_{n_{(1,2)} \geq 0} \,\,
    \sum_{n_{(1,3)} \geq 0} \,\,
    \sum_{n_{(2,3)} \geq 0} \,\,
    \sum_{n_{(2,4)} \geq 0} \,\,
    \sum_{n_{(3,4)} \geq 0} 
\no \\[1ex]
    &\quad
    \times
    \tilde{a}_{(1,2)}^{n_{(1,2)}} \,\,
    \tilde{a}_{(1,3)}^{n_{(1,3)}} \,\,
    \tilde{a}_{(2,3)}^{n_{(2,3)}} \,\,
    \tilde{a}_{(2,4)}^{n_{(2,4)}} \,\,
    \tilde{a}_{(3,4)}^{n_{(3,4)}}
\no \\
    &\quad
    \vphantom{a_{\mathstrut}^{\mathstrut}}
    \times
    \big(
      \delta_{n_{(1,2)},0}
    + \delta_{n_{(2,3)},0}
    - \delta_{n_{(1,2)},0} \delta_{n_{(2,3)},0}
    \big)
\no \\
    &\quad
    \vphantom{a_{\mathstrut}^{\mathstrut}}
    \times
    \big(
      \delta_{n_{(2,3)},0}
    + \delta_{n_{(3,4)},0}
    - \delta_{n_{(2,3)},0} \delta_{n_{(3,4)},0}
    \big)
\no \\
    &\quad
    \vphantom{a_{\mathstrut}^{\mathstrut}}
    \times
    \big(
      \delta_{n_{(3,4)},0}
    + \delta_{n_{(1,3)},0}
    - \delta_{n_{(3,4)},0} \delta_{n_{(1,3)},0}
    \big)
\no \\
    &\quad
    \vphantom{a_{\mathstrut}^{\mathstrut}}
    \times
    \big(
      \delta_{n_{(1,3)},0}
    + \delta_{n_{(2,4)},0}
    - \delta_{n_{(1,3)},0} \delta_{n_{(2,4)},0}
    \big)
\no \\
    &\quad
    \vphantom{a_{\mathstrut}^{\mathstrut}}
    \times
    \big(
      \delta_{n_{(2,4)},0}
    + \delta_{n_{(1,2)},0}
    - \delta_{n_{(2,4)},0} \delta_{n_{(1,2)},0}
    \big)
    \, ,
\end{align}
We can use $\delta_{n,0}^2 = \delta_{n,0}$ to simplify the string of Kronecker deltas. After this simplification, it is straightforward to perform the sums, and we find
\begin{alignat}{2}
\label{eq:Aqinf5}
    \lim_{q \to \infty}
    \cA^{(5)}_q(1, 2, 3, 4, 5)
    =
    - \tilde{g}^3
    \Lambda^{5-3d/2}
    \Bigg(
      \frac{1}{ \mfs_{12} \mfs_{34} }
    &
    + \frac{1}{ \mfs_{23} \mfs_{45} }
    + \frac{1}{ \mfs_{34} \mfs_{51} }
    + \frac{1}{ \mfs_{45} \mfs_{12} }
    + \frac{1}{ \mfs_{51} \mfs_{23} }    
\no \\
    &
    + \frac{1}{ \mfs_{12} }
    + \frac{1}{ \mfs_{23} }
    + \frac{1}{ \mfs_{34} }
    + \frac{1}{ \mfs_{45} }
    + \frac{1}{ \mfs_{51} }
    + 1
    \Bigg)
    \, ,
\end{alignat}
where $\mfs_{I} = (s_I - m_0^2)/\Lambda^2$. This expression is precisely a five-point partial amplitude computed from an adjoint scalar theory with cubic, quartic, and quintic interaction terms. Again, we revisit this observation in~\autoref{sec:AS}.

\subsubsection{$N=6$}

For $N=6$, we again begin with~\eqref{eq:Aqlim}. We first use the set of overlapping pairs of planar channels $\cO^{(6)}$ tabulated in~\autoref{sec:app} to explicitly write out the string of Kronecker deltas. We then use $\delta_{n,0}^2 = \delta_{n,0}$ to simplify the string. Finally, we perform the remaining sums and find
\begin{align}
\label{eq:Aqinf6}
    \lim_{q \to \infty}
    \cA^{(6)}_q
    (1,2,3,4,5,6)
    =
    - \tilde{g}^4
    \Lambda^{6-2d}
    \Bigg(
    &
    \eqmakebox[1][c]{$\displaystyle \frac{1}{ \mfs_{12} \mfs_{34} \mfs_{345} }$}
    + ( 5 \text{ cyclic perms.} )
\no \\
    \vphantom{\Bigg(}
    {}+{}
    &
    \eqmakebox[1][c]{$\displaystyle \frac{1}{ \mfs_{12} \mfs_{45} \mfs_{345} }$}
    + ( 5 \text{ cyclic perms.} )
\no \\
    \vphantom{\Bigg(}
    {}+{}
    &
    \eqmakebox[1][c]{$\displaystyle \frac{1}{ \mfs_{12} \mfs_{34} \mfs_{56} }$}
    + ( 1 \text{ cyclic perm.} )
\no \\
    \vphantom{\Bigg(}
    {}+{}
    &
    \eqmakebox[2][c]{$\displaystyle \frac{1}{ \mfs_{12} \mfs_{123} }$}
    + ( 5 \text{ cyclic perms.} )
\no \\
    \vphantom{\Bigg(}
    {}+{}
    &
    \eqmakebox[2][c]{$\displaystyle \frac{1}{ \mfs_{12} \mfs_{345} }$}
    + ( 5 \text{ cyclic perms.} )
\no \\
    \vphantom{\Bigg(}
    {}+{}
    &
    \eqmakebox[2][c]{$\displaystyle \frac{1}{ \mfs_{12} \mfs_{34} }$}
    + ( 5 \text{ cyclic perms.} )
\no \\
    \vphantom{\Bigg(}
    {}+{}
    &
    \eqmakebox[2][c]{$\displaystyle \frac{1}{ \mfs_{12} \mfs_{45} }$}
    + ( 2 \text{ cyclic perms.} )
\no \\
    \vphantom{\Bigg(}
    {}+{}
    &
    \eqmakebox[3][c]{$\displaystyle \frac{1}{ \mfs_{12} }$}
    + ( 5 \text{ cyclic perms.} )
\no \\
    \vphantom{\Bigg(}
    {}+{}
    &
    \eqmakebox[3][c]{$\displaystyle \frac{1}{ \mfs_{123} }$}
    + ( 2 \text{ cyclic perms.} )
    + 1
    \Bigg)
    \, ,
\end{align}
where $\mfs_{I} = (s_I - m_0^2)/\Lambda^2$ and ``$(m \text{ cyclic perms.})$" denotes the $m$ unique terms obtained by cyclically permuting the particles labels of the preceding term. As before, this expression is precisely a six-point partial amplitude computed from an adjoint scalar theory with cubic, quartic, quintic, and now sextic interaction terms. Once more, we revisit this observation in~\autoref{sec:AS}.

\subsubsection{General $N$}

We now return to the case of general $N$. We begin with~\eqref{eq:Aqlim}. The trickiest part of this calculation is making sense of the following product of Kronecker deltas,
\begin{align}
    \Delta^{(N)}
    =
    \prod_{\{K,L\} \in \cO^{(N)}}
    \delta_{n_K n_L , 0}
    \, ,
\end{align}
where the product runs over all unique pairs of overlapping planar channels. The factor $\Delta^{(N)}$ is equal to a sum of products of deltas $\delta_{n_I}$ which vanishes unless the summation variables satisfy $n_K = 0$, $n_L = 0$, or $n_K = n_L = 0$ for every pair of overlapping planar channels $K$ and $L$. In other words, only summation variables corresponding to mutually non-overlapping channels can simultaneously be non-zero. At most $N-3$ planar channels can be mutually non-overlapping in $N$-point scattering, so $\Delta^{(N)}$ is equal to a sum of terms with at least ${\frac{1}{2} N(N-3) - (N-3) = \frac{1}{2}(N-2)(N-3)}$ and at most $\frac{1}{2} N (N-3)$ factors of unique deltas $\delta_{n_I}$. The terms with the fewest $\delta_{n_I}$ (i.e. the terms which can support the largest number $\frac{1}{2}(N-2)(N-3)$ of non-zero $n_I$) are given by
\begin{align}
\label{eq:step1}
    \Delta^{(N)}
    =
    \sum_{X \in \cN^{(N)}_{[N-3]}}
    \prod_{I \in \cC^{(N)} / X \vphantom{\cN^{(N)}_{[N-3]}}}
    \delta_{n_I,0}
    + \cdots
    \, ,
\end{align}
where we have used the fact that elements of the set \smash{$\cN^{(N)}_{[N-3]}$} are $(N-3)$-element subsets of the set of planar channels $\cC^{(N)}$. The string of Kronecker deltas in~\eqref{eq:step1} double counts contributions to the sum with ${\frac{1}{2}(N-2)(N-3) - 1}$ non-zero $n_I$, so we must subtract the appropriate sums of products of deltas with all $(N-4)$-tuples of mutually non-overlapping planar channels excluded,
\begin{align}
\label{eq:step2}
    \Delta^{(N)}
    =
    \sum_{X \in \cN^{(N)}_{[N-3]}}
    \prod_{I \in \cC^{(N)} / X \vphantom{\cN^{(N)}_{[N-3]}}}
    \delta_{n_I,0}
    -
    \sum_{X \in \cN^{(N)}_{[N-4]}}
    \prod_{I \in \cC^{(N)} / X \vphantom{\cN^{(N)}_{[N-4]}}}
    \delta_{n_I,0}
    + \cdots
    \, .
\end{align}
The expression~\eqref{eq:step2} then double counts contributions with ${\frac{1}{2}(N-2)(N-3) - 2}$ non-zero summation variables, so we must add a similar set of terms. This pattern continues until we add or subtract the product of all $\frac{1}{2} N(N-3)$ Kronecker deltas. After the dust settles, we find
\begin{align}
    \Delta^{(N)}
    =
    (-)^{N-1}
    \prod_{I \in \cC^{(N)} \vphantom{\cN^{(N)}_{[n]}}}
    \delta_{n_I,0}
    +
    \sum_{n=2}^{N-3}
    (-)^{N-1-n}
    \sum_{X \in \cN^{(N)}_{[n]}}
    \prod_{I \in \cC^{(N)} / X \vphantom{\cN^{(N)}_{[n]}}}
    \delta_{n_I,0}
    \, ,
\end{align}
which we then use in~\eqref{eq:Aqlim} to find
\begin{align}
\label{eq:AqinfN}
    \lim_{q \to \infty}
    \cA^{(N)}_q(1, 2, \dots, N)
    &=
    - \tilde{g}^{N-3}
    \Lambda^{N+d-\frac{1}{2}Nd}
    \Bigg(
    1
    + \sum_{n=1}^{N-3}
    \sum_{ \{I_1, \dots, I_n\} \in \cN^{(N)}_{[n]}}
    \frac{1}{ \mfs_{I_1} \dots \mfs_{I_n} }
    \Bigg)
    \, ,
\end{align}
where $\mfs_{I} = (s_I - m_0^2)/\Lambda^2$. The expression~\eqref{eq:AqinfN} is our final result for the $q \to \infty$ limit of the $N$-point Baker-Coon-Romans partial amplitude. In~\autoref{sec:AS}, we prove that the adjoint scalar field theory described in~\autoref{sec:intro} has precisely the same tree-level amplitudes.


\section{Taking \texorpdfstring{$q < 1$}{q < 1}}
\label{sec:q<1}

In this section, we discuss the subtleties involved with extending the Baker-Coon-Romans formula~\eqref{eq:BCRN} to ${q < 1}$. Although the Baker-Coon-Romans formula only converges for $q > 1$, there is a straightforward procedure to extend the four-point formula to ${q < 1}$ using the $q$-gamma function. At five points and higher, however, we encounter various difficulties.

\subsection{Four points}

We begin our discussion with the four-point Baker-Coon-Romans formula~\eqref{eq:BCR4}. For now, we assume that ${q > 1}$ so that the sums in~\eqref{eq:BCR4} converge. The sum over $n_{(2,3)}$ can be performed using the $q$-binomial theorem~\eqref{eq:qbinom} with ${a = a_{(2,3)} q^{-n_{(1,2)}}}$ and ${b = 0}$. The remaining sum over $n_{(1,2)}$ can then be performed using the $q$-binomial theorem~\eqref{eq:qbinom} with ${a = a_{(1,2)}}$ and ${b = a_{(2,3)}}$, yielding the following compact expression,
\begin{align}
\label{eq:BCR4summed}
    \cA^{(4)}_q
    (1,2,3,4)
    &=
    - g^2
    \mu^{4-d}
    (1-q) 
    \frac{ (q^{-1};q^{-1})_\infty
           ( a_{(1,2)} a_{(2,3)} ; q^{-1} )_\infty }
         { ( a_{(1,2)} ; q^{-1} )_\infty
           ( a_{(2,3)} ; q^{-1} )_\infty }
    \, .
\end{align}
This expression is manifestly duality invariant (i.e.\ invariant under $s \leftrightarrow t$), and the two $q$-Pochhammer symbols in the denominator respectively contain the $s$-channel and $t$-channels poles at each $m_n^2$. Now, is this expression valid for $q < 1$? As written, the $q$-Pochhammer symbols diverge for $q < 1$, but we may write this expression in terms of a particular special function, the so-called $q$-gamma function, which allows us to take $q < 1$.

\subsubsection{The $q$-gamma function}

In mathematics, a $q$-analog of a theorem, function, identity, or expression is a generalization involving a deformation parameter $q$ that returns the original mathematical object in the limit $q \to 1$. Many special functions and differential equations have well-studied $q$-analogs dating back to the nineteenth century~\cite{gasper_rahman_2004}. The $q$-analog of the gamma function $\Gamma(z)$ is called the $q$-gamma function $\Gamma_q(z)$~\cite{gasper_rahman_2004} and is defined for complex $q$ and $z$ as follows,
\begin{align}
\label{eq:qGam}
    \Gamma_q(z)
    &=
\begin{cases}
    \phantom{q^{\frac{z(z-1)}{2}}}
    (1-q)^{1-z}
    \displaystyle
    \frac{ (q^{\phantom{z}};q)_\infty } 
         { (q^z;q)_\infty }
    \qquad
    &
    |q| < 1
\\[3ex]
    q^{\frac{z(z-1)}{2}}
    (q-1)^{1-z}
    \displaystyle
    \frac{ (q^{-1};q^{-1})_\infty }
         { (q^{-z};q^{-1})_\infty }
    \qquad
    &
    |q| > 1
\end{cases}
    \, .
\end{align}
The $q$-gamma function becomes the ordinary gamma function as $\lim_{q \to 1^\pm} \Gamma_q(z) = \Gamma(z)$. The piecewise definition ensures that the $q$-gamma function obeys a functional equation,
\begin{align}
    \Gamma_q(z+1) = [z]_q \, \Gamma_q(z)
    \, ,
\end{align}
analogous to that of the gamma function, $\Gamma(z+1) = z \, \Gamma(z)$~\cite{8bd4b335-fdba-3059-b063-d29589c21a38}. The piecewise definition also implies the following relationship between $\Gamma_{q}(z)$ and $\Gamma_{q^{-1}}(z)$,
\begin{align}
\label{eq:qinvG}
    \Gamma_{q^{-1}}(z)
    =
    q^{-\frac{1}{2}(z-1)(z-2)}
    \,
    \Gamma_q(z)
    \, .
\end{align}
We now use the definition of the $q$-gamma function~\eqref{eq:qGam} to write~\eqref{eq:BCR4summed} in terms of $\Gamma_{q}(z)$. After some simple algebra, we find the following expression which is valid for all $q \geq 0$,
\begin{align}
\label{eq:BCR4q<1}
    \cA^{(4)}_q
    (1,2,3,4)
    &=
    g^2
    \mu^{4-d}
    \,
    q^{\sigma_{(1,2)} \sigma_{(2,3)} - 1}
    \,
    \frac{ \Gamma_q ( -\sigma_{(1,2)} ) 
           \Gamma_q ( -\sigma_{(2,3)} ) }
         { \Gamma_q ( {} - \sigma_{(1,2)} - \sigma_{(2,3)} ) }
    \, ,
\end{align}
where we have defined the $q$-deformed Regge trajectory,
\begin{align}
\label{eq:qRegge}
    \sigma_I
    = 
    \frac{ \ln a_I }{ \ln q_{\phantom{I}} }
    =
    \frac{ \ln [ 1 + (q-1)(s_I-m_0^2)/\mu^2 ] }
         { \ln q }
    \, .
\end{align}
Up to a normalization convention, this form of the four-point Coon amplitude is exactly the form analyzed in~\cite{Geiser:2022icl} for all $q \geq 0$. In the limit $q \to 1$, it reproduces the Veneziano amplitude~\cite{Veneziano:1968yb} exactly as expected,
\begin{align}
    \lim_{q \to 1}
    \cA^{(4)}_q
    (1,2,3,4)
    &=
    g^2
    \mu^{4-d}
    \,
    \frac{ \Gamma \big( {-(s-m_0^2)/\mu^2} \big)
           \Gamma \big( {-(t-m_0^2)/\mu^2} \big) }
         { \Gamma \big( {-(s+t-2m_0^2)/\mu^2} \big) }
    \, .
\end{align}
Many other properties of~\eqref{eq:BCR4q<1} are discussed in~\cite{Geiser:2022icl}. Although we began with the four-point Baker-Coon-Romans formula which is only valid for $q>1$, we have ultimately arrived at an expression~\eqref{eq:BCR4q<1} valid for all $q \geq 0$. The trick was the $q$-gamma function, which provided a continuation of sorts in $q$.

\subsubsection{The loss of meromorphicity}

The continuation of the four-point Coon amplitude to ${q<1}$ introduces a peculiar property. Conventional wisdom says that tree-level scattering amplitudes should be meromorphic functions of the relevant Mandelstam variables (in this case $s$ and~$t$). However, the prefactor $\smash{q^{\sigma_{(1,2)} \sigma_{(2,3)}}}$ in~\eqref{eq:BCR4q<1} is explicitly non-meromorphic in $s$ and $t$. This factor introduces branch cuts in the complex $s$-plane and $t$-plane starting at the accumulation point of the Coon spectrum ${m_\infty^2 = (1-q)^{-1} \mu^2 + m_0^2}$. When ${q \geq 1}$, the Coon spectrum is unbounded, the accumulation point vanishes, and the four-point Coon amplitude is meromorphic. The non-meromorphicity of the four-point Coon amplitude with $q<1$ (and the interplay of this property with unitarity) has been discussed at length in~\cite{Figueroa:2022onw, Geiser:2022icl, Jepsen:2023sia}. The four-point Coon amplitude with $q<1$ may be unitary, but it is non-meromorphic. On the other hand, the four-point Coon amplitude with~${q>1}$ is meromorphic but non-unitary. Only the Veneziano amplitude at $q=1$ is both meromorphic and unitary. In any case, we have demonstrated how this non-meromorphicity arises through the definition of the $q$-gamma function when continuing the four-point Baker-Coon-Romans formula from $q > 1$ to $q < 1$. In contrast, Coon originally introduced the factor $\smash{q^{\sigma_{(1,2)} \sigma_{(2,3)}}}$ by hand to ensure that the four-point Coon amplitude with $q<1$ had polynomial residues on its poles~\cite{Coon:1972qz, Baker:1976en}.

\subsubsection{$q \to 0$}

With~\eqref{eq:BCR4q<1} in hand, we may now compute the $q \to 0$ limit of the four-point Coon amplitude. Taking $q \to 0$ alone collapses the Coon spectrum~\eqref{eq:spectrum} to a scalar of mass-squared $m_0^2$ plus an infinite number of particles with mas-squared $m_0^2 + \mu^2$. To interpret $q \to 0$ as a field theory limit, we must also take $\mu \to \infty$ (with $m_0^2$ fixed) so that these heavy particles are effectively integrated out, leaving only the lowest-lying scalar with mass-squared $m_0^2$. From this perspective, we expect the resulting amplitudes to correspond to a field theory describing the self-interactions of these scalars. In close analogy with the $q \to \infty$ limit, we define a new scale $\Lambda^2 = q \mu^2$ and a new dimensionless coupling $\tilde{g} = q^{(d-6)/4} g$ which we keep fixed in the limit $q \to 0$ so that the combination ${\tilde{g} \Lambda^{3-d/2} = g \mu^{3-d/2}}$ (and thus the three-point Coon amplitude~\eqref{eq:BCR3}) remains finite. For $d < 6$, this limit corresponds to a weak coupling limit in $g$. For $d > 6$, this limit corresponds to a strong coupling limit in $g$.

In summary, we are considering the limit,
\begin{align}
    q \to 0
    \text{ with fixed }
    m_0^2,
    \text{ }
    \Lambda^2 = q \mu^2,
    \text{ and }
    \tilde{g} = q^{(d-6)/4} g
    \, .
\end{align}
In this limit, the non-meromorphic prefactor is simply $\smash{q^{\sigma_{(1,2)} \sigma_{(2,3)}}} \to 1$. To correctly compute the limit of the $q$-gamma functions, we first write them as infinite products using~\eqref{eq:qGam}. After some straightforward manipulations, we find
\begin{align}
    \cA^{(4)}_q
    (1,2,3,4)
    &=
    - \tilde{g}^2
    \Lambda^{4-d}
    \,
    q^{\sigma_{(1,2)} \sigma_{(2,3)}}
    \bigg(
      \frac{\mu^2}{s-m_0^2}
    + \frac{\mu^2}{t-m_0^2}
    - (1-q)
    \bigg)
\no \\
    & \quad
    \times
    \prod_{n \geq 1}
    \frac{ (1-q^{n+1})
           \Big( 1 - \Big[ 1+(q-1)\frac{s-m_0^2}{\mu^2} \Big]
                     \Big[ 1+(q-1)\frac{t-m_0^2}{\mu^2} \Big] q^n \Big) }
         { \Big( 1 - \Big[ 1+(q-1)\frac{s-m_0^2}{\mu^2} \Big] q^n \Big)
           \Big( 1 - \Big[ 1+(q-1)\frac{t-m_0^2}{\mu^2} \Big] q^n \Big) }
    \, .
\end{align}
The entire second line of this expression becomes one as $q \to 0$. The first line, however, requires some hand-waving. If we assume that the combinations ${\mu^2 / (s-m_0^2)}$ and ${\mu^2 / (t-m_0^2)}$ remain finite (even though we are taking $\mu \to \infty$), then we have
\begin{align}
\label{eq:Aq04}
    \lim_{q \to 0}
    \cA^{(4)}_q
    (1,2,3,4)
    &=
    - \tilde{g}^2
    \Lambda^{4-d}
    \bigg(
      \frac{\mu^2}{s-m_0^2}
    + \frac{\mu^2}{t-m_0^2}
    - 1
    \bigg)
    \, .
\end{align}
The resulting amplitude is essentially the same as the $q \to \infty$ limit of the four-point Coon amplitude~\eqref{eq:AqinfN} but with the sign of the four-point term reversed. We revisit this observation in~\autoref{sec:disc}. For now, let us turn to five points.

\subsection{Five points}

We begin our discussion with the five-point Baker-Coon-Romans formula~\eqref{eq:BCR5}. For the sake of clarity, we label the five five-point planar scattering channels using the manifestly cyclic invariant notation $\cC^{(5)}=\{12,23,34,45,51\}$. Following the four-point strategy above, our goal is to write~\eqref{eq:BCR5} in terms of known special functions. Again, we first assume that ${q > 1}$ so that the sums in~\eqref{eq:BCR5} converge. We can perform the sums over $n_{23}$, $n_{34}$, and $n_{51}$ using the $q$-binomial theorem~\eqref{eq:qbinom} three times in succession to find
\begin{align}
    \cA^{(5)}_q
    (1,2,3,4,5)
    &=
    - g^3
    \mu^{5-3d/2}
    (1-q)^2
    \big[ (q^{-1};q^{-1})_\infty \big]^2
    \sum_{n_{12} \geq 0} \,\,
    \sum_{n_{45} \geq 0} 
\no \\
    & \quad
    \times
    \frac{ ( a_{23} a_{34} q^{-n_{12}-n_{45}} ; q^{-1})_\infty }
         { ( a_{23} q^{-n_{12}} ; q^{-1})_\infty
           ( a_{34} q^{-n_{45}} ; q^{-1})_\infty
           ( a_{51} q^{-n_{12}-n_{45}} ; q^{-1})_\infty }
\no \\
    & \quad
    \times
    \frac{ a_{12}^{n_{12}} }{ (q^{-1};q^{-1})_{n_{12}} }
    \frac{ a_{45}^{n_{45}} }{ (q^{-1};q^{-1})_{n_{45}} }
    \, .
\end{align}
Because the two remaining summation variables $n_{12}$ and $n_{45}$ each appear in four different $q$-Pochhammer symbols, we cannot use the $q$-binomial theorem in a straightforward way. Instead, we can use the $q$-binomial theorem in reverse to reintroduce the sum over $n_{51}$,
\begin{align}
    \frac{ ( a_{23} a_{34} q^{-n_{12}-n_{45}} ; q^{-1})_\infty }
         { ( a_{51} q^{-n_{12}-n_{45}} ; q^{-1})_\infty }
    &=
    \sum_{n_{51} \geq 0}
    \frac{ a_{51}^{n_{51}} }
         { (q^{-1};q^{-1})_{n_{51}} }
    ( a_{23} a_{34} / a_{51} ; q^{-1})_{n_{51}} 
    q^{-n_{12}n_{51}-n_{45}n_{51}} 
    \, .
\end{align}
We can then use the $q$-binomial theorem to perform the sums over $n_{12}$ and $n_{23}$, leaving only the sum over $n_{51}$. After some straightforward algebra, we find
\begin{align}
\label{eq:BCR5summed}
    \cA^{(5)}_q
    (1,2,3,4,5)
    &=
    - g^3
    \mu^{5-3d/2}
    (1-q)^2
\no \\
    & \quad
    \times
    \frac{ (q^{-1};q^{-1})_\infty
           ( a_{12} a_{23} ; q^{-1} )_\infty }
         { ( a_{12} ; q^{-1} )_\infty
           ( a_{23} ; q^{-1} )_\infty }
    \frac{ (q^{-1};q^{-1})_\infty
           ( a_{34} a_{45} ; q^{-1} )_\infty }
         { ( a_{34} ; q^{-1} )_\infty
           ( a_{45} ; q^{-1} )_\infty }
\no \\
    & \quad
    \times
    \sum_{n_{51} \geq 0}
    \frac{ ( a_{12} ; q^{-1} )_{n_{51}} 
           ( a_{45} ; q^{-1} )_{n_{51}} 
           ( a_{23} a_{34} / a_{51} ; q^{-1} )_{n_{51}} }
         { ( a_{12} a_{23} ; q^{-1} )_{n_{51}} 
           ( a_{34} a_{45} ; q^{-1} )_{n_{51}} 
           ( q^{-1} ; q^{-1} )_{n_{51}} } 
    \,
    a_{51}^{n_{51}}
    \, .
\end{align}
Now, is this expression valid for $q < 1$? The two combinations of $q$-Pochhammer symbols in the second line are exactly equal to those which appear in the fully-summed expression for the four-point Coon amplitude, and we know how to continue these factors to $q<1$ using the $q$-gamma function. The third line (i.e.\ the sum over $n_{51}$) can similarly be written in terms of another special function, the so-called $q$-hypergeometric (or basic hypergeometric) function $_3 \Phi_2$.

\subsubsection{The $q$-hypergeometric function}

The $q$-hypergeometric functions $_r \Phi_s$ are the $q$-analogs of the hypergeometric functions $_r F_s$. Like their undeformed $q=1$ counterparts, these special functions have a long history in the mathematics literature with entire books devoted to tabulating their properties~\cite{Romans:1988qs, gasper_rahman_2004}.

The $q$-hypergeometric function $_r \Phi_s$ has $r+s+1$ arguments ${(a_1, \dots, a_r, b_1, \dots, b_s, z)}$ (in addition to the deformation parameter $q$) and is defined by the following series when convergent (and elsewhere by analytic continuation),
\begin{align}
\label{eq:qPhi}
    {}_r\Phi_s
    \Big[
    \begin{smallmatrix}
    a_1 \,\, a_2 \,\, \cdots \,\, a_r
    \\
    b_1 \,\, b_2 \,\, \cdots \,\, b_s
    \end{smallmatrix}
    ; \, q \, 
    ; \, z
    \Big]
    &=
    \sum_{n=0}^\infty
    \frac{ (a_1;q)_n \cdots (a_r;q)_n }
         { (b_1;q)_n \cdots (b_s;q)_n (q;q)_n }
    \Big(
    (-)^n \, q^{\binom{n}{2}}
    \Big)^{1+s-r}
    \,
    z^n 
    \, .
\end{align}
If $0 < |q| < 1$ and $r \leq s$, then the series converges absolutely for all $z$. If $0 < |q| < 1$ and $r=s+1$, then the series converges absolutely for $|z| < 1$. If $0 < |q| < 1$ and $r>s+1$, then the series diverges for all $z \neq 0$ (unless it terminates). If $|q| > 1$, then the series converges absolutely for ${|z| < |b_1 \cdots b_s q / a_1 \cdots a_r|}$ and diverges for ${|z| > |b_1 \cdots b_s q / a_1 \cdots a_r|}$ (unless it terminates).

The usual hypergeometric function $_r F_s$ has $r+s+1$ arguments ${(a_1, \dots, a_r, b_1, \dots, b_s, z)}$ and is defined by the following series when convergent (and elsewhere by analytic continuation),
\begin{align}
    {}_r F_s
    \Big[
    \begin{smallmatrix}
    a_1 \,\, a_2 \,\, \cdots \,\, a_r
    \\
    b_1 \,\, b_2 \,\, \cdots \,\, b_s
    \end{smallmatrix}
    ; \, z
    \Big]
    &=
    \sum_{n=0}^\infty
    \frac{ \Gamma(a_1+n) }{ \Gamma(a_1) }
    \cdots
    \frac{ \Gamma(a_r+n) }{ \Gamma(a_r) }
    \frac{ \Gamma(b_1) }{ \Gamma(b_1+n) }
    \cdots
    \frac{ \Gamma(b_s) }{ \Gamma(b_s+n) }
    \frac{z^n}{n!} 
    \, .
\end{align}
If $r \leq s$, then the series converges absolutely for all $z$. If $r=s+1$, then the series converges absolutely for $|z| < 1$ and for $|z|=1$ if $\Re(b_1+\cdots+b_s-a_1-\cdots-a_r)>0$.  If $r=s+1$ and $|z| > 1$ or $r>s+1$ and $z \neq 0$, then the series diverges (unless it terminates). The hypergeometric and $q$-hypergeometric functions functions are related by
\begin{align}
    \lim_{q \to 1}
    {}_r\Phi_s
    \Big[
    \begin{smallmatrix}
    q^{a_1} \,\, q^{a_2} \,\, \cdots \,\, q^{a_r}
    \\
    q^{b_1} \,\, q^{b_2} \,\, \cdots \,\, q^{b_s}
    \end{smallmatrix}
    ; \, q \, 
    ; \, (q-1)^{1+s-r} \, z
    \Big]
    &=
    {}_r F_s
    \Big[
    \begin{smallmatrix}
    a_1 \,\, a_2 \,\, \cdots \,\, a_r
    \\
    b_1 \,\, b_2 \,\, \cdots \,\, b_s
    \end{smallmatrix}
    ; \, z
    \Big]
    \, .
\end{align}
The analytic continuations of their series definitions can be explicitly defined using various contour integrals~\cite{gasper_rahman_2004, doi:10.1080/10652469.2016.1231674}.

When $r=s+1$, these functions have several nice properties. For example, the $q$\nobreakdash-hypergeometric function $_{r}\Phi_{r-1}$ with deformation parameter $q^{-1}$ is related to $_{r}\Phi_{r-1}$ with deformation parameter $q$ as follows,
\begin{align}
\label{eq:qinvPhi}
    {}_r\Phi_{r-1}
    \Big[
    \begin{smallmatrix}
    a_1 \,\, a_2 \,\, \cdots \,\, a_{r\phantom{{}-1}}
    \\
    b_1 \,\, b_2 \,\, \cdots \,\, b_{r-1}
    \end{smallmatrix}
    ; \, q \, 
    ; \, z
    \Big]
    &=
    {}_r\Phi_{r-1}
    \Big[
    \begin{smallmatrix}
    a_1^{-1} \,\, a_2^{-1} \,\, \cdots \,\, a_{r\phantom{{}-1}}^{-1}
    \\
    b_1^{-1} \,\, b_2^{-1} \,\, \cdots \,\, b_{r-1}^{-1}
    \end{smallmatrix}
    ; \, q^{-1} \, 
    ; \,
    \big(
    \tfrac{ a_1 \cdots a_{r\phantom{{}-1}} \mathstrut }
          { b_1 \cdots b_{r-1} \mathstrut }
    \big)
    \, q^{-1} \, z
    \Big]
    \, .
\end{align}
Many more properties can be found in~\cite{gasper_rahman_2004}.

Using the definition~\eqref{eq:qGam} of the $q$-gamma function, the definition~\eqref{eq:qPhi} of $_3 \Phi_2$, and the identity~\eqref{eq:qinvPhi} we can now rewrite the partially-summed five-point Baker-Coon-Romans formula~\eqref{eq:BCR5summed} as follows,
\begin{align}
\label{eq:BCR5q<1}
    \cA^{(5)}_q
    (1,2,3,4,5)
    &=
    - g^3
    \mu^{5-3d/2}
    q^{\sigma_{12} \sigma_{23} + \sigma_{34} \sigma_{45} - 2}
    \,
    \frac{ \Gamma_q ( -\sigma_{12} ) 
           \Gamma_q ( -\sigma_{23} ) }
         { \Gamma_q ( {} - \sigma_{12} - \sigma_{23} ) }
    \frac{ \Gamma_q ( -\sigma_{34} ) 
           \Gamma_q ( -\sigma_{45} ) }
         { \Gamma_q ( {} - \sigma_{34} - \sigma_{45} ) }
\no \\[1ex]
    & \quad
    \times
    {}_3\Phi_2
    \Big[
    \begin{smallmatrix}
    q^{-\sigma_{12}}
    \quad
    q^{-\sigma_{45}}
    \quad
    q^{-\sigma_{23}-\sigma_{34}+\sigma_{51}}
    \\
    q^{-\sigma_{12}-\sigma_{23}}
    \quad
    q^{-\sigma_{34}-\sigma_{45}}
    \end{smallmatrix}
    ; \, q \, 
    ; \, q
    \Big]
    \, ,
\end{align}
where we have reintroduced the $q$-deformed Regge trajectories $\sigma_I$ defined in~\eqref{eq:qRegge}. This expression now provides a clear definition for the five-point Coon amplitude with $q < 1$ since the $q$-gamma function and the $q$-hypergeometric function are defined for all $q$ through the relations~\eqref{eq:qinvG} and~\eqref{eq:qinvPhi}.

The five-point Coon amplitude was first written in the form~\eqref{eq:BCR5q<1}, i.e.\ in terms of~$_3 \Phi_2$, by Romans~\cite{Romans:1988qs}. However, Romans did not arrive at this equation from the Baker-Coon-Romans formula~\eqref{eq:BCR5}. Instead, Romans postulated the five-point expression~\eqref{eq:BCR5q<1} by simply $q$-deforming the five-point tree-level string amplitude, which can similarly be written using the gamma function and the hypergeometric function $_3 F_2$. Examining the four-point Coon amplitude and this five-point amplitude then led Romans to the general $N$-point Baker-Coon-Romans formula~\eqref{eq:BCRN} which reproduced~\eqref{eq:BCR4q<1} at four points and~\eqref{eq:BCR5q<1} at five points. To see this logic in reverse, we can take the $q \to 1$ limit of~\eqref{eq:BCR5q<1}. We find
\begin{align}
\label{eq:A5string}
    \lim_{q \to 1}
    \cA^{(5)}_q
    (1,2,3,4,5)
    &=
    - g^3
    \mu^{5-3d/2}
    \frac{ \Gamma( -\mfs_{12} )
           \Gamma( -\mfs_{23} ) }
         { \Gamma( -\mfs_{12}-\mfs_{23} ) }
    \frac{ \Gamma( -\mfs_{34} )
           \Gamma( -\mfs_{45} ) }
         { \Gamma( -\mfs_{34}-\mfs_{45} ) }
\no \\[1ex]
    & \quad
    \times
    {}_3 F_2
    \Big[
    \begin{smallmatrix}
    -\mfs_{12}
    \quad
    -\mfs_{45}
    \quad
    -\mfs_{23} - \mfs_{34} + \mfs_{51}
    \\
    -\mfs_{12} - \mfs_{23} 
    \quad
    -\mfs_{34} - \mfs_{45} 
    \end{smallmatrix}
    ; \,
    1
    \Big]
    \, ,
\end{align}
where we have used the dimensionless Mandelstam variables $\mfs_I = (s_I - m_0^2)/\mu^2$ to keep the expression compact. As expected,~\eqref{eq:A5string} is precisely the five-point tree-level open string amplitude~\cite{Bardakci:1968rse, Virasoro:1969pd, Nakanishi:1971ve, Medina:2002nk, Hanson_2006, Boels:2013jua}. 

Although we began with the five-point Baker-Coon-Romans formula which is only valid for $q>1$, we have ultimately arrived at an expression~\eqref{eq:BCR5q<1} valid for all $q \geq 0$. This time, the trick was the $q$-hypergeometric function, which provided a continuation of sorts in $q$.

\subsubsection{Relation to Cheung-Remmen's $\cA^{(4)}_{r}$ and $\cA^{(4)}_{q,r}$}

Before we examine the properties of~\eqref{eq:BCR5q<1}, let us note a curious relation between the five-point Coon amplitude and the four-point ``hypergeometric" and \mbox{``basic hypergeometric"} (or ``hypergeometric Coon") amplitudes recently discovered by Cheung and Remmen~\cite{Cheung:2023adk, Rigatos:2023asb}. In~\cite{Cheung:2023adk}, Cheung and Remmen bootstrapped four-point scattering amplitudes directly from an input mass spectrum and several physical constraints.

For the string spectrum $m_n^2 = m_0^2 + \mu^2 n$, their procedure yields the hypergeometric amplitude $\cA^{(4)}_{r}$, a one-parameter generalization of the Veneziano amplitude with a new real-valued deformation parameter $r$. In the limit $r \to 0$, the hypergeometric amplitude becomes the Veneziano amplitude.

For the Coon spectrum~\eqref{eq:spectrum}, the same procedure yields the basic hypergeometric amplitude $\cA^{(4)}_{q,r}$, a one-parameter generalization of the four-point Coon amplitude with a new real-valued deformation parameter $r$. In the limit $r \to 0$, the basic hypergeometric amplitude becomes the four-point Coon amplitude. Since the Coon amplitude is itself a one-parameter deformation of the Veneziano amplitude, the basic hypergeometric amplitude can also be thought of as two-parameter generalization of the Veneziano amplitude.

Like the Coon amplitude, the hypergeometric and basic hypergeometric amplitudes are more precisely partial amplitudes which depend on an ordering of the external states. For the canonical ordering $(1,2,3,4)$, these two amplitudes are given by
\begin{align}
    \cA^{(4)}_{r}
    (1,2,3,4)
    &=
    g^2
    \mu^{4-d} 
    \,
    \frac{ \Gamma( -\mfs_{12} )
           \Gamma( -\mfs_{23} ) }
         { \Gamma( -\mfs_{12}-\mfs_{23} ) }
    \,
    {}_3 F_2
    \Big[
    \begin{smallmatrix}
    -\mfs_{12}
    \quad
    -\mfs_{23}
    &
    r
    \\
    -\mfs_{12} - \mfs_{23} 
    &
    r+1
    \end{smallmatrix}
    ; \,
    1
    \Big]
    \, ,
\no \\[2ex]
    \cA^{(4)}_{q, r}
    (1,2,3,4)
    &=
    g^2
    \mu^{4-d}
    q^{\sigma_{12} \sigma_{23} - 1}
    \,
    \frac{ \Gamma_q ( -\sigma_{12} ) 
           \Gamma_q ( -\sigma_{23} ) }
         { \Gamma_q ( {} - \sigma_{12} - \sigma_{23} ) }
    \,
    {}_3\Phi_2
    \Big[
    \begin{smallmatrix}
    q^{-\sigma_{12}}
    \quad
    q^{-\sigma_{23}}
    &
    q^{r\phantom{{}+1}}
    \\
    q^{-\sigma_{12}-\sigma_{23}}
    &
    q^{r+1}
    \end{smallmatrix}
    ; \, q \, 
    ; \, q
    \Big]
    \, ,
\end{align}
Our expressions differ from those in~\cite{Cheung:2023adk} by a few trivial normalization factors. With our conventions, we can immediately identify a set of relations between the four-point hypergeometric and basic hypergeometric amplitudes on one hand and the five-point string and Coon amplitudes on the other. The hypergeometric and basic hypergeometric amplitudes are simply the five-point string and Coon amplitudes evaluated at particular values of the kinematic variables $s_{34}$, $s_{45}$, and $s_{51}$. Specifically, we have
\begin{align}
\label{eq:specialkin}
    \cA^{(4)}_{r}
    (1,2,3,4)
    &=
    {}\phantom{q \, }{}
    g^{-1} \mu^{d/2-1}
    \,
    \mfs_{45}
    \times
    \cA^{(5)}
    (1,2,3,4,5)
    \bigg|_{\substack{ \mfs_{34}=-1 \\
                       \mfs_{45}=-r \\ 
                       \mfs_{51}=-1 }}
    \, ,
\no \\
    \cA^{(4)}_{q,r}
    (1,2,3,4)
    &=
    q \, g^{-1} \mu^{d/2-1}
    \,
    \mfs_{45}
    \times
    \cA^{(5)}_{q}
    (1,2,3,4,5)
    \bigg|_{\substack{ \sigma_{34}=-1 \\
                       \sigma_{45}=-r \\ 
                       \sigma_{51}=-1 }}
    \, ,                   
\end{align}
where $\mfs_I = (s_I - m_0^2)/\mu^2$ and $\sigma_I$ is defined in~\eqref{eq:qRegge}. Here $\cA^{(5)} = \lim_{q \to 1} \cA^{(5)}_{q}$ is the five-point tree-level open string amplitude. The relation between the four-point hypergeometric amplitude and the five-point string amplitude was first observed in~\cite{Cheung:2023adk}, but the relation between the four-point basic hypergeometric amplitude and the five-point Coon amplitude is a new result which firmly places $\cA^{(4)}_{q,r}$ within the broader family
of Coon amplitudes.

The relationships between these six amplitudes (the five-point Coon amplitude $\cA^{(5)}_{q}$, the five-point tree-level open string amplitude $\cA^{(5)}$, the four-point basic hypergeometric amplitude $\cA^{(4)}_{q,r}$, the four-point hypergeometric amplitude $\cA^{(4)}_{r}$, the four-point Coon amplitude $\cA^{(4)}_{q}$, and the four-point tree-level open string amplitude $\cA^{(4)}$) can be summarized in the following commutative diagram:
\begin{equation}
    \begin{tikzcd}
    \cA^{(5)}_{q}   \arrow[r, "\text{s.k.}"] \arrow[d, "q \to 1"]  &
    \cA^{(4)}_{q,r} \arrow[r, "r \to 0"] \arrow[d, "q \to 1"] &
    \cA^{(4)}_{q}   \arrow[d, "q \to 1"]
    \\
    \cA^{(5)}       \arrow[r, "\text{s.k.}"] &
    \cA^{(4)}_{r}   \arrow[r, "r \to 0"] &
    \cA^{(4)}
    \end{tikzcd}
\end{equation}
Here $\text{s.k.}$ refers to the special kinematic values given in~\eqref{eq:specialkin}. The commutativity of this diagram simply follows from the expressions for each amplitude. These relations clarify some of the properties of the hypergeometric amplitude described in~\cite{Cheung:2023adk, Rigatos:2023asb}. For example, in~\cite{Rigatos:2023asb}, the critical dimension of the hypergeometric amplitude with $m_0^2 = - 1/\alpha'$ was found to be equal to that of bosonic string theory $d_c = 26$, but this fact simply follows from the relation between $\cA^{(4)}_{r}$ and the five-point tree-level string amplitude. Similarly, the transcendental structure of the low-energy expansion of $\cA^{(4)}_{r}$ observed in~\cite{Cheung:2023adk} simply follows from the well-known transcendental properties of tree-level string amplitudes~\cite{Schlotterer:2012ny, DHoker:2019blr, DHoker:2021ous}.

\subsubsection{The loss of meromorphicity}

The continuation of the five-point Coon amplitude to ${q<1}$ again introduces the peculiar property of non-meromorphicity. The two $q^{\sigma \sigma}$ factors on the first line of~\eqref{eq:BCR5q<1} are non-meromorphic in the Mandelstam variables, but these non-meromorphicities only appear for $q < 1$. When $q > 1$, these factors cancel against similar non-meromorphic factors in the definition of the $q$-gamma function so that the full amplitude is meromorphic. The meromorphicity properties are thus similar to those of the four-point Coon amplitude discussed above.

\subsubsection{The loss of factorization}

Unfortunately, meromorphicity is not the only property which disappears when $q < 1$. The factorization (and duality) properties of the five-point Coon amplitude are also a problem. Duality invariance implies that five-point Coon amplitude should have simple poles in all five planar scattering channels. The $s_{12}$, $s_{23}$, $s_{34}$, and $s_{45}$ poles are clearly exhibited by the four gamma functions in~\eqref{eq:BCR5q<1}, but the $s_{51}$ poles seem to be nowhere in sight! For $q > 1$, the $s_{51}$-channel poles are hiding within $_3\Phi_2$. In fact, we can use a particular identity~\cite{https://doi.org/10.1112/jlms/s1-11.4.276} obeyed by the $_3\Phi_2$ functions to explicitly exhibit the poles in any four of the planar scattering channels, with the remaining channel's poles hidden within the $_3\Phi_2$ function. In~\cite{Romans:1988qs}, Romans used this identity to prove the duality-invariance of the five-point Coon amplitude with $q>1$ written in the form~\eqref{eq:BCR5q<1}. Moreover, we derived~\eqref{eq:BCR5q<1} from the five-point Baker-Coon-Romans formula which exhibited factorization and duality invariance for all $q > 1$, so there should be no problem with~\eqref{eq:BCR5q<1} in this regime. The trouble only arises with $q < 1$. In this case, we can simply take ${s_{51} \to m_0^2}$ (and thus ${\sigma_{51} \to 0}$) in~\eqref{eq:BCR5q<1}. We expect to find a simple pole $\sim 1 /(s_{51}-m_0^2)$, but the result is finite. We can double-check this result by reviewing the convergence conditions for the $q$-hypergeometric function, but there is no way to recover an $s_{51}$ pole when $q < 1$.

We conclude that the five-point Coon amplitude with $q<1$ no longer factorizes on the $s_{51}$ channel and is no longer duality-invariant. We suspect that this phenomenon arises because the various sums and limits do not commute. In other words, it must be the case that the radius of convergence of some intermediate sum that led to the expression~\eqref{eq:BCR5q<1} was incompatible with $q<1$. Perhaps some piecewise definition of the $q$-hypergeometric function (in analogy with the $q$-gamma function) is instead needed. In any case, it does not seem like our simple special function approach suffices to define a fully consistent five-point Coon amplitude with $q < 1$. What worked at four points does not extend to five.

\subsubsection{$q \to 0$}

We conclude this subsection by computing the $q \to 0$ limit (with fixed $m_0^2$, $\Lambda^2 = q \mu^2$, and $\tilde{g} = q^{(d-6)/4} g$) of the expression~\eqref{eq:BCR5q<1} for the five-point Coon amplitude to exhibit the problems with factorization described above. We begin by rewriting~\eqref{eq:BCR5q<1} in terms of two four-point Coon amplitudes~\eqref{eq:BCR4q<1},
\begin{align}
    \cA^{(5)}_q
    (1,2,3,4,5)
    &=
    -
    \big( g  \mu^{3-d/2} \big)^{-1}
    \times
    \cA^{(4)}_q
    (1,2,3,4)
    \times
    \cA^{(4)}_q
    (3,4,5,1)
\no \\[1ex]
    & \quad
    \times
    {}_3\Phi_2
    \Big[
    \begin{smallmatrix}
    q^{-\sigma_{12}}
    \quad
    q^{-\sigma_{45}}
    \quad
    q^{-\sigma_{23}-\sigma_{34}+\sigma_{51}}
    \\
    q^{-\sigma_{12}-\sigma_{23}}
    \quad
    q^{-\sigma_{34}-\sigma_{45}}
    \end{smallmatrix}
    ; \, q \, 
    ; \, q
    \Big]
    \, .
\end{align}
The $q \to 0$ limit of the four-point Coon amplitude is given in~\eqref{eq:Aq04}, the $q \to 0$ limit of ${}_3\Phi_2$ is simply one, and the combination $g  \mu^{3-d/2} = \tilde{g} \Lambda^{3-d/2}$ is finite in this limit. Hence,
\begin{align}
    \lim_{q \to 0}
    \cA^{(5)}_q
    (1,2,3,4,5)
    =
    - \tilde{g}^3
    \Lambda^{5-3d/2}
    &
    \bigg(
      \frac{\mu^2}{s_{12}-m_0^2}
    + \frac{\mu^2}{s_{23}-m_0^2}
    - 1
    \bigg)
\no \\
    {} \times {}
    &
    \bigg(
      \frac{\mu^2}{s_{34}-m_0^2}
    + \frac{\mu^2}{s_{45}-m_0^2}
    - 1
    \bigg)
    \, .
\end{align}
As we predicted, factorizaiton is violated, and the $s_{51}$ poles are nowhere to be found.

\subsection{Higher points}

Even if we ignore the problems with factorization at five-points, there seems to be no hope of extending this special function procedure to $N \geq 6$ points. The six-point tree-level open string amplitude cannot be cleanly written in terms of a well-known special function, so a $q$-deformed special function representation is unlikely to exist~\cite{Romans:1988qs}. This hiccup does not mean that a continuation of the Coon amplitude to $q < 1$ does not exist. Instead, we have learned that such a continuation requires a new formulation.


\section{\texorpdfstring{$\mathrm{U}(N_F)$}{U(NF)} adjoint scalar amplitudes}
\label{sec:AS}

In this section, we compute the $N$-point tree-level amplitudes of the $\mathrm{U}(N_F)$ adjoint scalar theory introduced in~\autoref{sec:intro} and defined by the following Lagrangian,
\begin{align}
\label{eq:LagAS2}
    \cL_{\text{AS}}
    =
    - \frac{1}{2} \Tr \p_\mu \phi \, \p^\mu \phi
    - \frac{1}{2} \, m_0^2 \Tr \phi^2
    - \sum_{n \geq 3}
    \frac{1}{n} \,
    \tilde{g}^{n-2} \,
    \Lambda^{n+d-\frac{1}{2}nd} 
    \Tr \phi^n
    \, .
\end{align}
The scalar field $\phi = T^a \phi^a$ transforms in the adjoint representation of the global flavor symmetry group $\mathrm{U}(N_F)$. The matrices $T^{a}$ (with adjoint index ${a = 0, 1, 2, \dots N_F^2-1}$) are the generators of $\mathrm{U}(N_F)$ in the defining representation, normalized by ${\Tr(T^a T^b) = \delta^{ab}}$. The coupling constant $\tilde{g}$ is dimensionless, and $\Lambda$ is an arbitrary mass scale which gives the $n$\nobreakdash-point couplings the usual mass dimensions ${[g_n] = n + d - \half n d}$ in $d$ spacetime dimensions. As described in~\eqref{eq:AqAAS1}, the $N$-point tree-level amplitudes of this adjoint scalar theory are exactly equal to the $N$-point Baker-Coon-Romans amplitudes in the limit $q \to \infty$. We demonstrate this equality explicitly.

\subsection{Feynman rules}

We begin by writing out the Feynman rules for the scalar field $\phi^a$. We first rewrite the Lagrangian~\eqref{eq:LagAS2} as
\begin{align}
\label{eq:LagAS22}
    \cL_{\text{AS}}
    &=
    - \frac{1}{2} \delta^{ab}
        ( \p_\mu \phi^{a} )
        ( \p^\mu \phi^{b} )
    - \frac{1}{2} m_0^2 \, \delta^{ab}
        \phi^{a} \phi^{b}
\no \\
    &\quad
    - \sum_{n \geq 3}
    \frac{1}{n} \,
    \tilde{g}^{n-2} \,
    \Lambda^{n+d-\frac{1}{2}nd} 
    \Tr(T^{a_1} \cdots T^{a_n}) \,
    \phi^{a_1}
    \cdots
    \phi^{a_n}
    \, .
\end{align}
The propagator is given by
\begin{align}
    \begin{tikzpicture}[baseline=-0.5ex]
    \node (a) at (-1,0) [left] {$a$};
    \node (b) at (1,0) [right] {$b$};
    \node (p) at (0,0.25) [above] {$p$};
    \draw (-1,0)--(1,0);
    \draw[->] (-0.25,0.25)--(0.25,0.25);
    \end{tikzpicture}
    &=
    \frac{i \delta^{ab}}{-p^2 - m_0^2 + i\e}
    \, .
\end{align}
The $n$-point vertex is given by $-i \frac{1}{n} \tilde{g}^{n-2} \Lambda^{n+d-\frac{1}{2}nd} \Tr(T^{a_1} \cdots T^{a_n})$ summed over all the $n!$ permutations of the indices $a_1, \dots, a_n$. Collecting the cyclic permutations together cancels the factor of $\frac{1}{n}$, so we have
\begin{align}
    \begin{tikzpicture}[baseline=-0.5ex]
    \node (a1) at (0*360/4+135:1) {$a_1$};
    \node (a2) at (1*360/4+135:1) {$a_2$};
    \node (a3) at (2*360/4+135:1) {$a_3$};
    \node (a4) at (3*360/4+135:1) {$a_n$};
    \node (dots1) at ($(a3)!0.4!(a4)$) {\rotatebox{90}{$.$}};
    \node (dots2) at ($(a3)!0.5!(a4)$) {\rotatebox{90}{$.$}};
    \node (dots3) at ($(a3)!0.6!(a4)$) {\rotatebox{90}{$.$}};
    \node (o) at (0,0) [circle,fill=black,inner sep=0pt,minimum size=3pt] {}; 
    \draw (o)--(a1);
    \draw (o)--(a2);
    \draw (o)--(a3);
    \draw (o)--(a4);
    \end{tikzpicture}
    &=
    -i \tilde{g}^{n-2} \Lambda^{n+d-\frac{1}{2}nd}
    \sum_{\sigma \in S_n / \bbZ_n }
    \Tr
    (
    T^{a_{\sigma(1)}} T^{a_{\sigma(2)}} \cdots T^{a_{\sigma(n)}}
    )
    \, ,
\end{align}
where the sum is over elements $\sigma$ of the permutation group $S_n$ modulo the group of cyclic permutations $\bbZ_n$ acting on the labels $(1, 2, \dots, n)$ (or equivalently over elements of the permutation group $S_{n-1}$ acting on the labels $(2, \dots, n)$ with the first label fixed). For concreteness, we list the first few vertices.
\begin{itemize}
\item The three-point vertex is given by a sum over $(3-1)! = 2$ terms,
\begin{align}
    \begin{tikzpicture}[baseline=-0.5ex]
    \node (a1) at (0*360/3+90:1) {$a_1$};
    \node (a2) at (1*360/3+90:1) {$a_2$};
    \node (a3) at (2*360/3+90:1) {$a_3$};
    \node (o) at (0,0) [circle,fill=black,inner sep=0pt,minimum size=3pt] {}; 
    \draw (o)--(a1);
    \draw (o)--(a2);
    \draw (o)--(a3);
    \end{tikzpicture}
    &=
    -i \tilde{g} \Lambda^{3-d/2}
    \big[
      \Tr( T^{a_1} T^{a_2} T^{a_3} )
    + \Tr( T^{a_1} T^{a_3} T^{a_2} )
    \big]
    \, .
\end{align}
\item The four-point vertex is given by a sum over $(4-1)! = 6$ terms,
\begin{alignat}{2}
    \smash[b]{
    \begin{tikzpicture}[baseline=-0.5ex]
    \node (a1) at (0*360/4+135:1) {$a_1$};
    \node (a2) at (1*360/4+135:1) {$a_2$};
    \node (a3) at (2*360/4+135:1) {$a_3$};
    \node (a4) at (3*360/4+135:1) {$a_4$};
    \node (o) at (0,0) [circle,fill=black,inner sep=0pt,minimum size=3pt] {}; 
    \draw (o)--(a1);
    \draw (o)--(a2);
    \draw (o)--(a3);
    \draw (o)--(a4);
    \end{tikzpicture}
    }
    =
    -i \tilde{g}^2 \Lambda^{4-d}
    \big[
      \Tr( T^{a_1} T^{a_2} T^{a_3} T^{a_4} )
    &
    + \Tr( T^{a_1} T^{a_2} T^{a_4} T^{a_3} )
\no \\
    \vphantom{ -i \tilde{g}^2 \Lambda^{4-d}}
    + \Tr( T^{a_1} T^{a_3} T^{a_2} T^{a_4} )
    &
    + \Tr( T^{a_1} T^{a_3} T^{a_4} T^{a_2} )
\no \\
    \vphantom{ -i \tilde{g}^2 \Lambda^{4-d}}
    + \Tr( T^{a_1} T^{a_4} T^{a_2} T^{a_3} )
    &
    + \Tr( T^{a_1} T^{a_4} T^{a_3} T^{a_2} )
    \big]
    \, .
\end{alignat}
\end{itemize}

Any tree-level amplitude in this theory is given by a sum of products of traces with factors from the appropriate propagators and couplings. It is not difficult to compute these amplitudes by hand. However, because our symmetry group is $\mathrm{U}(N_F)$, there is a remarkable simplification. Using the completeness relation for the $\mathrm{U}(N_F)$ generators $T^a$ (which form a complete set of $N_F \times N_F$ Hermitian matrices), we may combine any products of traces connected by a propagator as follows,
\begin{align}
    {(T^a)_i}^j
    {(T^a)_k}^\ell
    =
    {\delta_i}^\ell
    {\delta_k}^j
    \qquad
    \implies
    \qquad
    \Tr(X T^a) \Tr (T^a Y) 
    =
    \Tr(XY)
    \, .
\end{align}
In the end, only single traces remain. These single traces can be stripped off, and the tree-level amplitudes can be written as a sum over partial amplitudes which depend on a particular cyclic ordering of the external states~\cite{Dixon:1996wi, Dixon:2013uaa, Cheung:2017pzi, Elvang:2013cua},
\begin{align}
    \cA^{(N) \, \text{tree}}
       _{\text{AS}; \, \mathrm{U}(N_F)}
    =
    \sum_{\sigma \in S_N / \bbZ_N }
    \Tr
    (
    T^{a_{\sigma(1)}} T^{a_{\sigma(2)}} \cdots T^{a_{\sigma(N)}}
    )
    \,
    \cA^{(N)}_{\text{AS}}
    (
    \sigma(1), \sigma(2), \dots, \sigma(N)
    )
    \, .
\end{align}
The partial amplitude for a given ordering can then be computed from planar ordered Feynman diagrams using the following ordered Feynman rules,
\begin{align}
    \begin{tikzpicture}[baseline=-0.5ex]
    \node (a) at (-1,0) [left] {};
    \node (b) at (1,0) [right] {};
    \node (p) at (0,0.25) [above] {$p$};
    \draw (-1,0)--(1,0);
    \draw[->] (-0.25,0.25)--(0.25,0.25);
    \end{tikzpicture}
    &=
    \frac{i}{-p^2 - m_0^2 + i\e}
    \, ,
&
    \begin{tikzpicture}[baseline=-0.5ex]
    \node (a1) at (0*360/4+135:1) {$1$};
    \node (a2) at (1*360/4+135:1) {$2$};
    \node (a3) at (2*360/4+135:1) {$3$};
    \node (a4) at (3*360/4+135:1) {$n$};
    \node (dots1) at ($(a3)!0.4!(a4)$) {\rotatebox{90}{$.$}};
    \node (dots2) at ($(a3)!0.5!(a4)$) {\rotatebox{90}{$.$}};
    \node (dots3) at ($(a3)!0.6!(a4)$) {\rotatebox{90}{$.$}};
    \node (o) at (0,0) [circle,fill=black,inner sep=0pt,minimum size=3pt] {}; 
    \draw (o)--(a1);
    \draw (o)--(a2);
    \draw (o)--(a3);
    \draw (o)--(a4);
    \end{tikzpicture}
    &=
    -i \tilde{g}^{n-2} \Lambda^{n+d-\frac{1}{2}nd}
    \, .
\end{align}
Since the cyclic order of the external states is fixed, we can ignore symmetry factors when computing these partial amplitudes. Each unique topology contributes a single term.

In the remainder of this section, we explicitly compute the four-point, five-point, and six-point partial amplitudes before deriving a general expression for the $N$-point case.

\subsection{Four points}

The four-point partial amplitude for the canonical ordering $(1,2,3,4)$ is given by a sum of three ordered Feynman diagrams,
\begin{align}
    i \cA^{(4)}_{\text{AS}}
    (1,2,3,4)
    &=
    \begin{tikzpicture}[baseline=-0.5ex, scale=0.5, every node/.style={font=\scriptsize}]
    \node (1) at (0,1)  {$1$};
    \node (2) at (0,-1) {$2$};
    \node (3) at (3,-1) {$3$};
    \node (4) at (3,1) {$4$};
    \node (a) at (1,0) [circle,fill=black,inner sep=0pt,minimum size=3pt] {}; 
    \node (b) at (2,0) [circle,fill=black,inner sep=0pt,minimum size=3pt] {}; 
    \draw (1)--(a);
    \draw (2)--(a);
    \draw (3)--(b);
    \draw (4)--(b);
    \draw (a)--(b);
    \end{tikzpicture}
    +
    \begin{tikzpicture}[baseline=-0.5ex, scale=0.5, every node/.style={font=\scriptsize}]
    \node (1) at (0,1) {$2$};
    \node (2) at (0,-1) {$3$};
    \node (3) at (3,-1) {$4$};
    \node (4) at (3,1) {$1$};
    \node (a) at (1,0) [circle,fill=black,inner sep=0pt,minimum size=3pt] {}; 
    \node (b) at (2,0) [circle,fill=black,inner sep=0pt,minimum size=3pt] {}; 
    \draw (1)--(a);
    \draw (2)--(a);
    \draw (3)--(b);
    \draw (4)--(b);
    \draw (a)--(b);
    \end{tikzpicture}
    +
    \begin{tikzpicture}[baseline=-0.5ex, scale=0.5, every node/.style={font=\scriptsize}]
    \node (1) at (0,1) {$1$};
    \node (2) at (0,-1) {$2$};
    \node (3) at (2,-1) {$3$};
    \node (4) at (2,1) {$4$};
    \node (a) at (1,0) [circle,fill=black,inner sep=0pt,minimum size=3pt] {}; 
    \draw (1)--(a);
    \draw (2)--(a);
    \draw (3)--(a);
    \draw (4)--(a);
    \end{tikzpicture}
    \, .
\end{align}
The first diagram is an $s$-channel diagram, the second is a $t$-channel diagram, and the third is a four-point contact term. Using the ordered Feynman rules, we find
\begin{align}
    \cA^{(4)}_{\text{AS}}
    (1,2,3,4)
    &=
    - \tilde{g}^2
    \Lambda^{4-d}
    \bigg(
      \frac{\Lambda^2}{s-m_0^2}
    + \frac{\Lambda^2}{t-m_0^2}
    + 1
    \bigg)
    \, ,
\end{align}
which is precisely the expression for the four-point Baker-Coon-Romans partial amplitude in the limit $q \to \infty$ given in~\eqref{eq:Aqinf4}.

\subsection{Five points}

The five-point partial amplitude for the canonical ordering $(1,2,3,4,5)$ is given by a sum of eleven ordered Feynman diagrams,
\begin{align}
    i \cA^{(5)}_{\text{AS}}
    (1,2,3,4,5)
    &=
    \begin{tikzpicture}[baseline=-0.5ex, scale=0.5, every node/.style={font=\scriptsize}]
    \node (1) at (-0.5,1) {$1$};
    \node (2) at (-0.5,-1) {$2$};
    \node (3) at (2.5,-1) {$3$};
    \node (4) at (2.5,1) {$4$};
    \node (5) at (1,1) {$5$};
    \node (a) at (0,0) [circle,fill=black,inner sep=0pt,minimum size=3pt] {}; 
    \node (b) at (1,0) [circle,fill=black,inner sep=0pt,minimum size=3pt] {}; 
    \node (c) at (2,0) [circle,fill=black,inner sep=0pt,minimum size=3pt] {}; 
    \draw (1)--(a);
    \draw (2)--(a);
    \draw (3)--(c);
    \draw (4)--(c);
    \draw (5)--(b);
    \draw (a)--(b);
    \draw (b)--(c);
    \end{tikzpicture}
    +
    \begin{tikzpicture}[baseline=-0.5ex, scale=0.5, every node/.style={font=\scriptsize}]
    \node (1) at (-0.5,1) {$2$};
    \node (2) at (-0.5,-1) {$3$};
    \node (3) at (2.5,-1) {$4$};
    \node (4) at (2.5,1) {$5$};
    \node (5) at (1,1) {$1$};
    \node (a) at (0,0) [circle,fill=black,inner sep=0pt,minimum size=3pt] {}; 
    \node (b) at (1,0) [circle,fill=black,inner sep=0pt,minimum size=3pt] {}; 
    \node (c) at (2,0) [circle,fill=black,inner sep=0pt,minimum size=3pt] {}; 
    \draw (1)--(a);
    \draw (2)--(a);
    \draw (3)--(c);
    \draw (4)--(c);
    \draw (5)--(b);
    \draw (a)--(b);
    \draw (b)--(c);
    \end{tikzpicture}
    +
    \begin{tikzpicture}[baseline=-0.5ex, scale=0.5, every node/.style={font=\scriptsize}]
    \node (1) at (-0.5,1) {$3$};
    \node (2) at (-0.5,-1) {$4$};
    \node (3) at (2.5,-1) {$5$};
    \node (4) at (2.5,1) {$1$};
    \node (5) at (1,1) {$2$};
    \node (a) at (0,0) [circle,fill=black,inner sep=0pt,minimum size=3pt] {}; 
    \node (b) at (1,0) [circle,fill=black,inner sep=0pt,minimum size=3pt] {}; 
    \node (c) at (2,0) [circle,fill=black,inner sep=0pt,minimum size=3pt] {}; 
    \draw (1)--(a);
    \draw (2)--(a);
    \draw (3)--(c);
    \draw (4)--(c);
    \draw (5)--(b);
    \draw (a)--(b);
    \draw (b)--(c);
    \end{tikzpicture}
    +
    \begin{tikzpicture}[baseline=-0.5ex, scale=0.5, every node/.style={font=\scriptsize}]
    \node (1) at (-0.5,1) {$4$};
    \node (2) at (-0.5,-1) {$5$};
    \node (3) at (2.5,-1) {$1$};
    \node (4) at (2.5,1) {$2$};
    \node (5) at (1,1) {$3$};
    \node (a) at (0,0) [circle,fill=black,inner sep=0pt,minimum size=3pt] {}; 
    \node (b) at (1,0) [circle,fill=black,inner sep=0pt,minimum size=3pt] {}; 
    \node (c) at (2,0) [circle,fill=black,inner sep=0pt,minimum size=3pt] {}; 
    \draw (1)--(a);
    \draw (2)--(a);
    \draw (3)--(c);
    \draw (4)--(c);
    \draw (5)--(b);
    \draw (a)--(b);
    \draw (b)--(c);
    \end{tikzpicture}
    +
    \begin{tikzpicture}[baseline=-0.5ex, scale=0.5, every node/.style={font=\scriptsize}]
    \node (1) at (-0.5,1) {$5$};
    \node (2) at (-0.5,-1) {$1$};
    \node (3) at (2.5,-1) {$2$};
    \node (4) at (2.5,1) {$3$};
    \node (5) at (1,1) {$4$};
    \node (a) at (0,0) [circle,fill=black,inner sep=0pt,minimum size=3pt] {}; 
    \node (b) at (1,0) [circle,fill=black,inner sep=0pt,minimum size=3pt] {}; 
    \node (c) at (2,0) [circle,fill=black,inner sep=0pt,minimum size=3pt] {}; 
    \draw (1)--(a);
    \draw (2)--(a);
    \draw (3)--(c);
    \draw (4)--(c);
    \draw (5)--(b);
    \draw (a)--(b);
    \draw (b)--(c);
    \end{tikzpicture}
\no \\
    &\quad
    +
    \begin{tikzpicture}[baseline=-0.5ex, scale=0.5, every node/.style={font=\scriptsize}]
    \node (1) at (-0.5,1) {$1$};
    \node (2) at (-0.5,-1) {$2$};
    \node (3) at (2,-1) {$3$};
    \node (4) at (2,0) {$4$};
    \node (5) at (2,1) {$5$};
    \node (a) at (0,0) [circle,fill=black,inner sep=0pt,minimum size=3pt] {}; 
    \node (b) at (1,0) [circle,fill=black,inner sep=0pt,minimum size=3pt] {}; 
    \draw (1)--(a);
    \draw (2)--(a);
    \draw (3)--(b);
    \draw (4)--(b);
    \draw (5)--(b);
    \draw (a)--(b);
    \end{tikzpicture}
    +
    \begin{tikzpicture}[baseline=-0.5ex, scale=0.5, every node/.style={font=\scriptsize}]
    \node (1) at (-0.5,1) {$2$};
    \node (2) at (-0.5,-1) {$3$};
    \node (3) at (2,-1) {$4$};
    \node (4) at (2,0) {$5$};
    \node (5) at (2,1) {$1$};
    \node (a) at (0,0) [circle,fill=black,inner sep=0pt,minimum size=3pt] {}; 
    \node (b) at (1,0) [circle,fill=black,inner sep=0pt,minimum size=3pt] {}; 
    \draw (1)--(a);
    \draw (2)--(a);
    \draw (3)--(b);
    \draw (4)--(b);
    \draw (5)--(b);
    \draw (a)--(b);
    \end{tikzpicture}
    +
    \begin{tikzpicture}[baseline=-0.5ex, scale=0.5, every node/.style={font=\scriptsize}]
    \node (1) at (-0.5,1) {$3$};
    \node (2) at (-0.5,-1) {$4$};
    \node (3) at (2,-1) {$5$};
    \node (4) at (2,0) {$1$};
    \node (5) at (2,1) {$2$};
    \node (a) at (0,0) [circle,fill=black,inner sep=0pt,minimum size=3pt] {}; 
    \node (b) at (1,0) [circle,fill=black,inner sep=0pt,minimum size=3pt] {}; 
    \draw (1)--(a);
    \draw (2)--(a);
    \draw (3)--(b);
    \draw (4)--(b);
    \draw (5)--(b);
    \draw (a)--(b);
    \end{tikzpicture}
    +
    \begin{tikzpicture}[baseline=-0.5ex, scale=0.5, every node/.style={font=\scriptsize}]
    \node (1) at (-0.5,1) {$4$};
    \node (2) at (-0.5,-1) {$5$};
    \node (3) at (2,-1) {$1$};
    \node (4) at (2,0) {$2$};
    \node (5) at (2,1) {$3$};
    \node (a) at (0,0) [circle,fill=black,inner sep=0pt,minimum size=3pt] {}; 
    \node (b) at (1,0) [circle,fill=black,inner sep=0pt,minimum size=3pt] {}; 
    \draw (1)--(a);
    \draw (2)--(a);
    \draw (3)--(b);
    \draw (4)--(b);
    \draw (5)--(b);
    \draw (a)--(b);
    \end{tikzpicture}
    +
    \begin{tikzpicture}[baseline=-0.5ex, scale=0.5, every node/.style={font=\scriptsize}]
    \node (1) at (-0.5,1) {$5$};
    \node (2) at (-0.5,-1) {$1$};
    \node (3) at (2,-1) {$2$};
    \node (4) at (2,0) {$3$};
    \node (5) at (2,1) {$4$};
    \node (a) at (0,0) [circle,fill=black,inner sep=0pt,minimum size=3pt] {}; 
    \node (b) at (1,0) [circle,fill=black,inner sep=0pt,minimum size=3pt] {}; 
    \draw (1)--(a);
    \draw (2)--(a);
    \draw (3)--(b);
    \draw (4)--(b);
    \draw (5)--(b);
    \draw (a)--(b);
    \end{tikzpicture}
\no \\
    &\quad
    +
    \begin{tikzpicture}[baseline=-0.5ex, scale=0.5, every node/.style={font=\scriptsize}]
    \node (1) at (1*360/5+90:1.2) {$1$};
    \node (2) at (2*360/5+90:1.2) {$2$};
    \node (3) at (3*360/5+90:1.2) {$3$};
    \node (4) at (4*360/5+90:1.2) {$4$};
    \node (5) at (5*360/5+90:1.2) {$5$};
    \node (a) at (0,0) [circle,fill=black,inner sep=0pt,minimum size=3pt] {}; 
    \draw (1)--(a);
    \draw (2)--(a);
    \draw (3)--(a);
    \draw (4)--(a);
    \draw (5)--(a);
    \end{tikzpicture}
    \, .
\end{align}
Each line contains diagrams related by cyclic permutations of the external labels. The five diagrams on the first line each have two three-point vertices and two propagators. The five diagrams on the second line each have one three-point vertex, one four-point vertex, and one propagator. The final diagram is a five-point contact term. Using the ordered Feynman rules, we find
\begin{alignat}{2}
    \cA^{(5)}_{\text{AS}}
    (1,2,3,4,5)
    =
    - \tilde{g}^3
    \Lambda^{5-\frac{3}{2}d}
    \Bigg(
      \frac{1}{ \mfs_{12} \mfs_{34} }
    &
    + \frac{1}{ \mfs_{23} \mfs_{45} }
    + \frac{1}{ \mfs_{34} \mfs_{51} }
    + \frac{1}{ \mfs_{45} \mfs_{12} }
    + \frac{1}{ \mfs_{51} \mfs_{23} }    
\no \\
    &
    + \frac{1}{ \mfs_{12} }
    + \frac{1}{ \mfs_{23} }
    + \frac{1}{ \mfs_{34} }
    + \frac{1}{ \mfs_{45} }
    + \frac{1}{ \mfs_{51} }
    + 1
    \Bigg)
    \, ,
\end{alignat}
where $\mfs_{I} = (s_I - m_0^2)/\Lambda^2$. This is precisely the expression for the five-point Baker-Coon-Romans partial amplitude in the limit $q \to \infty$ given in~\eqref{eq:Aqinf5}.

\subsection{Six points}

The six-point partial amplitude for the canonical ordering $(1,2,3,4,5,6)$ is given by a sum of forty-five ordered Feynman diagrams (which thankfully can be typeset on a single page),
\begin{align}
    i \cA^{(6)}_{\text{AS}}
    (1,2,3,4,5,6)
    &=
    \begin{tikzpicture}[baseline=-0.5ex, scale=0.5, every node/.style={font=\scriptsize}]
    \node (1) at (-0.2, 1.0) {$1$};
    \node (2) at (-0.2,-1.0) {$2$};
    \node (3) at ( 1.7,-1.0) {$3$};
    \node (4) at ( 1.7, 1.0) {$4$};
    \node (5) at ( 1.1, 1.0) {$5$};
    \node (6) at ( 0.4, 1.0) {$6$};
    \node (a) at ( 0.0, 0.0) [circle,fill=black,inner sep=0pt,minimum size=3pt] {}; 
    \node (b) at ( 0.5, 0.0) [circle,fill=black,inner sep=0pt,minimum size=3pt] {};
    \node (c) at ( 1.0, 0.0) [circle,fill=black,inner sep=0pt,minimum size=3pt] {};
    \node (d) at ( 1.5, 0.0) [circle,fill=black,inner sep=0pt,minimum size=3pt] {}; 
    \draw (1)--(a);
    \draw (2)--(a);
    \draw (3)--(d);
    \draw (4)--(d);
    \draw (5)--(c);
    \draw (6)--(b);
    \draw (a)--(b);
    \draw (b)--(c);
    \draw (c)--(d);
    \end{tikzpicture}
    +
    \begin{tikzpicture}[baseline=-0.5ex, scale=0.5, every node/.style={font=\scriptsize}]
    \node (1) at (-0.2, 1.0) {$2$};
    \node (2) at (-0.2,-1.0) {$3$};
    \node (3) at ( 1.7,-1.0) {$4$};
    \node (4) at ( 1.7, 1.0) {$5$};
    \node (5) at ( 1.1, 1.0) {$6$};
    \node (6) at ( 0.4, 1.0) {$1$};
    \node (a) at ( 0.0, 0.0) [circle,fill=black,inner sep=0pt,minimum size=3pt] {}; 
    \node (b) at ( 0.5, 0.0) [circle,fill=black,inner sep=0pt,minimum size=3pt] {};
    \node (c) at ( 1.0, 0.0) [circle,fill=black,inner sep=0pt,minimum size=3pt] {};
    \node (d) at ( 1.5, 0.0) [circle,fill=black,inner sep=0pt,minimum size=3pt] {}; 
    \draw (1)--(a);
    \draw (2)--(a);
    \draw (3)--(d);
    \draw (4)--(d);
    \draw (5)--(c);
    \draw (6)--(b);
    \draw (a)--(b);
    \draw (b)--(c);
    \draw (c)--(d);
    \end{tikzpicture}
    +
    \begin{tikzpicture}[baseline=-0.5ex, scale=0.5, every node/.style={font=\scriptsize}]
    \node (1) at (-0.2, 1.0) {$3$};
    \node (2) at (-0.2,-1.0) {$4$};
    \node (3) at ( 1.7,-1.0) {$5$};
    \node (4) at ( 1.7, 1.0) {$6$};
    \node (5) at ( 1.1, 1.0) {$1$};
    \node (6) at ( 0.4, 1.0) {$2$};
    \node (a) at ( 0.0, 0.0) [circle,fill=black,inner sep=0pt,minimum size=3pt] {}; 
    \node (b) at ( 0.5, 0.0) [circle,fill=black,inner sep=0pt,minimum size=3pt] {};
    \node (c) at ( 1.0, 0.0) [circle,fill=black,inner sep=0pt,minimum size=3pt] {};
    \node (d) at ( 1.5, 0.0) [circle,fill=black,inner sep=0pt,minimum size=3pt] {}; 
    \draw (1)--(a);
    \draw (2)--(a);
    \draw (3)--(d);
    \draw (4)--(d);
    \draw (5)--(c);
    \draw (6)--(b);
    \draw (a)--(b);
    \draw (b)--(c);
    \draw (c)--(d);
    \end{tikzpicture}
    +
    \begin{tikzpicture}[baseline=-0.5ex, scale=0.5, every node/.style={font=\scriptsize}]
    \node (1) at (-0.2, 1.0) {$4$};
    \node (2) at (-0.2,-1.0) {$5$};
    \node (3) at ( 1.7,-1.0) {$6$};
    \node (4) at ( 1.7, 1.0) {$1$};
    \node (5) at ( 1.1, 1.0) {$2$};
    \node (6) at ( 0.4, 1.0) {$3$};
    \node (a) at ( 0.0, 0.0) [circle,fill=black,inner sep=0pt,minimum size=3pt] {}; 
    \node (b) at ( 0.5, 0.0) [circle,fill=black,inner sep=0pt,minimum size=3pt] {};
    \node (c) at ( 1.0, 0.0) [circle,fill=black,inner sep=0pt,minimum size=3pt] {};
    \node (d) at ( 1.5, 0.0) [circle,fill=black,inner sep=0pt,minimum size=3pt] {}; 
    \draw (1)--(a);
    \draw (2)--(a);
    \draw (3)--(d);
    \draw (4)--(d);
    \draw (5)--(c);
    \draw (6)--(b);
    \draw (a)--(b);
    \draw (b)--(c);
    \draw (c)--(d);
    \end{tikzpicture}
    +
    \begin{tikzpicture}[baseline=-0.5ex, scale=0.5, every node/.style={font=\scriptsize}]
    \node (1) at (-0.2, 1.0) {$5$};
    \node (2) at (-0.2,-1.0) {$6$};
    \node (3) at ( 1.7,-1.0) {$1$};
    \node (4) at ( 1.7, 1.0) {$2$};
    \node (5) at ( 1.1, 1.0) {$3$};
    \node (6) at ( 0.4, 1.0) {$4$};
    \node (a) at ( 0.0, 0.0) [circle,fill=black,inner sep=0pt,minimum size=3pt] {}; 
    \node (b) at ( 0.5, 0.0) [circle,fill=black,inner sep=0pt,minimum size=3pt] {};
    \node (c) at ( 1.0, 0.0) [circle,fill=black,inner sep=0pt,minimum size=3pt] {};
    \node (d) at ( 1.5, 0.0) [circle,fill=black,inner sep=0pt,minimum size=3pt] {}; 
    \draw (1)--(a);
    \draw (2)--(a);
    \draw (3)--(d);
    \draw (4)--(d);
    \draw (5)--(c);
    \draw (6)--(b);
    \draw (a)--(b);
    \draw (b)--(c);
    \draw (c)--(d);
    \end{tikzpicture}
    +
    \begin{tikzpicture}[baseline=-0.5ex, scale=0.5, every node/.style={font=\scriptsize}]
    \node (1) at (-0.2, 1.0) {$6$};
    \node (2) at (-0.2,-1.0) {$1$};
    \node (3) at ( 1.7,-1.0) {$2$};
    \node (4) at ( 1.7, 1.0) {$3$};
    \node (5) at ( 1.1, 1.0) {$4$};
    \node (6) at ( 0.4, 1.0) {$5$};
    \node (a) at ( 0.0, 0.0) [circle,fill=black,inner sep=0pt,minimum size=3pt] {}; 
    \node (b) at ( 0.5, 0.0) [circle,fill=black,inner sep=0pt,minimum size=3pt] {};
    \node (c) at ( 1.0, 0.0) [circle,fill=black,inner sep=0pt,minimum size=3pt] {};
    \node (d) at ( 1.5, 0.0) [circle,fill=black,inner sep=0pt,minimum size=3pt] {}; 
    \draw (1)--(a);
    \draw (2)--(a);
    \draw (3)--(d);
    \draw (4)--(d);
    \draw (5)--(c);
    \draw (6)--(b);
    \draw (a)--(b);
    \draw (b)--(c);
    \draw (c)--(d);
    \end{tikzpicture}
\no \\
    &\quad
    +
    \begin{tikzpicture}[baseline=-0.5ex, scale=0.5, every node/.style={font=\scriptsize}]
    \node (1) at (-0.2, 1.0) {$1$};
    \node (2) at (-0.2,-1.0) {$2$};
    \node (3) at ( 1.0,-1.0) {$3$};
    \node (4) at ( 1.7,-1.0) {$4$};
    \node (5) at ( 1.7, 1.0) {$5$};
    \node (6) at ( 0.5, 1.0) {$6$};
    \node (a) at ( 0.0, 0.0) [circle,fill=black,inner sep=0pt,minimum size=3pt] {}; 
    \node (b) at ( 0.5, 0.0) [circle,fill=black,inner sep=0pt,minimum size=3pt] {};
    \node (c) at ( 1.0, 0.0) [circle,fill=black,inner sep=0pt,minimum size=3pt] {};
    \node (d) at ( 1.5, 0.0) [circle,fill=black,inner sep=0pt,minimum size=3pt] {}; 
    \draw (1)--(a);
    \draw (2)--(a);
    \draw (3)--(c);
    \draw (4)--(d);
    \draw (5)--(d);
    \draw (6)--(b);
    \draw (a)--(b);
    \draw (b)--(c);
    \draw (c)--(d);
    \end{tikzpicture}
    +
    \begin{tikzpicture}[baseline=-0.5ex, scale=0.5, every node/.style={font=\scriptsize}]
    \node (1) at (-0.2, 1.0) {$2$};
    \node (2) at (-0.2,-1.0) {$3$};
    \node (3) at ( 1.0,-1.0) {$4$};
    \node (4) at ( 1.7,-1.0) {$5$};
    \node (5) at ( 1.7, 1.0) {$6$};
    \node (6) at ( 0.5, 1.0) {$1$};
    \node (a) at ( 0.0, 0.0) [circle,fill=black,inner sep=0pt,minimum size=3pt] {}; 
    \node (b) at ( 0.5, 0.0) [circle,fill=black,inner sep=0pt,minimum size=3pt] {};
    \node (c) at ( 1.0, 0.0) [circle,fill=black,inner sep=0pt,minimum size=3pt] {};
    \node (d) at ( 1.5, 0.0) [circle,fill=black,inner sep=0pt,minimum size=3pt] {}; 
    \draw (1)--(a);
    \draw (2)--(a);
    \draw (3)--(c);
    \draw (4)--(d);
    \draw (5)--(d);
    \draw (6)--(b);
    \draw (a)--(b);
    \draw (b)--(c);
    \draw (c)--(d);
    \end{tikzpicture}
    +
    \begin{tikzpicture}[baseline=-0.5ex, scale=0.5, every node/.style={font=\scriptsize}]
    \node (1) at (-0.2, 1.0) {$3$};
    \node (2) at (-0.2,-1.0) {$4$};
    \node (3) at ( 1.0,-1.0) {$5$};
    \node (4) at ( 1.7,-1.0) {$6$};
    \node (5) at ( 1.7, 1.0) {$1$};
    \node (6) at ( 0.5, 1.0) {$2$};
    \node (a) at ( 0.0, 0.0) [circle,fill=black,inner sep=0pt,minimum size=3pt] {}; 
    \node (b) at ( 0.5, 0.0) [circle,fill=black,inner sep=0pt,minimum size=3pt] {};
    \node (c) at ( 1.0, 0.0) [circle,fill=black,inner sep=0pt,minimum size=3pt] {};
    \node (d) at ( 1.5, 0.0) [circle,fill=black,inner sep=0pt,minimum size=3pt] {}; 
    \draw (1)--(a);
    \draw (2)--(a);
    \draw (3)--(c);
    \draw (4)--(d);
    \draw (5)--(d);
    \draw (6)--(b);
    \draw (a)--(b);
    \draw (b)--(c);
    \draw (c)--(d);
    \end{tikzpicture}
    +
    \begin{tikzpicture}[baseline=-0.5ex, scale=0.5, every node/.style={font=\scriptsize}]
    \node (1) at (-0.2, 1.0) {$4$};
    \node (2) at (-0.2,-1.0) {$5$};
    \node (3) at ( 1.0,-1.0) {$6$};
    \node (4) at ( 1.7,-1.0) {$1$};
    \node (5) at ( 1.7, 1.0) {$2$};
    \node (6) at ( 0.5, 1.0) {$3$};
    \node (a) at ( 0.0, 0.0) [circle,fill=black,inner sep=0pt,minimum size=3pt] {}; 
    \node (b) at ( 0.5, 0.0) [circle,fill=black,inner sep=0pt,minimum size=3pt] {};
    \node (c) at ( 1.0, 0.0) [circle,fill=black,inner sep=0pt,minimum size=3pt] {};
    \node (d) at ( 1.5, 0.0) [circle,fill=black,inner sep=0pt,minimum size=3pt] {}; 
    \draw (1)--(a);
    \draw (2)--(a);
    \draw (3)--(c);
    \draw (4)--(d);
    \draw (5)--(d);
    \draw (6)--(b);
    \draw (a)--(b);
    \draw (b)--(c);
    \draw (c)--(d);
    \end{tikzpicture}
    +
    \begin{tikzpicture}[baseline=-0.5ex, scale=0.5, every node/.style={font=\scriptsize}]
    \node (1) at (-0.2, 1.0) {$5$};
    \node (2) at (-0.2,-1.0) {$6$};
    \node (3) at ( 1.0,-1.0) {$1$};
    \node (4) at ( 1.7,-1.0) {$2$};
    \node (5) at ( 1.7, 1.0) {$3$};
    \node (6) at ( 0.5, 1.0) {$4$};
    \node (a) at ( 0.0, 0.0) [circle,fill=black,inner sep=0pt,minimum size=3pt] {}; 
    \node (b) at ( 0.5, 0.0) [circle,fill=black,inner sep=0pt,minimum size=3pt] {};
    \node (c) at ( 1.0, 0.0) [circle,fill=black,inner sep=0pt,minimum size=3pt] {};
    \node (d) at ( 1.5, 0.0) [circle,fill=black,inner sep=0pt,minimum size=3pt] {}; 
    \draw (1)--(a);
    \draw (2)--(a);
    \draw (3)--(c);
    \draw (4)--(d);
    \draw (5)--(d);
    \draw (6)--(b);
    \draw (a)--(b);
    \draw (b)--(c);
    \draw (c)--(d);
    \end{tikzpicture}
    +
    \begin{tikzpicture}[baseline=-0.5ex, scale=0.5, every node/.style={font=\scriptsize}]
    \node (1) at (-0.2, 1.0) {$6$};
    \node (2) at (-0.2,-1.0) {$1$};
    \node (3) at ( 1.0,-1.0) {$2$};
    \node (4) at ( 1.7,-1.0) {$3$};
    \node (5) at ( 1.7, 1.0) {$4$};
    \node (6) at ( 0.5, 1.0) {$5$};
    \node (a) at ( 0.0, 0.0) [circle,fill=black,inner sep=0pt,minimum size=3pt] {}; 
    \node (b) at ( 0.5, 0.0) [circle,fill=black,inner sep=0pt,minimum size=3pt] {};
    \node (c) at ( 1.0, 0.0) [circle,fill=black,inner sep=0pt,minimum size=3pt] {};
    \node (d) at ( 1.5, 0.0) [circle,fill=black,inner sep=0pt,minimum size=3pt] {}; 
    \draw (1)--(a);
    \draw (2)--(a);
    \draw (3)--(c);
    \draw (4)--(d);
    \draw (5)--(d);
    \draw (6)--(b);
    \draw (a)--(b);
    \draw (b)--(c);
    \draw (c)--(d);
    \end{tikzpicture}
\no \\
    &\quad
    +
    \begin{tikzpicture}[baseline=-0.5ex, scale=0.5, every node/.style={font=\scriptsize}]
    \node (1) at (1*360/3+90-30:1.5) {$1$};
    \node (2) at (1*360/3+90+30:1.5) {$2$};
    \node (3) at (2*360/3+90-30:1.5) {$3$};
    \node (4) at (2*360/3+90+30:1.5) {$4$};
    \node (5) at (3*360/3+90-30:1.5) {$5$};
    \node (6) at (3*360/3+90+30:1.5) {$6$};
    \node (o) at (0,0) [circle,fill=black,inner sep=0pt,minimum size=3pt] {};
    \node (a) at (1*360/3+90:0.6) [circle,fill=black,inner sep=0pt,minimum size=3pt] {};
    \node (b) at (2*360/3+90:0.6) [circle,fill=black,inner sep=0pt,minimum size=3pt] {};
    \node (c) at (3*360/3+90:0.6) [circle,fill=black,inner sep=0pt,minimum size=3pt] {};
    \draw (1)--(a);
    \draw (2)--(a);
    \draw (3)--(b);
    \draw (4)--(b);
    \draw (5)--(c);
    \draw (6)--(c);
    \draw (o)--(a);
    \draw (o)--(b);
    \draw (o)--(c);
    \end{tikzpicture}
    +
    \begin{tikzpicture}[baseline=-0.5ex, scale=0.5, every node/.style={font=\scriptsize}]
    \node (1) at (1*360/3+90-30:1.5) {$2$};
    \node (2) at (1*360/3+90+30:1.5) {$3$};
    \node (3) at (2*360/3+90-30:1.5) {$4$};
    \node (4) at (2*360/3+90+30:1.5) {$5$};
    \node (5) at (3*360/3+90-30:1.5) {$6$};
    \node (6) at (3*360/3+90+30:1.5) {$1$};
    \node (o) at (0,0) [circle,fill=black,inner sep=0pt,minimum size=3pt] {};
    \node (a) at (1*360/3+90:0.6) [circle,fill=black,inner sep=0pt,minimum size=3pt] {};
    \node (b) at (2*360/3+90:0.6) [circle,fill=black,inner sep=0pt,minimum size=3pt] {};
    \node (c) at (3*360/3+90:0.6) [circle,fill=black,inner sep=0pt,minimum size=3pt] {};
    \draw (1)--(a);
    \draw (2)--(a);
    \draw (3)--(b);
    \draw (4)--(b);
    \draw (5)--(c);
    \draw (6)--(c);
    \draw (o)--(a);
    \draw (o)--(b);
    \draw (o)--(c);
    \end{tikzpicture}
\no \\
    &\quad
    +
    \begin{tikzpicture}[baseline=-0.5ex, scale=0.5, every node/.style={font=\scriptsize}]
    \node (1) at (-0.2, 1.0) {$1$};
    \node (2) at (-0.2,-1.0) {$2$};
    \node (3) at (0.75,-1.0) {$3$};
    \node (4) at ( 1.7,-1.0) {$4$};
    \node (5) at ( 1.7, 1.0) {$5$};
    \node (6) at (0.75, 1.0) {$6$};
    \node (a) at ( 0.0, 0.0) [circle,fill=black,inner sep=0pt,minimum size=3pt] {}; 
    \node (b) at (0.75, 0.0) [circle,fill=black,inner sep=0pt,minimum size=3pt] {};
    \node (c) at ( 1.5, 0.0) [circle,fill=black,inner sep=0pt,minimum size=3pt] {};
    \draw (1)--(a);
    \draw (2)--(a);
    \draw (3)--(b);
    \draw (4)--(c);
    \draw (5)--(c);
    \draw (6)--(c);
    \draw (a)--(b);
    \draw (b)--(c);
    \end{tikzpicture}
    +
    \begin{tikzpicture}[baseline=-0.5ex, scale=0.5, every node/.style={font=\scriptsize}]
    \node (1) at (-0.2, 1.0) {$2$};
    \node (2) at (-0.2,-1.0) {$3$};
    \node (3) at (0.75,-1.0) {$4$};
    \node (4) at ( 1.7,-1.0) {$5$};
    \node (5) at ( 1.7, 1.0) {$6$};
    \node (6) at (0.75, 1.0) {$1$};
    \node (a) at ( 0.0, 0.0) [circle,fill=black,inner sep=0pt,minimum size=3pt] {}; 
    \node (b) at (0.75, 0.0) [circle,fill=black,inner sep=0pt,minimum size=3pt] {};
    \node (c) at ( 1.5, 0.0) [circle,fill=black,inner sep=0pt,minimum size=3pt] {};
    \draw (1)--(a);
    \draw (2)--(a);
    \draw (3)--(b);
    \draw (4)--(c);
    \draw (5)--(c);
    \draw (6)--(c);
    \draw (a)--(b);
    \draw (b)--(c);
    \end{tikzpicture}
    +
    \begin{tikzpicture}[baseline=-0.5ex, scale=0.5, every node/.style={font=\scriptsize}]
    \node (1) at (-0.2, 1.0) {$3$};
    \node (2) at (-0.2,-1.0) {$4$};
    \node (3) at (0.75,-1.0) {$5$};
    \node (4) at ( 1.7,-1.0) {$6$};
    \node (5) at ( 1.7, 1.0) {$1$};
    \node (6) at (0.75, 1.0) {$2$};
    \node (a) at ( 0.0, 0.0) [circle,fill=black,inner sep=0pt,minimum size=3pt] {}; 
    \node (b) at (0.75, 0.0) [circle,fill=black,inner sep=0pt,minimum size=3pt] {};
    \node (c) at ( 1.5, 0.0) [circle,fill=black,inner sep=0pt,minimum size=3pt] {};
    \draw (1)--(a);
    \draw (2)--(a);
    \draw (3)--(b);
    \draw (4)--(c);
    \draw (5)--(c);
    \draw (6)--(c);
    \draw (a)--(b);
    \draw (b)--(c);
    \end{tikzpicture}
    +
    \begin{tikzpicture}[baseline=-0.5ex, scale=0.5, every node/.style={font=\scriptsize}]
    \node (1) at (-0.2, 1.0) {$4$};
    \node (2) at (-0.2,-1.0) {$5$};
    \node (3) at (0.75,-1.0) {$6$};
    \node (4) at ( 1.7,-1.0) {$1$};
    \node (5) at ( 1.7, 1.0) {$2$};
    \node (6) at (0.75, 1.0) {$3$};
    \node (a) at ( 0.0, 0.0) [circle,fill=black,inner sep=0pt,minimum size=3pt] {}; 
    \node (b) at (0.75, 0.0) [circle,fill=black,inner sep=0pt,minimum size=3pt] {};
    \node (c) at ( 1.5, 0.0) [circle,fill=black,inner sep=0pt,minimum size=3pt] {};
    \draw (1)--(a);
    \draw (2)--(a);
    \draw (3)--(b);
    \draw (4)--(c);
    \draw (5)--(c);
    \draw (6)--(c);
    \draw (a)--(b);
    \draw (b)--(c);
    \end{tikzpicture}
    +
    \begin{tikzpicture}[baseline=-0.5ex, scale=0.5, every node/.style={font=\scriptsize}]
    \node (1) at (-0.2, 1.0) {$5$};
    \node (2) at (-0.2,-1.0) {$6$};
    \node (3) at (0.75,-1.0) {$1$};
    \node (4) at ( 1.7,-1.0) {$2$};
    \node (5) at ( 1.7, 1.0) {$3$};
    \node (6) at (0.75, 1.0) {$4$};
    \node (a) at ( 0.0, 0.0) [circle,fill=black,inner sep=0pt,minimum size=3pt] {}; 
    \node (b) at (0.75, 0.0) [circle,fill=black,inner sep=0pt,minimum size=3pt] {};
    \node (c) at ( 1.5, 0.0) [circle,fill=black,inner sep=0pt,minimum size=3pt] {};
    \draw (1)--(a);
    \draw (2)--(a);
    \draw (3)--(b);
    \draw (4)--(c);
    \draw (5)--(c);
    \draw (6)--(c);
    \draw (a)--(b);
    \draw (b)--(c);
    \end{tikzpicture}
    +
    \begin{tikzpicture}[baseline=-0.5ex, scale=0.5, every node/.style={font=\scriptsize}]
    \node (1) at (-0.2, 1.0) {$6$};
    \node (2) at (-0.2,-1.0) {$1$};
    \node (3) at (0.75,-1.0) {$2$};
    \node (4) at ( 1.7,-1.0) {$3$};
    \node (5) at ( 1.7, 1.0) {$4$};
    \node (6) at (0.75, 1.0) {$5$};
    \node (a) at ( 0.0, 0.0) [circle,fill=black,inner sep=0pt,minimum size=3pt] {}; 
    \node (b) at (0.75, 0.0) [circle,fill=black,inner sep=0pt,minimum size=3pt] {};
    \node (c) at ( 1.5, 0.0) [circle,fill=black,inner sep=0pt,minimum size=3pt] {};
    \draw (1)--(a);
    \draw (2)--(a);
    \draw (3)--(b);
    \draw (4)--(c);
    \draw (5)--(c);
    \draw (6)--(c);
    \draw (a)--(b);
    \draw (b)--(c);
    \end{tikzpicture}
\no \\
    &\quad
    +
    \begin{tikzpicture}[baseline=-0.5ex, scale=0.5, every node/.style={font=\scriptsize}]
    \node (1) at (-0.2, 1.0) {$1$};
    \node (2) at (-0.2,-1.0) {$2$};
    \node (3) at (0.75,-1.0) {$3$};
    \node (4) at ( 1.7,-1.0) {$4$};
    \node (5) at ( 1.7, 1.0) {$5$};
    \node (6) at (0.75, 1.0) {$6$};
    \node (a) at ( 0.0, 0.0) [circle,fill=black,inner sep=0pt,minimum size=3pt] {}; 
    \node (b) at (0.75, 0.0) [circle,fill=black,inner sep=0pt,minimum size=3pt] {};
    \node (c) at ( 1.5, 0.0) [circle,fill=black,inner sep=0pt,minimum size=3pt] {};
    \draw (1)--(a);
    \draw (2)--(a);
    \draw (3)--(c);
    \draw (4)--(c);
    \draw (5)--(c);
    \draw (6)--(b);
    \draw (a)--(b);
    \draw (b)--(c);
    \end{tikzpicture}
    +
    \begin{tikzpicture}[baseline=-0.5ex, scale=0.5, every node/.style={font=\scriptsize}]
    \node (1) at (-0.2, 1.0) {$2$};
    \node (2) at (-0.2,-1.0) {$3$};
    \node (3) at (0.75,-1.0) {$4$};
    \node (4) at ( 1.7,-1.0) {$5$};
    \node (5) at ( 1.7, 1.0) {$6$};
    \node (6) at (0.75, 1.0) {$1$};
    \node (a) at ( 0.0, 0.0) [circle,fill=black,inner sep=0pt,minimum size=3pt] {}; 
    \node (b) at (0.75, 0.0) [circle,fill=black,inner sep=0pt,minimum size=3pt] {};
    \node (c) at ( 1.5, 0.0) [circle,fill=black,inner sep=0pt,minimum size=3pt] {};
    \draw (1)--(a);
    \draw (2)--(a);
    \draw (3)--(c);
    \draw (4)--(c);
    \draw (5)--(c);
    \draw (6)--(b);
    \draw (a)--(b);
    \draw (b)--(c);
    \end{tikzpicture}
    +
    \begin{tikzpicture}[baseline=-0.5ex, scale=0.5, every node/.style={font=\scriptsize}]
    \node (1) at (-0.2, 1.0) {$3$};
    \node (2) at (-0.2,-1.0) {$4$};
    \node (3) at (0.75,-1.0) {$5$};
    \node (4) at ( 1.7,-1.0) {$6$};
    \node (5) at ( 1.7, 1.0) {$1$};
    \node (6) at (0.75, 1.0) {$2$};
    \node (a) at ( 0.0, 0.0) [circle,fill=black,inner sep=0pt,minimum size=3pt] {}; 
    \node (b) at (0.75, 0.0) [circle,fill=black,inner sep=0pt,minimum size=3pt] {};
    \node (c) at ( 1.5, 0.0) [circle,fill=black,inner sep=0pt,minimum size=3pt] {};
    \draw (1)--(a);
    \draw (2)--(a);
    \draw (3)--(c);
    \draw (4)--(c);
    \draw (5)--(c);
    \draw (6)--(b);
    \draw (a)--(b);
    \draw (b)--(c);
    \end{tikzpicture}
    +
    \begin{tikzpicture}[baseline=-0.5ex, scale=0.5, every node/.style={font=\scriptsize}]
    \node (1) at (-0.2, 1.0) {$4$};
    \node (2) at (-0.2,-1.0) {$5$};
    \node (3) at (0.75,-1.0) {$6$};
    \node (4) at ( 1.7,-1.0) {$1$};
    \node (5) at ( 1.7, 1.0) {$2$};
    \node (6) at (0.75, 1.0) {$3$};
    \node (a) at ( 0.0, 0.0) [circle,fill=black,inner sep=0pt,minimum size=3pt] {}; 
    \node (b) at (0.75, 0.0) [circle,fill=black,inner sep=0pt,minimum size=3pt] {};
    \node (c) at ( 1.5, 0.0) [circle,fill=black,inner sep=0pt,minimum size=3pt] {};
    \draw (1)--(a);
    \draw (2)--(a);
    \draw (3)--(c);
    \draw (4)--(c);
    \draw (5)--(c);
    \draw (6)--(b);
    \draw (a)--(b);
    \draw (b)--(c);
    \end{tikzpicture}
    +
    \begin{tikzpicture}[baseline=-0.5ex, scale=0.5, every node/.style={font=\scriptsize}]
    \node (1) at (-0.2, 1.0) {$5$};
    \node (2) at (-0.2,-1.0) {$6$};
    \node (3) at (0.75,-1.0) {$1$};
    \node (4) at ( 1.7,-1.0) {$2$};
    \node (5) at ( 1.7, 1.0) {$3$};
    \node (6) at (0.75, 1.0) {$4$};
    \node (a) at ( 0.0, 0.0) [circle,fill=black,inner sep=0pt,minimum size=3pt] {}; 
    \node (b) at (0.75, 0.0) [circle,fill=black,inner sep=0pt,minimum size=3pt] {};
    \node (c) at ( 1.5, 0.0) [circle,fill=black,inner sep=0pt,minimum size=3pt] {};
    \draw (1)--(a);
    \draw (2)--(a);
    \draw (3)--(c);
    \draw (4)--(c);
    \draw (5)--(c);
    \draw (6)--(b);
    \draw (a)--(b);
    \draw (b)--(c);
    \end{tikzpicture}
    +
    \begin{tikzpicture}[baseline=-0.5ex, scale=0.5, every node/.style={font=\scriptsize}]
    \node (1) at (-0.2, 1.0) {$6$};
    \node (2) at (-0.2,-1.0) {$1$};
    \node (3) at (0.75,-1.0) {$2$};
    \node (4) at ( 1.7,-1.0) {$3$};
    \node (5) at ( 1.7, 1.0) {$4$};
    \node (6) at (0.75, 1.0) {$5$};
    \node (a) at ( 0.0, 0.0) [circle,fill=black,inner sep=0pt,minimum size=3pt] {}; 
    \node (b) at (0.75, 0.0) [circle,fill=black,inner sep=0pt,minimum size=3pt] {};
    \node (c) at ( 1.5, 0.0) [circle,fill=black,inner sep=0pt,minimum size=3pt] {};
    \draw (1)--(a);
    \draw (2)--(a);
    \draw (3)--(c);
    \draw (4)--(c);
    \draw (5)--(c);
    \draw (6)--(b);
    \draw (a)--(b);
    \draw (b)--(c);
    \end{tikzpicture}
\no \\
    &\quad
    +
    \begin{tikzpicture}[baseline=-0.5ex, scale=0.5, every node/.style={font=\scriptsize}]
    \node (1) at (-0.2, 1.0) {$1$};
    \node (2) at (-0.2,-1.0) {$2$};
    \node (3) at ( 1.7,-1.0) {$3$};
    \node (4) at ( 1.7, 1.0) {$4$};
    \node (5) at ( 1.1, 1.0) {$5$};
    \node (6) at ( 0.4, 1.0) {$6$};
    \node (a) at ( 0.0, 0.0) [circle,fill=black,inner sep=0pt,minimum size=3pt] {}; 
    \node (b) at ( 0.75,0.0) [circle,fill=black,inner sep=0pt,minimum size=3pt] {};
    \node (c) at ( 1.5, 0.0) [circle,fill=black,inner sep=0pt,minimum size=3pt] {};
    \draw (1)--(a);
    \draw (2)--(a);
    \draw (3)--(c);
    \draw (4)--(c);
    \draw (5)--(b);
    \draw (6)--(b);
    \draw (a)--(b);
    \draw (b)--(c);
    \end{tikzpicture}
    +
    \begin{tikzpicture}[baseline=-0.5ex, scale=0.5, every node/.style={font=\scriptsize}]
    \node (1) at (-0.2, 1.0) {$2$};
    \node (2) at (-0.2,-1.0) {$3$};
    \node (3) at ( 1.7,-1.0) {$4$};
    \node (4) at ( 1.7, 1.0) {$5$};
    \node (5) at ( 1.1, 1.0) {$6$};
    \node (6) at ( 0.4, 1.0) {$1$};
    \node (a) at ( 0.0, 0.0) [circle,fill=black,inner sep=0pt,minimum size=3pt] {}; 
    \node (b) at ( 0.75,0.0) [circle,fill=black,inner sep=0pt,minimum size=3pt] {};
    \node (c) at ( 1.5, 0.0) [circle,fill=black,inner sep=0pt,minimum size=3pt] {};
    \draw (1)--(a);
    \draw (2)--(a);
    \draw (3)--(c);
    \draw (4)--(c);
    \draw (5)--(b);
    \draw (6)--(b);
    \draw (a)--(b);
    \draw (b)--(c);
    \end{tikzpicture}
    +
    \begin{tikzpicture}[baseline=-0.5ex, scale=0.5, every node/.style={font=\scriptsize}]
    \node (1) at (-0.2, 1.0) {$3$};
    \node (2) at (-0.2,-1.0) {$4$};
    \node (3) at ( 1.7,-1.0) {$5$};
    \node (4) at ( 1.7, 1.0) {$6$};
    \node (5) at ( 1.1, 1.0) {$1$};
    \node (6) at ( 0.4, 1.0) {$2$};
    \node (a) at ( 0.0, 0.0) [circle,fill=black,inner sep=0pt,minimum size=3pt] {}; 
    \node (b) at ( 0.75,0.0) [circle,fill=black,inner sep=0pt,minimum size=3pt] {};
    \node (c) at ( 1.5, 0.0) [circle,fill=black,inner sep=0pt,minimum size=3pt] {};
    \draw (1)--(a);
    \draw (2)--(a);
    \draw (3)--(c);
    \draw (4)--(c);
    \draw (5)--(b);
    \draw (6)--(b);
    \draw (a)--(b);
    \draw (b)--(c);
    \end{tikzpicture}
    +
    \begin{tikzpicture}[baseline=-0.5ex, scale=0.5, every node/.style={font=\scriptsize}]
    \node (1) at (-0.2, 1.0) {$4$};
    \node (2) at (-0.2,-1.0) {$5$};
    \node (3) at ( 1.7,-1.0) {$6$};
    \node (4) at ( 1.7, 1.0) {$1$};
    \node (5) at ( 1.1, 1.0) {$2$};
    \node (6) at ( 0.4, 1.0) {$3$};
    \node (a) at ( 0.0, 0.0) [circle,fill=black,inner sep=0pt,minimum size=3pt] {}; 
    \node (b) at ( 0.75,0.0) [circle,fill=black,inner sep=0pt,minimum size=3pt] {};
    \node (c) at ( 1.5, 0.0) [circle,fill=black,inner sep=0pt,minimum size=3pt] {};
    \draw (1)--(a);
    \draw (2)--(a);
    \draw (3)--(c);
    \draw (4)--(c);
    \draw (5)--(b);
    \draw (6)--(b);
    \draw (a)--(b);
    \draw (b)--(c);
    \end{tikzpicture}
    +
    \begin{tikzpicture}[baseline=-0.5ex, scale=0.5, every node/.style={font=\scriptsize}]
    \node (1) at (-0.2, 1.0) {$5$};
    \node (2) at (-0.2,-1.0) {$6$};
    \node (3) at ( 1.7,-1.0) {$1$};
    \node (4) at ( 1.7, 1.0) {$2$};
    \node (5) at ( 1.1, 1.0) {$3$};
    \node (6) at ( 0.4, 1.0) {$4$};
    \node (a) at ( 0.0, 0.0) [circle,fill=black,inner sep=0pt,minimum size=3pt] {}; 
    \node (b) at ( 0.75,0.0) [circle,fill=black,inner sep=0pt,minimum size=3pt] {};
    \node (c) at ( 1.5, 0.0) [circle,fill=black,inner sep=0pt,minimum size=3pt] {};
    \draw (1)--(a);
    \draw (2)--(a);
    \draw (3)--(c);
    \draw (4)--(c);
    \draw (5)--(b);
    \draw (6)--(b);
    \draw (a)--(b);
    \draw (b)--(c);
    \end{tikzpicture}
    +
    \begin{tikzpicture}[baseline=-0.5ex, scale=0.5, every node/.style={font=\scriptsize}]
    \node (1) at (-0.2, 1.0) {$6$};
    \node (2) at (-0.2,-1.0) {$1$};
    \node (3) at ( 1.7,-1.0) {$2$};
    \node (4) at ( 1.7, 1.0) {$3$};
    \node (5) at ( 1.1, 1.0) {$4$};
    \node (6) at ( 0.4, 1.0) {$5$};
    \node (a) at ( 0.0, 0.0) [circle,fill=black,inner sep=0pt,minimum size=3pt] {}; 
    \node (b) at ( 0.75,0.0) [circle,fill=black,inner sep=0pt,minimum size=3pt] {};
    \node (c) at ( 1.5, 0.0) [circle,fill=black,inner sep=0pt,minimum size=3pt] {};
    \draw (1)--(a);
    \draw (2)--(a);
    \draw (3)--(c);
    \draw (4)--(c);
    \draw (5)--(b);
    \draw (6)--(b);
    \draw (a)--(b);
    \draw (b)--(c);
    \end{tikzpicture}
\no \\
    &\quad
    +
    \begin{tikzpicture}[baseline=-0.5ex, scale=0.5, every node/.style={font=\scriptsize}]
    \node (1) at (-0.2, 1.0) {$1$};
    \node (2) at (-0.2,-1.0) {$2$};
    \node (3) at (0.75,-1.0) {$3$};
    \node (4) at ( 1.7,-1.0) {$4$};
    \node (5) at ( 1.7, 1.0) {$5$};
    \node (6) at (0.75, 1.0) {$6$};
    \node (a) at ( 0.0, 0.0) [circle,fill=black,inner sep=0pt,minimum size=3pt] {}; 
    \node (b) at (0.75, 0.0) [circle,fill=black,inner sep=0pt,minimum size=3pt] {};
    \node (c) at ( 1.5, 0.0) [circle,fill=black,inner sep=0pt,minimum size=3pt] {};
    \draw (1)--(a);
    \draw (2)--(a);
    \draw (3)--(b);
    \draw (4)--(c);
    \draw (5)--(c);
    \draw (6)--(b);
    \draw (a)--(b);
    \draw (b)--(c);
    \end{tikzpicture}
    +
    \begin{tikzpicture}[baseline=-0.5ex, scale=0.5, every node/.style={font=\scriptsize}]
    \node (1) at (-0.2, 1.0) {$2$};
    \node (2) at (-0.2,-1.0) {$3$};
    \node (3) at (0.75,-1.0) {$4$};
    \node (4) at ( 1.7,-1.0) {$5$};
    \node (5) at ( 1.7, 1.0) {$6$};
    \node (6) at (0.75, 1.0) {$1$};
    \node (a) at ( 0.0, 0.0) [circle,fill=black,inner sep=0pt,minimum size=3pt] {}; 
    \node (b) at (0.75, 0.0) [circle,fill=black,inner sep=0pt,minimum size=3pt] {};
    \node (c) at ( 1.5, 0.0) [circle,fill=black,inner sep=0pt,minimum size=3pt] {};
    \draw (1)--(a);
    \draw (2)--(a);
    \draw (3)--(b);
    \draw (4)--(c);
    \draw (5)--(c);
    \draw (6)--(b);
    \draw (a)--(b);
    \draw (b)--(c);
    \end{tikzpicture}
    +
    \begin{tikzpicture}[baseline=-0.5ex, scale=0.5, every node/.style={font=\scriptsize}]
    \node (1) at (-0.2, 1.0) {$3$};
    \node (2) at (-0.2,-1.0) {$4$};
    \node (3) at (0.75,-1.0) {$5$};
    \node (4) at ( 1.7,-1.0) {$6$};
    \node (5) at ( 1.7, 1.0) {$1$};
    \node (6) at (0.75, 1.0) {$2$};
    \node (a) at ( 0.0, 0.0) [circle,fill=black,inner sep=0pt,minimum size=3pt] {}; 
    \node (b) at (0.75, 0.0) [circle,fill=black,inner sep=0pt,minimum size=3pt] {};
    \node (c) at ( 1.5, 0.0) [circle,fill=black,inner sep=0pt,minimum size=3pt] {};
    \draw (1)--(a);
    \draw (2)--(a);
    \draw (3)--(b);
    \draw (4)--(c);
    \draw (5)--(c);
    \draw (6)--(b);
    \draw (a)--(b);
    \draw (b)--(c);
    \end{tikzpicture}
\no \\
    &\quad
    +
    \begin{tikzpicture}[baseline=-0.5ex, scale=0.5, every node/.style={font=\scriptsize}]
    \node (1) at (-0.2, 1.0) {$1$};
    \node (2) at (-0.2,-1.0) {$2$};
    \node (3) at (0.75,-1.0) {$3$};
    \node (4) at ( 1.7,-1.0) {$4$};
    \node (5) at ( 1.7, 1.0) {$5$};
    \node (6) at (0.75, 1.0) {$6$};
    \node (a) at ( 0.0, 0.0) [circle,fill=black,inner sep=0pt,minimum size=3pt] {}; 
    \node (b) at (1.225,0.0) [circle,fill=black,inner sep=0pt,minimum size=3pt] {};
    \draw (1)--(a);
    \draw (2)--(a);
    \draw (3)--(b);
    \draw (4)--(b);
    \draw (5)--(b);
    \draw (6)--(b);
    \draw (a)--(b);
    \end{tikzpicture}
    +
    \begin{tikzpicture}[baseline=-0.5ex, scale=0.5, every node/.style={font=\scriptsize}]
    \node (1) at (-0.2, 1.0) {$2$};
    \node (2) at (-0.2,-1.0) {$3$};
    \node (3) at (0.75,-1.0) {$4$};
    \node (4) at ( 1.7,-1.0) {$5$};
    \node (5) at ( 1.7, 1.0) {$6$};
    \node (6) at (0.75, 1.0) {$1$};
    \node (a) at ( 0.0, 0.0) [circle,fill=black,inner sep=0pt,minimum size=3pt] {}; 
    \node (b) at (1.225,0.0) [circle,fill=black,inner sep=0pt,minimum size=3pt] {};
    \draw (1)--(a);
    \draw (2)--(a);
    \draw (3)--(b);
    \draw (4)--(b);
    \draw (5)--(b);
    \draw (6)--(b);
    \draw (a)--(b);
    \end{tikzpicture}
    +
    \begin{tikzpicture}[baseline=-0.5ex, scale=0.5, every node/.style={font=\scriptsize}]
    \node (1) at (-0.2, 1.0) {$3$};
    \node (2) at (-0.2,-1.0) {$4$};
    \node (3) at (0.75,-1.0) {$5$};
    \node (4) at ( 1.7,-1.0) {$6$};
    \node (5) at ( 1.7, 1.0) {$1$};
    \node (6) at (0.75, 1.0) {$2$};
    \node (a) at ( 0.0, 0.0) [circle,fill=black,inner sep=0pt,minimum size=3pt] {}; 
    \node (b) at (1.225,0.0) [circle,fill=black,inner sep=0pt,minimum size=3pt] {};
    \draw (1)--(a);
    \draw (2)--(a);
    \draw (3)--(b);
    \draw (4)--(b);
    \draw (5)--(b);
    \draw (6)--(b);
    \draw (a)--(b);
    \end{tikzpicture}
    +
    \begin{tikzpicture}[baseline=-0.5ex, scale=0.5, every node/.style={font=\scriptsize}]
    \node (1) at (-0.2, 1.0) {$4$};
    \node (2) at (-0.2,-1.0) {$5$};
    \node (3) at (0.75,-1.0) {$6$};
    \node (4) at ( 1.7,-1.0) {$1$};
    \node (5) at ( 1.7, 1.0) {$2$};
    \node (6) at (0.75, 1.0) {$3$};
    \node (a) at ( 0.0, 0.0) [circle,fill=black,inner sep=0pt,minimum size=3pt] {}; 
    \node (b) at (1.225,0.0) [circle,fill=black,inner sep=0pt,minimum size=3pt] {};
    \draw (1)--(a);
    \draw (2)--(a);
    \draw (3)--(b);
    \draw (4)--(b);
    \draw (5)--(b);
    \draw (6)--(b);
    \draw (a)--(b);
    \end{tikzpicture}
    +
    \begin{tikzpicture}[baseline=-0.5ex, scale=0.5, every node/.style={font=\scriptsize}]
    \node (1) at (-0.2, 1.0) {$5$};
    \node (2) at (-0.2,-1.0) {$6$};
    \node (3) at (0.75,-1.0) {$1$};
    \node (4) at ( 1.7,-1.0) {$2$};
    \node (5) at ( 1.7, 1.0) {$3$};
    \node (6) at (0.75, 1.0) {$4$};
    \node (a) at ( 0.0, 0.0) [circle,fill=black,inner sep=0pt,minimum size=3pt] {}; 
    \node (b) at (1.225,0.0) [circle,fill=black,inner sep=0pt,minimum size=3pt] {};
    \draw (1)--(a);
    \draw (2)--(a);
    \draw (3)--(b);
    \draw (4)--(b);
    \draw (5)--(b);
    \draw (6)--(b);
    \draw (a)--(b);
    \end{tikzpicture}
    +
    \begin{tikzpicture}[baseline=-0.5ex, scale=0.5, every node/.style={font=\scriptsize}]
    \node (1) at (-0.2, 1.0) {$6$};
    \node (2) at (-0.2,-1.0) {$1$};
    \node (3) at (0.75,-1.0) {$2$};
    \node (4) at ( 1.7,-1.0) {$3$};
    \node (5) at ( 1.7, 1.0) {$4$};
    \node (6) at (0.75, 1.0) {$5$};
    \node (a) at ( 0.0, 0.0) [circle,fill=black,inner sep=0pt,minimum size=3pt] {}; 
    \node (b) at (1.225,0.0) [circle,fill=black,inner sep=0pt,minimum size=3pt] {};
    \draw (1)--(a);
    \draw (2)--(a);
    \draw (3)--(b);
    \draw (4)--(b);
    \draw (5)--(b);
    \draw (6)--(b);
    \draw (a)--(b);
    \end{tikzpicture}
\no \\
    &\quad
    +
    \begin{tikzpicture}[baseline=-0.5ex, scale=0.5, every node/.style={font=\scriptsize}]
    \node (1) at (-0.2, 1.0) {$1$};
    \node (2) at (-0.2,-1.0) {$2$};
    \node (3) at (0.75,-1.0) {$3$};
    \node (4) at ( 1.7,-1.0) {$4$};
    \node (5) at ( 1.7, 1.0) {$5$};
    \node (6) at (0.75, 1.0) {$6$};
    \node (a) at (0.375, 0.0) [circle,fill=black,inner sep=0pt,minimum size=3pt] {}; 
    \node (b) at (1.225,0.0) [circle,fill=black,inner sep=0pt,minimum size=3pt] {};
    \draw (1)--(a);
    \draw (2)--(a);
    \draw (3)--(a);
    \draw (4)--(b);
    \draw (5)--(b);
    \draw (6)--(b);
    \draw (a)--(b);
    \end{tikzpicture}
    +
    \begin{tikzpicture}[baseline=-0.5ex, scale=0.5, every node/.style={font=\scriptsize}]
    \node (1) at (-0.2, 1.0) {$2$};
    \node (2) at (-0.2,-1.0) {$3$};
    \node (3) at (0.75,-1.0) {$4$};
    \node (4) at ( 1.7,-1.0) {$5$};
    \node (5) at ( 1.7, 1.0) {$6$};
    \node (6) at (0.75, 1.0) {$1$};
    \node (a) at (0.375, 0.0) [circle,fill=black,inner sep=0pt,minimum size=3pt] {}; 
    \node (b) at (1.225,0.0) [circle,fill=black,inner sep=0pt,minimum size=3pt] {};
    \draw (1)--(a);
    \draw (2)--(a);
    \draw (3)--(a);
    \draw (4)--(b);
    \draw (5)--(b);
    \draw (6)--(b);
    \draw (a)--(b);
    \end{tikzpicture}
    +
    \begin{tikzpicture}[baseline=-0.5ex, scale=0.5, every node/.style={font=\scriptsize}]
    \node (1) at (-0.2, 1.0) {$3$};
    \node (2) at (-0.2,-1.0) {$4$};
    \node (3) at (0.75,-1.0) {$5$};
    \node (4) at ( 1.7,-1.0) {$6$};
    \node (5) at ( 1.7, 1.0) {$1$};
    \node (6) at (0.75, 1.0) {$2$};
    \node (a) at (0.375, 0.0) [circle,fill=black,inner sep=0pt,minimum size=3pt] {}; 
    \node (b) at (1.225,0.0) [circle,fill=black,inner sep=0pt,minimum size=3pt] {};
    \draw (1)--(a);
    \draw (2)--(a);
    \draw (3)--(a);
    \draw (4)--(b);
    \draw (5)--(b);
    \draw (6)--(b);
    \draw (a)--(b);
    \end{tikzpicture}
\no \\
    &\quad
    +
    \begin{tikzpicture}[baseline=-0.5ex, scale=0.5, every node/.style={font=\scriptsize}]
    \node (1) at (1*360/6+90:1.2) {$1$};
    \node (2) at (2*360/6+90:1.2) {$2$};
    \node (3) at (3*360/6+90:1.2) {$3$};
    \node (4) at (4*360/6+90:1.2) {$4$};
    \node (5) at (5*360/6+90:1.2) {$5$};
    \node (6) at (6*360/6+90:1.2) {$6$};
    \node (a) at (0,0) [circle,fill=black,inner sep=0pt,minimum size=3pt] {}; 
    \draw (1)--(a);
    \draw (2)--(a);
    \draw (3)--(a);
    \draw (4)--(a);
    \draw (5)--(a);
    \draw (6)--(a);
    \end{tikzpicture}
    \, .
\end{align}
Again, each line contains diagrams related by cyclic permutations of the external labels. The fourteen diagrams on the first three lines each have three three-point vertices and three propagators. The twenty-four diagrams on the fourth through seventh lines each have two three-point vertices, one four-point vertex, and two propagators. The six diagrams on the eighth line have one five-point vertex, one three-point vertex, and a single propagator. The three diagrams on the ninth line have two four-point vertices and a single propagator. The final diagram is a six-point contact term. Using the ordered Feynman rules, we find
\begin{align}
    \cA^{(6)}_{\text{AS}}
    (1,2,3,4,5,6)
    =
    - \tilde{g}^4
    \Lambda^{6-2d}
    \Bigg(
    &
    \eqmakebox[1][c]{$\displaystyle \frac{1}{ \mfs_{12} \mfs_{34} \mfs_{345} }$}
    + ( 5 \text{ cyclic perms.} )
\no \\
    \vphantom{\Bigg(}
    {}+{}
    &
    \eqmakebox[1][c]{$\displaystyle \frac{1}{ \mfs_{12} \mfs_{45} \mfs_{345} }$}
    + ( 5 \text{ cyclic perms.} )
\no \\
    \vphantom{\Bigg(}
    {}+{}
    &
    \eqmakebox[1][c]{$\displaystyle \frac{1}{ \mfs_{12} \mfs_{34} \mfs_{56} }$}
    + ( 1 \text{ cyclic perm.} )
\no \\
    \vphantom{\Bigg(}
    {}+{}
    &
    \eqmakebox[2][c]{$\displaystyle \frac{1}{ \mfs_{12} \mfs_{123} }$}
    + ( 5 \text{ cyclic perms.} )
\no \\
    \vphantom{\Bigg(}
    {}+{}
    &
    \eqmakebox[2][c]{$\displaystyle \frac{1}{ \mfs_{12} \mfs_{345} }$}
    + ( 5 \text{ cyclic perms.} )
\no \\
    \vphantom{\Bigg(}
    {}+{}
    &
    \eqmakebox[2][c]{$\displaystyle \frac{1}{ \mfs_{12} \mfs_{34} }$}
    + ( 5 \text{ cyclic perms.} )
\no \\
    \vphantom{\Bigg(}
    {}+{}
    &
    \eqmakebox[2][c]{$\displaystyle \frac{1}{ \mfs_{12} \mfs_{45} }$}
    + ( 2 \text{ cyclic perms.} )
\no \\
    \vphantom{\Bigg(}
    {}+{}
    &
    \eqmakebox[3][c]{$\displaystyle \frac{1}{ \mfs_{12} }$}
    + ( 5 \text{ cyclic perms.} )
\no \\
    \vphantom{\Bigg(}
    {}+{}
    &
    \eqmakebox[3][c]{$\displaystyle \frac{1}{ \mfs_{123} }$}
    + ( 2 \text{ cyclic perms.} )
    + 1
    \Bigg)
    \, ,
\end{align}
where $\mfs_{I} = (s_I - m_0^2)/\Lambda^2$ and ``$(m \text{ cyclic perms.})$" denotes the $m$ unique terms obtained by cyclically permuting the particles labels of the preceding term. This is precisely the expression for the six-point Baker-Coon-Romans partial amplitude in the limit $q \to \infty$ given in~\eqref{eq:Aqinf6}.

\subsection{\texorpdfstring{$N$}{N} points}

It would be straightforward to continue with an explicit calculation of the seven-point partial amplitude, but typesetting the ordered Feynman diagrams and the final expression would be quite difficult. Instead, we turn to the general $N$-point case.

We begin with a simple observation from the four-point, five-point, and six-point calculations. In these three cases, the $N$-point partial amplitude is given by an overall factor of $-\tilde{g}^{N-2} \Lambda^{N+d-\frac{1}{2}Nd}$ multiplying $1$ plus a sum of (dimensionless) propagators. The term without any propagators corresponds to the $N$-point vertex. Each term in the sum of propagators correspond to a unique set of mutually non-overlapping planar channels, i.e.\ the elements of the set \smash{$\cN^{(N)}_{[n]}$} defined in \autoref{sec:N} for $1 \leq n \leq N-3$. At four points, we simply have a sum over planar channels. At five points, we have a sum over planar channels and pairs of non-overlapping planar channels. At six points, we have a sum over planar channels, pairs of non-overlapping planar channels, and triples of non-overlapping planar channels. At higher points, the pattern continues. Crucially, the ordered Feynman diagrams with the canonical ordering $(1,2,\dots,N)$ are in one-to-one correspondence with the sets of mutually non-overlapping planar channels. Given a set of mutually non-overlapping planar channels, one can uniquely construct an ordered Feynman diagram. Hence, we can write the $N$-point partial amplitude as a sum over elements of the sets \smash{$\cN^{(N)}_{[n]}$}. Collecting the powers of $i$, $\tilde{g}$, and $\Lambda$ from the ordered Feynman rules, we find
\begin{align}
    \cA^{(N)}_{\text{AS}}
    (1,2,\dots,N)
    &=
    - \tilde{g}^{N-2}
    \Lambda^{N+d-\frac{1}{2}Nd}
    \Bigg(
    1
    + \sum_{n=1}^{N-3}
    \sum_{ \{I_1, \dots, I_n\} \in \cN^{(N)}_{[n]}}
    \frac{1}{ \mfs_{I_1} \dots \mfs_{I_n} }
    \Bigg)
    \, ,
\end{align}
which is precisely the expression for the $N$-point Baker-Coon-Romans partial amplitude in the limit $q \to \infty$ given in~\eqref{eq:AqinfN}. Summing the partial amplitudes over trace structures then leads to the equality~\eqref{eq:AqAAS1} between the full amplitudes in either theory. In other words, we have shown that the $N$-point tree-level amplitudes of the $\mathrm{U}(N_F)$ adjoint scalar theory are exactly equal to the $N$-point Baker-Coon-Romans amplitudes in the limit $q \to \infty$. 

\subsection{\texorpdfstring{$\mathrm{SU}(N_F)$}{SU(NF)}}

Before concluding this section, we briefly discuss the adjoint scalar theory with $\mathrm{SU}(N_F)$ global symmetry group (with arbitrary $N_F \geq 2$). The unitary and special unitary groups differ only by $1/N_F$ corrections, so at large $N_F$ we expect to find a similar equality between the tree-level $\mathrm{SU}(N_F)$ adjoint scalar amplitudes and the Baker-Coon-Romans amplitudes in the limit $q \to \infty$.

The Feynman rules for the $\mathrm{SU}(N_F)$ theory are the same as those for the $\mathrm{U}(N_F)$ theory. Only the $\mathrm{U}(1) \subset \mathrm{U}(N_F)$ generator $T^0 \propto \mathds{1}$ needs to be thrown out. The remaining $\mathrm{SU}(N_F)$ generators $T^a$ (which form a complete set of $N_F \times N_F$ traceless Hermitian matrices), obey the completeness relation
\begin{align}
    {(T^a)_i}^j
    {(T^a)_k}^\ell
    &=
    {\delta_i}^\ell
    {\delta_k}^j
    -
    \frac{1}{N_F}
    {\delta_i}^j
    {\delta_k}^\ell
    \, , 
\end{align}
which allows us to combine any products of traces connected by a propagator as follows,
\begin{align}
    \Tr(X T^a) \Tr (T^a Y) 
    &=
    \Tr(XY)
    - \frac{1}{N_F}
    \Tr(X) \Tr(Y)
    \, .
\end{align}
In the end, both single and multi-trace structures remain in the tree-level amplitudes, but multi-traces are suppressed by $1/N_F$. At leading order in $1/N_F$, the single traces can be stripped off, and the tree-level amplitudes can be written as a sum over ordered partial amplitudes just as in the $\mathrm{U}(N_F)$ case. These partial amplitudes are precisely equal to those calculated in the $\mathrm{U}(N_F)$ theory. Hence,
\begin{align}
    \cA^{(N) \, \text{tree}}
       _{\text{AS}; \, \mathrm{SU}(N_F)}
    &=
    \cA^{(N) \, \text{tree}}
       _{\text{AS}; \, \mathrm{U}(N_F)}
    +
    \cO \big( N_F^{-1} \big)
    \, .
\end{align}
Summing the partial amplitudes over trace structures and then using~\eqref{eq:AqAAS1} leads to the equality~\eqref{eq:AqAAS2} (at leading order in $1/N_F$) between the full $\mathrm{SU}(N_F)$ adjoint scalar amplitudes and the Baker-Coon-Romans amplitudes in the limit $q \to \infty$.


\section{Discussion}
\label{sec:disc}

In this paper, we studied the $N$-point Baker-Coon-Romans amplitude. We reviewed its convergence, duality, and factorization properties, and we computed its $q \to \infty$ limit. Although the Baker-Coon-Romans formula is only valid for ${q > 1}$, we showed that the four-point case admits a straightforward extension to all ${q \geq 0}$ which reproduces the usual expression for the four-point Coon amplitude. At five points, we carried out a similar procedure but found inconsistencies with duality and factorization when ${q < 1}$. Despite these issues, we found a new relation between the five-point Baker-Coon-Romans amplitude and the four-point basic hypergeometric amplitude, analogous to the known relation between the five-point tree-level open string amplitude and the four-point hypergeometric amplitude. Finally, we discovered an exact correspondence between the $q \to \infty$ limit of the Baker-Coon-Romans amplitudes and the field theory amplitudes of a scalar transforming in the adjoint representation of a global symmetry group with an infinite set of non-derivative single-trace interaction terms. This correspondence at $q = \infty$ is the first definitive realization of the Coon amplitude (in any limit) from a field theory described by a Lagrangian.

Despite our results and the recent burst of interest in Coon amplitudes, there are still many open problems and possible future directions in the study of Coon amplitudes.

For instance, there is still no consistent formulation of the general $N$-point Coon amplitude with ${q < 1}$. Although there are no problems with the Baker-Coon-Romans formulation at four-points, we failed to derive a duality-invariant five-point Coon amplitude with $q < 1$ from the five-point Baker-Coon-Romans formula. There is, however, an old proposal in~\cite{GonzalezMestres:1975ord} for an $N$-point Coon amplitude with $q < 1$, but it is unclear whether this formulation is consistent. We hope to analyze this old proposal in future work.

Orthogonally, we have only scratched the surface of the field theory limit of the Coon amplitude. Although we have identified the field theory in the strict $q \to \infty$ limit, we have not discussed any of the higher derivative interactions which should arise from integrating out the higher mass particles in the Coon spectrum. Moreover, we have said nothing of the field theory limit at finite $q$. Fortunately, we have data at $q \to \infty$, $q = 1$, and $q \to 0$. The $q \to \infty$ Lagrangian is given in~\eqref{eq:LagAS1} and is valid up to $\cO(q^{-1})$ and higher-derivative corrections. At $q = 1$, the Coon amplitude reduces to a tree-level open string amplitude, and it has been long known that the field theory limit (i.e.\ the low-energy or $\alpha' \to 0$ limit) of the relevant dual resonance model is a $\phi^3$ theory with no higher-point non-derivative interactions~\cite{Scherk:1971xy}. At $q \to 0$, we only have reliable data from the four-point amplitude~\eqref{eq:Aq04}, which has the same structure as the four-point $q \to \infty$ amplitude but with the sign of the four-point interaction reversed. If we restore the hidden powers of $q$ within the coupling constant $\tilde{g}$ and the scale $\Lambda$ in favor of the original coupling $g$ and scale $\mu$, then we can rewrite the Lagrangians at $q \to \infty$, $q = 1$, and $q \to 0$ as follows,
\begin{align}
    \cL_{q \to \infty}
    &=
    \cL_{\text{free}}
    - \sum_{n \geq 3}
    \frac{1}{n}
    q^{n-3}
    \big[
    1 + \cO(q^{-1})
    \big]
    g^{n-2}
    \mu^{n+d-\frac{1}{2}nd} 
    \Tr \phi^n
    + \text{(h.d.)}
    + \cO(q^{-1})
    \, ,
\no \\[1ex]
    \cL_{q=1}
    &=
    \cL_{\text{free}}
    - \frac{1}{3}
    g
    \mu^{3-d/2}
    \Tr \phi^3
    \vphantom{\sum_{n = 3}^{4}}
    + \text{(h.d.)}
    \, ,
\\[1ex] \no
    \cL_{q \to 0}
    &=
    \cL_{\text{free}}
    - \sum_{n = 3}^{4}
    \frac{1}{n}
    (-q^{-1})^{n-3}
    \big[
    1 + \cO(q)
    \big]
    g^{n-2}
    \mu^{n+d-\frac{1}{2}nd} 
    \Tr \phi^n
    + \cO(\phi^5)
    + \text{(h.d.)}
    + \cO(q)
    \, ,
\end{align}
where (h.d.) refers to higher-derivative corrections and
\begin{align}
    \cL_{\text{free}}
    =
    - \frac{1}{2} \Tr \p_\mu \phi \, \p^\mu \phi
    - \frac{1}{2} \, m_0^2 \Tr \phi^2
    \, .
\end{align}
From this evidence, we conjecture that the low-energy field theory limit of the Coon amplitude with generic $q$ is given by the following Lagrangian,
\begin{align}
    \cL_{q}
    &=
    \cL_{\text{free}}
    - \sum_{n \geq 3}
    \frac{1}{n}
    (q-q^{-1})^{n-3}
    g^{n-2}
    \mu^{n+d-\frac{1}{2}nd} 
    \Tr \phi^n
    + \text{(h.d.)}
    \, .
\end{align}
This conjecture matches all of our data at $q \to \infty$, $q = 1$, and $q \to 0$. Just like the $q \to \infty$ Lagrangian~\eqref{eq:LagAS1}, this Lagrangian can be resummed, yielding a logarithmic interaction term which may be amenable to a semiclassical analysis. We leave this analysis to future work. Perhaps this conjecture can even lead to a consistent formulation of the Coon amplitude with $q < 1$. We hope to address these questions in future work. 

Even if our conjecture is wrong or if there is no consistent Coon amplitude with $q < 1$, the correspondence at $q \to \infty$ is an exact result. Thus, it would be interesting to further study the adjoint scalar field theory~\eqref{eq:LagAS1} which lives at $q = \infty$. To this end, we present a brief classical analysis of this theory in \autoref{sec:app2}.

It would also be interesting to study the transcendental properties of the low-energy expansion of the $N$-point Coon amplitudes at generic $q$ as was done for four-point string amplitudes in~\cite{Schlotterer:2012ny, DHoker:2019blr, DHoker:2021ous}, for the four-point Coon amplitude in~\cite{Geiser:2022icl}, and for the four-point hypergeometric amplitude in~\cite{Cheung:2023adk}.

In many ways, the Coon amplitude remain mysterious. We hope the current flurry of interest answers the most pressing open questions before the Coon amplitude fades back into the literature.


\subsection*{Acknowledgements}

We are grateful to L.~Lindwasser for collaboration on related work. We thank H.~Elvang, L.~Lindwasser, and G.~Remmen for useful comments. We also thank our JHEP reviewer for helpful comments and the recommendation to include \autoref{sec:app2}. NG is supported by a Leinweber Postdoctoral Research Fellowship and a Van Loo Postdoctoral Research Fellowship along with partial support from Department of Energy grant DE-SC0007859.


\appendix

\section{Overlapping and non-overlapping sets}
\label{sec:app}

In this appendix, we derive explicit expressions for the sets of overlapping and non-overlapping planar channels defined at the end of \autoref{sec:N}. 

\subsection{Ordering the planar channels}

We begin by recalling the set of planar channels $\cC^{(N)}$ given in~\eqref{eq:CN}. For ${N = 3,4,5,6}$ (the cases which we explicitly consider in this paper), the sets of planar channels are given by
\begin{align}
    \cC^{(3)}
    &=
    \emptyset
    \, ,
\no \\
    \cC^{(4)}
    &=
    \{
    (1,2),
    (2,3)
    \}
    \, ,
\no \\
    \cC^{(5)}
    &=
    \{
    (1,2),
    (1,3),
    (2,3),
    (2,4),
    (3,4)
    \}
    \, ,
\no \\
    \cC^{(6)}
    &=
    \{
    (1,2),
    (1,3),
    (1,4),
    (2,3),
    (2,4),
    (2,5),
    (3,4),
    (3,5),
    (4,5)
    \}
    \, .
\end{align}
For general $N$, the elements of $\cC^{(N)}$ may be ordered as follows:
\begin{align}
\label{eq:order}
    (i,j)
    &=
    (k,\ell)
    \qquad 
    \Longleftrightarrow
    \qquad 
    i = k
    \text{ and }
    j = \ell
\no \\ 
    (i,j)
    &<
    (k,\ell)
    \qquad 
    \Longleftrightarrow
    \qquad 
    i < k
    \text{ or }
    (
    i = k
    \text{ and }
    j < \ell
    )
\no \\ 
    (i,j)
    &>
    (k,\ell)
    \qquad 
    \Longleftrightarrow
    \qquad 
    i > k
    \text{ or }
    (
    i = k
    \text{ and }
    j > \ell
    )
\end{align}
It will also be convenient to visualize this set as follows:
\begin{align}
\label{eq:CNvisual}
\begin{matrix*}[l]
    (1,2) \\
    (1,3) & (2,3)\\
    \phantom{(1} \vdots & \phantom{(2} \vdots & \ddots\\
    (1,N\!-\!2) & (2,N\!-\!2) & \cdots & (N\!-\!3,N\!-\!2)\\
    & (2,N\!-\!1) & \cdots & (N\!-\!3,N\!-\!1) & (N\!-\!2,N\!-\!1)
\end{matrix*}
\end{align}
The ordering defined in~\eqref{eq:order} proceeds down each column of~\eqref{eq:CNvisual}, from left to right. 

\subsection{Overlapping pairs}

To construct an explicit expression for the set $\cO^{(N)}$ whose elements are the two-element subsets $\{ (i,j), (k,\ell) \} \subset \cC^{(N)}$ with  $(i,j),(k,\ell)$ overlapping, we begin by considering a single planar channel. Given any planar channel $(i,j) \in \cC^{(N)}$, we may partition the full set of planar channels into the disjoint sets of its overlapping channels \smash{$\cO^{(N)}_{(i,j)}$} and non-overlapping channels \smash{$\cN^{(N)}_{(i,j)}$},
\begin{align}
    \cC^{(N)}
    =
    \{ (i,j) \}
    \cup
    \cO^{(N)}_{(i,j)}
    \cup
    \cN^{(N)}_{(i,j)}
    \, .
\end{align}
The sets of overlapping and non-overlapping planar channels can be further partitioned into sets of channels ``less than" and ``greater than" $(i,j)$ using the ordering defined in~\eqref{eq:order},
\begin{align}
    \cO^{(N)}_{(i,j)}
    &=
    \cO^{(N)}_{<(i,j)}
    \cup
    \cO^{(N)}_{>(i,j)}
    \, ,
    &
    \cN^{(N)}_{(i,j)}
    &=
    \cN^{(N)}_{<(i,j)}
    \cup
    \cN^{(N)}_{>(i,j)}
    \, .
\end{align}
The four sets $\cO^{(N)}_{<(i,j)}$, $\cO^{(N)}_{>(i,j)}$, $\cN^{(N)}_{<(i,j)}$, $\cN^{(N)}_{>(i,j)}$ are all mutually disjoint. We may write down explicit expressions for these sets using the visualization of $\cC^{(N)}$ given in~\eqref{eq:CNvisual} (now with some additional detail): 
\begin{align}
\scriptscriptstyle
\begin{smallmatrix*}[l]
    \color{blue}{(1,2)}
\\
    \color{blue}{(1,3)}
    &
    \color{blue}{(2,3)}
\\
    \color{blue}{\phantom{(1} \tvdots}
    &
    \color{blue}{\phantom{(2} \tvdots}
    &
    \color{blue}{\tddots}
\\
    \color{blue}{(1,i\!-\!1)}
    &
    \color{blue}{(2,i\!-\!1)}
    &
    \color{blue}{\cdots}
    &
    \color{blue}{(i\!-\!2,i\!-\!1)}
\\
    \color{red}{(1,i)}
    &
    \color{red}{(2,i)}
    &
    \color{red}{\cdots}
    &
    \color{red}{(i\!-\!2,i)}
    &
    \color{red}{(i\!-\!1,i)}
\\
    \color{red}{(1,i\!+\!1)}
    &
    \color{red}{(2,i\!+\!1)}
    &
    \color{red}{\cdots}
    &
    \color{red}{(i\!-\!2,i\!+\!1)}
    &
    \color{red}{(i\!-\!1,i\!+\!1)}
    &
    \color{blue}{(i,i\!+\!1)}
\\
    \color{red}{(1,i\!+\!2)}
    &
    \color{red}{(2,i\!+\!2)}
    &
    \color{red}{\cdots}
    &
    \color{red}{(i\!-\!2,i\!+\!2)}
    &
    \color{red}{(i\!-\!1,i\!+\!2)}
    &
    \color{blue}{(i,i\!+\!2)}
    &
    \color{blue}{(i\!+\!1,i\!+\!2)}
\\
    \color{red}{\phantom{(1} \tvdots}
    &
    \color{red}{\phantom{(2} \tvdots}
    &
    \color{red}{\phantom{.} \tvdots}
    &
    \color{red}{\phantom{(i\!-\!2} \tvdots}
    &
    \color{red}{\phantom{(i\!-\!1} \tvdots}
    &
    \color{blue}{\phantom{(i} \tvdots}
    &
    \color{blue}{\phantom{(i\!+\!1} \tvdots}
    &
    \color{blue}{\tddots}
\\
    \color{red}{(1,j\!-\!1)}
    &
    \color{red}{(2,j\!-\!1)}
    &
    \color{red}{\cdots}
    &
    \color{red}{(i\!-\!2,j\!-\!1)}
    &
    \color{red}{(i\!-\!1,j\!-\!1)}
    &
    \color{blue}{(i,j\!-\!1)}
    &
    \color{blue}{(i\!+\!1,j\!-\!1)}
    &
    \color{blue}{\cdots}
    &
    \color{blue}{(j\!-\!2,j\!-\!1)}
\\
    \color{blue}{(1,j)}
    &
    \color{blue}{(2,j)}
    &
    \color{blue}{\cdots}
    &
    \color{blue}{(i\!-\!2,j)}
    &
    \color{blue}{(i\!-\!1,j)}
    &
    \boldsymbol{(i,j)}
    &
    \color{blue}{(i\!+\!1,j)}
    &
    \color{blue}{\cdots}
    &
    \color{blue}{(j\!-\!2,j)}
    &
    \color{blue}{(j\!-\!1,j)}
\\
    \color{blue}{(1,j\!+\!1)}
    &
    \color{blue}{(2,j\!+\!1)}
    &
    \color{blue}{\cdots}
    &
    \color{blue}{(i\!-\!2,j\!+\!1)}
    &
    \color{blue}{(i\!-\!1,j\!+\!1)}
    &
    \color{blue}{(i,j\!+\!1)}
    &
    \color{red}{(i\!+\!1,j\!+\!1)}
    &
    \color{red}{\cdots}
    &
    \color{red}{(j\!-\!2,j\!+\!1)}
    &
    \color{red}{(j\!-\!1,j\!+\!1)}
    &
    \color{red}{(j,j\!+\!1)}
\\
    \color{blue}{(1,j\!+\!2)}
    &
    \color{blue}{(2,j\!+\!2)}
    &
    \color{blue}{\cdots}
    &
    \color{blue}{(i\!-\!2,j\!+\!2)}
    &
    \color{blue}{(i\!-\!1,j\!+\!2)}
    &
    \color{blue}{(i,j\!+\!2)}
    &
    \color{red}{(i\!+\!1,j\!+\!2)}
    &
    \color{red}{\cdots}
    &
    \color{red}{(j\!-\!2,j\!+\!2)}
    &
    \color{red}{(j\!-\!1,j\!+\!2)}
    &
    \color{red}{(j,j\!+\!2)}
    &
    \color{blue}{(j\!+\!1,j\!+\!2)}
\\
    \color{blue}{\phantom{(1} \tvdots}
    &
    \color{blue}{\phantom{(2} \tvdots}
    &
    \color{blue}{\phantom{.} \tvdots}
    &
    \color{blue}{\phantom{(i\!-\!2} \tvdots}
    &
    \color{blue}{\phantom{(i\!-\!1} \tvdots}
    &
    \color{blue}{\phantom{(i} \tvdots}
    &
    \color{red}{\phantom{(i\!+\!1} \tvdots}
    &
    \color{red}{\phantom{.} \tvdots}
    &
    \color{red}{\phantom{(j\!-\!2} \tvdots}
    &
    \color{red}{\phantom{(j\!-\!1} \tvdots}
    &
    \color{red}{\phantom{(j} \tvdots}
    &
    \color{blue}{\phantom{(j\!+\!1} \tvdots}
    &
    \color{blue}{\tddots}
\\
    \color{blue}{(1,N\!-\!2)} 
    &
    \color{blue}{(2,N\!-\!2)}
    &
    \color{blue}{\cdots}
    &
    \color{blue}{(i\!-\!2,N\!-\!2)}
    &
    \color{blue}{(i\!-\!1,N\!-\!2)}
    &
    \color{blue}{(i,N\!-\!2)}
    &
    \color{red}{(i\!+\!1,N\!-\!2)}
    &
    \color{red}{\cdots}
    &
    \color{red}{(j\!-\!2,N\!-\!2)}
    &
    \color{red}{(j\!-\!1,N\!-\!2)}
    &
    \color{red}{(j,N\!-\!2)}
    &
    \color{blue}{(j\!+\!1,N\!-\!2)}
    &
    \color{blue}{\cdots}
    &
    \color{blue}{(N\!-\!3,N\!-\!2)}
\\
    &
    \color{blue}{(2,N\!-\!1)}
    &
    \color{blue}{\cdots}
    &
    \color{blue}{(i\!-\!2,N\!-\!1)}
    &
    \color{blue}{(i\!-\!1,N\!-\!1)}
    &
    \color{blue}{(i,N\!-\!1)}
    &
    \color{red}{(i\!+\!1,N\!-\!1)}
    &
    \color{red}{\cdots}
    &
    \color{red}{(j\!-\!2,N\!-\!1)}
    &
    \color{red}{(j\!-\!1,N\!-\!1)}
    &
    \color{red}{(j,N\!-\!1)}
    &
    \color{blue}{(j\!+\!1,N\!-\!1)}
    &
    \color{blue}{\cdots}
    &
    \color{blue}{(N\!-\!3,N\!-\!1)}
    &
    \color{blue}{(N\!-\!2,N\!-\!1)}
\end{smallmatrix*}
\no
\end{align}
Here the red channels are those which overlap with $(i,j)$, and the blue channels are those which do not. Examining the visualization above, we find the following explicit expressions for the two sets of overlapping channels,
\begin{align}
\label{eq:O<O>}
    \cO^{(N)}_{<(i,j)}
    &=
    \{
    (k,\ell)
    \, : \,
    1 \leq k \leq i-1
    \, , \,
    i \leq \ell \leq j-1
    \}
    \, ,
\no \\[1ex]
    \cO^{(N)}_{>(i,j)}
    &=
    \{
    (k,\ell)
    \, : \,
    i+1 \leq k \leq j
    \, , \,
    j+1 \leq \ell \leq N-1
    \}
    \, ,
\end{align}
and for the two sets of non-overlapping channels,
\begin{align}
\label{eq:N<N>}
    \cN^{(N)}_{<(i,j)}
    =
    \phantom{\,\cup}
    &
    \{
    (k,\ell)
    \, : \,
    1 \leq k \leq i-1
    \, , \,
    k+1 \leq \ell \leq i-1
    \}
\no \\
    \vphantom{\cN^{(N)}_{>(i,j)}}
    \,\cup\,
    &
    \{
    (k,\ell)
    \, : \,
    1 \leq k \leq i-1
    \, , \,
    j \leq \ell \leq N-1
    \, , \,
    (k,\ell) \neq (1,N-1)
    \}
\no \\
    \vphantom{\cN^{(N)}_{>(i,j)}}
    \,\cup\,
    &
    \{
    (k,\ell)
    \, : \,
    k=i
    \, , \,
    k+1 \leq \ell \leq j-1
    \}
    \, ,
\no \\[1ex]
    \cN^{(N)}_{>(i,j)}
    =
    \phantom{\,\cup}
    &
    \{
    (k,\ell)
    \, : \,
    k=i
    \, , \,
    j+1 \leq \ell \leq N-1
    \, , \,
    (k,\ell) \neq (1,N-1)
    \}
\no \\
    \vphantom{\cN^{(N)}_{>(i,j)}}
    \,\cup\,
    &
    \{
    (k,\ell)
    \, : \,
    i+1 \leq k \leq j
    \, , \,
    k+1 \leq \ell \leq j
    \}
\no \\
    \smash[b]{\vphantom{\cN^{(N)}_{>(i,j)}}}
    \,\cup\,
    &
    \{
    (k,\ell)
    \, : \,
    j+1 \leq k \leq N-2
    \, , \,
    k+1 \leq \ell \leq N-1
    \}
    \, .
\end{align}
The set of distinct pairs of overlapping channels $\cO^{(N)}$ can now be constructed by considering each channel $(i,j) \in \cC^{(N)}$ and forming the set of overlapping pairs $\{ (i,j), (k,\ell) \}$ with ${(k,\ell) > (i,j)}$ so that each pair is counted just once. In other words,
\begin{align}
\label{eq:ON}
   \cO^{(N)}
   &=
   \Big\{
   \{ (i,j), (k,\ell) \}
   \, : \,
   (i,j) \in \cC^{(N)}
   \, , \,
   (k,\ell) \in \cO^{(N)}_{>(i,j)}
   \Big\}
\no \\
    &=
    \Big\{
    \{ (i,j), (k,\ell) \}
    \, : \,
    1 \leq i < k \leq j < \ell \leq N-1
    \Big\}
    \, ,
\end{align}
where we have used~\eqref{eq:CN} and the second line of~\eqref{eq:O<O>} to write the second equality. The number of distinct pairs of overlapping channels is then given by,
\begin{align}
    |\cO^{(N)}|
    =
    \eqmakebox[0][c]{$\displaystyle\sum_{i=1      \mathstrut}^{N-2} \,$}
    \eqmakebox[0][c]{$\displaystyle\sum_{j=i+1    \mathstrut}^{N-1} \,$}
    \eqmakebox[0][c]{$\displaystyle\sum_{k=i+1    \mathstrut}^{j  } \,$}
    \eqmakebox[0][c]{$\displaystyle\sum_{\ell=j+1 \mathstrut}^{N-1} \,$}
    1
    =
    \frac{1}{24} N(N-1)(N-2)(N-3)
    \, , 
\end{align}
which grows more quickly than the number of planar channels~$|\cC^{(N)}| = \half N(N-3)$.

\subsection{Non-overlapping \texorpdfstring{$n$}{n}-tuples}

To construct an explicit expression for the set \smash{$\cN^{(N)}_{[n]}$} whose elements are the $n$-element subsets $\{ (i_1,j_1), \dots, (i_n,j_n) \} \subset \cC^{(N)}$ with all $(i_\ell,j_\ell)$ non-overlapping, we begin with the set of non-overlapping pairs of planar channels (i.e.\ the case with $n=2$). We can list each distinct pair of non-overlapping planar channels once and only once, just as in our construction of $\cO^{(N)}$ above, by
\begin{align}
    \cN^{(N)}_{[2]}
    &=
    \Big\{
    \{ (i_1,j_1), (i_2,j_2) \}
    \, : \,
    \{ (i_1,j_1) \} \in \cN^{(N)}_{[1]}
    \, , \,
    (i_2,j_2) \in \cN^{(N)}_{>(i_1,j_1)}
    \Big\}
    \, .
\end{align}
The sets of non-overlapping $n$-tuples with $n \geq 3$ are then defined recursively by,
\begin{align}
    \cN^{(N)}_{[n]}
    =
    \Big\{
    \{ (i_1,j_1), \dots, (i_n,j_n) \}
    \, : \,
    \{ (i_1,j_1), \dots, (i_{n-1},j_{n-1}) \}
        &\in \cN^{(N)}_{[n-1]}
    \, , \,
\no \\
    (i_n,j_n)
        &\in \bigcap_{\ell = 1}^{n-1}
        \cN^{(N)}_{>(i_\ell,j_\ell)}
    \Big\}
    \, .
\end{align}
The sets \smash{$\cN^{(N)}_{[n]}$} with $n \geq N-2$ are empty. In other words, there are at most $N-3$ mutually non-overlapping planar channels in $N$-point scattering.

\subsection{Examples}

We conclude this section by tabulating the non-empty sets defined above for $N=3,4,5,6$. Pairs of overlapping planar channels first appear at $N=4$:
\begin{align}
    \cO^{(4)}
    &=
    \Big\{
    \{ (1,2), (2,3) \}
    \Big\}
    \, ,
\no \\
    \cO^{(5)}
    &=
    \Big\{
    \{ (1,2), (2,3) \},
    \{ (1,2), (2,4) \},
    \{ (1,3), (2,4) \},
    \{ (1,3), (3,4) \},
    \{ (2,3), (3,4) \}
    \Big\}
    \, ,
\no \\
    \cO^{(6)}
    &=
    \Big\{
    \{ (1,2), (2,3) \},
    \{ (1,2), (2,4) \},
    \{ (1,2), (2,5) \},
    \{ (1,3), (2,4) \},
    \{ (1,3), (2,5) \},
\no \\
    & \quad
    \phantom{\Big\{} \,\,
    \{ (1,3), (3,4) \},
    \{ (1,3), (3,5) \},
    \{ (1,4), (2,5) \},
    \{ (1,4), (3,5) \},
    \{ (1,4), (4,5) \},
\no \\
    & \quad
    \phantom{\Big\{} \,\,
    \{ (2,3), (3,4) \},
    \{ (2,3), (3,5) \},
    \{ (2,4), (3,5) \},
    \{ (2,4), (4,5) \},
\no \\
    & \quad
    \phantom{\Big\{} \,\,
    \{ (3,4), (4,5) \}
    \Big\}
    \, .
\intertext{Pairs of non-overlapping planar channels first appear at $N=5$:}
    \cN^{(5)}_{[2]}
    &=
    \Big\{
    \{ (1,2), (1,3) \},
    \{ (1,2), (3,4) \},
    \{ (1,3), (2,3) \},
    \{ (2,3), (2,4) \},
    \{ (2,4), (3,4) \}
    \Big\}
    \, ,
\no \\
    \cN^{(6)}_{[2]}
    &=
    \Big\{
    \{ (1,2), (1,3) \},
    \{ (1,2), (1,4) \},
    \{ (1,2), (3,4) \},
    \{ (1,2), (3,5) \},
    \{ (1,2), (4,5) \}
\no \\
    & \quad
    \phantom{\Big\{} \,\,
    \{ (1,3), (1,4) \},
    \{ (1,3), (2,3) \},
    \{ (1,3), (4,5) \},
    \{ (1,4), (2,3) \},
    \{ (1,4), (2,4) \},
\no \\
    & \quad
    \phantom{\Big\{} \,\,
    \{ (1,4), (3,4) \},
    \{ (2,3), (2,4) \},
    \{ (2,3), (2,5) \},
    \{ (2,3), (4,5) \},
    \{ (2,4), (2,5) \},
\no \\
    & \quad
    \phantom{\Big\{} \,\,
    \{ (2,4), (3,4) \},
    \{ (2,5), (3,4) \},
    \{ (2,5), (3,5) \},
    \{ (2,5), (4,5) \},
    \{ (3,4), (3,5) \},
\no \\
    & \quad
    \phantom{\Big\{} \,\,
    \{ (3,5), (4,5) \}
    \Big\}
    \, .
\intertext{Triples of non-overlapping planar channels first appear at $N=6$:}
    \cN^{(6)}_{[3]}
    &=
    \Big\{
    \{ (1,2), (1,3), (1,4) \},
    \{ (1,2), (1,3), (4,5) \},
    \{ (1,2), (1,4), (3,4) \},
\no \\
    & \quad
    \phantom{\Big\{} \,\,    
    \{ (1,2), (3,4), (3,5) \},
    \{ (1,2), (3,5), (4,5) \},
    \{ (1,3), (1,4), (2,3) \},
\no \\
    & \quad
    \phantom{\Big\{} \,\,    
    \{ (1,3), (2,3), (4,5) \},
    \{ (1,4), (2,3), (2,4) \},
    \{ (1,4), (2,4), (3,4) \},
\no \\
    & \quad
    \phantom{\Big\{} \,\,
    \{ (2,3), (2,4), (2,5) \},
    \{ (2,3), (2,5), (4,5) \},
    \{ (2,4), (2,5), (3,4) \},
\no \\
    & \quad
    \phantom{\Big\{} \,\,
    \{ (2,5), (3,4), (3,5) \},
    \{ (2,5), (3,5), (4,5) \}
    \Big\}
    \, .
\end{align}
Quadruples of non-overlapping planar channels first appear at $N=7$, but we do not explicitly need these sets in this paper.

\section{Classical analysis of the adjoint scalar theory}
\label{sec:app2}

In this appendix, we briefly analyze the adjoint scalar theory Lagrangian~\eqref{eq:LagAS1} and show that it has a stable vacuum (at least classically). We begin by recalling the resummed expression for the Lagrangian, 
\begin{align}
    \cL_{\text{AS}}
    =
    - \frac{1}{2} \Tr \p_\mu \phi \, \p^\mu \phi
    - \frac{1}{2} (m_0^2 - \Lambda^2) \Tr \phi^2
    &
    + \frac{\Lambda^d}{\tilde{g}^{\phantom{2}}}
        \Tr \phi / \Lambda^{\frac{d-2}{2}}
\no \\
    &
    + \frac{\Lambda^d}{\tilde{g}^2}
        \Tr \ln (1 - \tilde{g} \, \phi / \Lambda^{\frac{d-2}{2}} )
    \, .
\end{align}
The scalar field $\phi = T^a \phi^a$ transforms in the adjoint representation of a global flavor symmetry group $G$. The coupling constant $\tilde{g}$ is dimensionless.

There are two distinct dimensionful parameters in this Lagrangian. The first parameter $m_0^2$ is the mass-squared of the lightest scalar in the corresponding Coon amplitude which may be tachyonic, massless, or massive. At tree-level, all three choices are consistent. The $N$-point amplitudes computed in \autoref{sec:AS} are valid for tachyonic, massless, and massive scalars. The second parameter $\Lambda^2$ is related to the parameters $q$ and~$\mu^2$ of the corresponding Coon amplitude by $\Lambda^2 = q^{-1} \mu^2$ in the limit $q, \mu \to \infty$ with $\Lambda^2$ fixed. Thus, $\Lambda^2$ is manifestly positive but is otherwise a free parameter introduced by the double-scaling limit of the Coon amplitude. More correctly, the dimensionless ratio $m_0^2/\Lambda^2$ is a free parameter which can naively be positive, negative, or zero, depending on the sign of~$m_0^2$. When considering only the tree-level adjoint scalar amplitudes, all three sign choices and all magnitudes appear consistent. Because we are working strictly at tree-level, there is no need to introduce a cut-off or to interpret $\Lambda^2$ as a cut-off scale (which would restrict ${|m_0^2| < \Lambda^2})$. Hence, we can safely consider the range ${|m_0^2| \geq \Lambda^2}$. Considering this theory at loop-level would necessarily change our conclusions, but such an analysis is beyond the scope of this work (in part because the Coon amplitude is only well-defined at tree-level). We now turn to a classical analysis of the Lagrangian to further probe the stability of the theory.

To simplify our discussion, we can consider the case without a global symmetry group.\footnote{Here we follow the precedent of~\cite{Brekke:1987ptq}, in which the authors study a field theory realization of $p$-adic string theory and in their field theory analysis consider only the case without Chan-Paton factors. We thank our JHEP reviewer for this suggestion.} The resultant amplitudes are simpler and do not decompose into partial amplitudes with trace factors (i.e.\ Chan-Paton factors) as described in \autoref{sec:intro}. The simplified Lagrangian is given by
\begin{align}
    \cL_\varphi
    =
    - \frac{1}{2} \p_\mu \varphi \, \p^\mu \varphi
    - \frac{1}{2} (m_0^2 - \Lambda^2) \varphi^2
    + \frac{\Lambda^d}{\tilde{g}^{\phantom{2}}}
        \varphi / \Lambda^{\frac{d-2}{2}}
    + \frac{\Lambda^d}{\tilde{g}^2}
        \ln (1 - \tilde{g} \, \varphi / \Lambda^{\frac{d-2}{2}} )
    \, ,
\end{align}
where $\varphi$ is now a real scalar field. To keep the Lagrangian real-valued, the scalar field can only take values in the range $\varphi \in (-\infty, \varphi_m)$ with $\varphi_m = \tilde{g}^{-1} \Lambda^{\frac{d-2}{2}}$. The equation of motion for $\varphi$ is highly non-linear but can be simply written in terms of the dimensionless combination $ \varphi /  \varphi_m = \tilde{g} \, \varphi / \Lambda^{\frac{d-2}{2}}$,
\begin{align}
    ( \p^2 - m_0^2 )
    ( \varphi /  \varphi_m )
    =
    \Lambda^2 \,
    \frac{ ( \varphi /  \varphi_m )^2 }
         { 1 - ( \varphi /  \varphi_m ) }
    \, .
\end{align}
The field potential $V(\varphi)$ is non-meromorphic (due to the logarithm) and given by
\begin{align}
    \frac{\tilde{g}^2}{\Lambda^d} \,
    V(\varphi)
    =
    \frac{1}{2}
    \big( m_0^2/\Lambda^2 - 1 \big)
    ( \varphi /  \varphi_m )^2
    - ( \varphi /  \varphi_m )
    -
    \ln (1 - \varphi /  \varphi_m )
    \, .
\end{align}
The potential has critical points at $\varphi = 0$ and $\varphi = \varphi_c = \varphi_m ( 1 - \Lambda^2 / m_0^2 )^{-1}$. The potential tends to $+\infty$ as $\varphi \to \varphi_m $ and to $\pm \infty$ as $\varphi \to -\infty$. The dimensionless ratio $m_0^2 / \Lambda^2$ controls the qualitative features of the potential (including the sign of that latter limit).

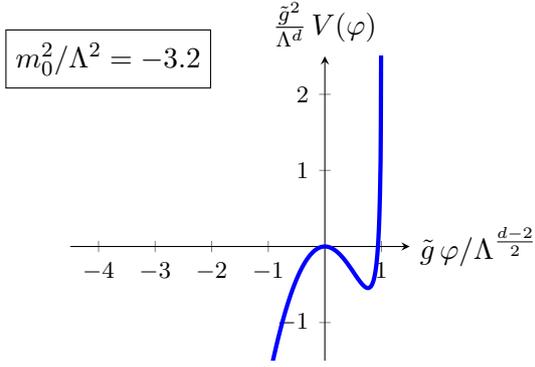
\begin{figure}
\centering

\begin{subfigure}[c]{0.4\textwidth}
\centering
\begin{tikzpicture}
\pgfmathsetmacro{\m}{-3.2};
\begin{axis}
[
    width = 1\textwidth,
    height = 0.25\textheight,
    axis x line=center,
    axis y line=center,
    clip xlimits = true,
    clip ylimits = true,
    xmin = -4.5,
    xmax = 1.5,
    ymin = -1.5,
    ymax = 2.5,
    xtick = {-4,-3,-2,-1,1},
    xticklabels = {$-4$,$-3$,$-2$,$-1$,$1$},
    ytick = {-1,1,2},
    yticklabels = {$-1$,$1$,$2$},
    xlabel = {$ \tilde{g} \, \varphi / \Lambda^{\frac{d-2}{2}}$},
    ylabel = {$\frac{\tilde{g}^2 \mathstrut}{\Lambda^d} \, V(\varphi)$},
    x label style = {anchor=west},
    y label style = {anchor=south},
    ticklabel style = {font=\footnotesize}
];
\addplot[domain= {-4.5}:{0.999}, smooth, blue, ultra thick, samples=600]
    { (1/2)*(\m-1)*x^2-x-ln(1-x) };
\end{axis}
\draw (0.5,4)
    node[rectangle, draw=black, align=center, minimum height=2em]
    {$m_0^2/\Lambda^2=\m$};
\end{tikzpicture}
\caption{If $m_0^2 / \Lambda^2 < 0$, then $V(\varphi)$ has a local minimum at $\varphi = \varphi_c > 0$ and a local maximum at $\varphi = 0$.}
\end{subfigure}
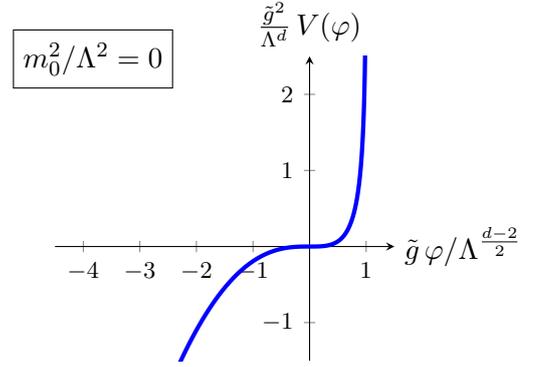
\hfill
\begin{subfigure}[c]{0.4\textwidth}
\centering
\begin{tikzpicture}
\pgfmathsetmacro{\m}{0};
\begin{axis}
[
    width = 1\textwidth,
    height = 0.25\textheight,
    axis x line=center,
    axis y line=center,
    clip xlimits = true,
    clip ylimits = true,
    xmin = -4.5,
    xmax = 1.5,
    ymin = -1.5,
    ymax = 2.5,
    xtick = {-4,-3,-2,-1,1},
    xticklabels = {$-4$,$-3$,$-2$,$-1$,$1$},
    ytick = {-1,1,2},
    yticklabels = {$-1$,$1$,$2$},
    xlabel = {$ \tilde{g} \, \varphi / \Lambda^{\frac{d-2}{2}}$},
    ylabel = {$\frac{\tilde{g}^2 \mathstrut}{\Lambda^d} \, V(\varphi)$},
    x label style = {anchor=west},
    y label style = {anchor=south},
    ticklabel style = {font=\footnotesize}
];
\addplot[domain= {-4.5}:{0.999}, smooth, blue, ultra thick, samples=600]
    { (1/2)*(\m-1)*x^2-x-ln(1-x) };
\end{axis}
\draw (0.5,4)
    node[rectangle, draw=black, align=center, minimum height=2em]
    {$m_0^2/\Lambda^2=\m$};
\end{tikzpicture}
\caption{If $m_0^2 / \Lambda^2 = 0$, then $V(\varphi)$ has no local minima or maxima and is flat at $\varphi = \varphi_c = 0$.}
\end{subfigure}

\bigskip

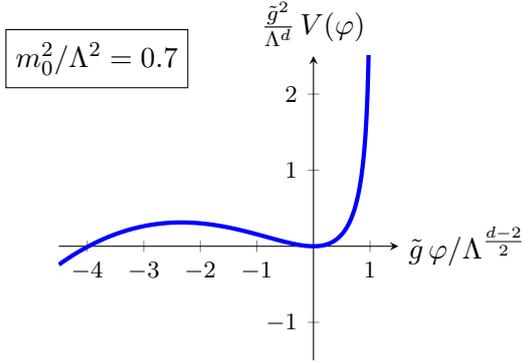
\begin{subfigure}[c]{0.4\textwidth}
\centering
\begin{tikzpicture}
\pgfmathsetmacro{\m}{0.7};
\begin{axis}
[
    width = 1\textwidth,
    height = 0.25\textheight,
    axis x line=center,
    axis y line=center,
    clip xlimits = true,
    clip ylimits = true,
    xmin = -4.5,
    xmax = 1.5,
    ymin = -1.5,
    ymax = 2.5,
    xtick = {-4,-3,-2,-1,1},
    xticklabels = {$-4$,$-3$,$-2$,$-1$,$1$},
    ytick = {-1,1,2},
    yticklabels = {$-1$,$1$,$2$},
    xlabel = {$ \tilde{g} \, \varphi / \Lambda^{\frac{d-2}{2}}$},
    ylabel = {$\frac{\tilde{g}^2 \mathstrut}{\Lambda^d} \, V(\varphi)$},
    x label style = {anchor=west},
    y label style = {anchor=south},
    ticklabel style = {font=\footnotesize}
];
\addplot[domain= {-4.5}:{0.999}, smooth, blue, ultra thick, samples=600]
    { (1/2)*(\m-1)*x^2-x-ln(1-x) };
\end{axis}
\draw (0.5,4)
    node[rectangle, draw=black, align=center, minimum height=2em]
    {$m_0^2/\Lambda^2=\m$};
\end{tikzpicture}
\caption{If $0 < m_0^2 / \Lambda^2 < 1$, then $V(\varphi)$ has a local minimum at $\varphi = 0$ and a local maximum at $\varphi = \varphi_c < 0$.}
\end{subfigure}
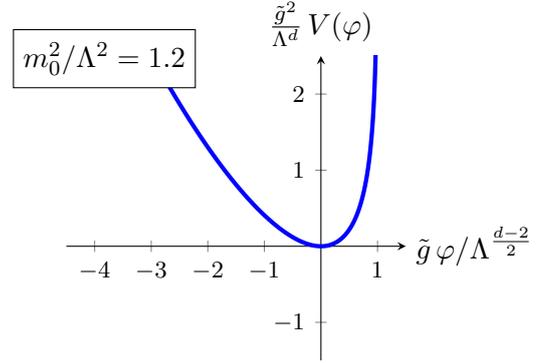
\hfill
\begin{subfigure}[c]{0.4\textwidth}
\centering
\begin{tikzpicture}
\pgfmathsetmacro{\m}{1.2};
\begin{axis}
[
    width = 1\textwidth,
    height = 0.25\textheight,
    axis x line=center,
    axis y line=center,
    clip xlimits = true,
    clip ylimits = true,
    xmin = -4.5,
    xmax = 1.5,
    ymin = -1.5,
    ymax = 2.5,
    xtick = {-4,-3,-2,-1,1},
    xticklabels = {$-4$,$-3$,$-2$,$-1$,$1$},
    ytick = {-1,1,2},
    yticklabels = {$-1$,$1$,$2$},
    xlabel = {$ \tilde{g} \, \varphi / \Lambda^{\frac{d-2}{2}}$},
    ylabel = {$\frac{\tilde{g}^2 \mathstrut}{\Lambda^d} \, V(\varphi)$},
    x label style = {anchor=west},
    y label style = {anchor=south},
    ticklabel style = {font=\footnotesize}
];
\addplot[domain= {-4.5}:{0.999}, smooth, blue, ultra thick, samples=600]
    { (1/2)*(\m-1)*x^2-x-ln(1-x) };
\end{axis}
\draw (0.5,4)
    node[rectangle, draw=black, align=center, minimum height=2em, fill = white]
    {$m_0^2/\Lambda^2=\m$};
\end{tikzpicture}
\caption{If $m_0^2 / \Lambda^2 \geq 1$, then $V(\varphi)$ has a global minimum at $\varphi = 0$. \\}
\end{subfigure}

\caption{Plots of $\frac{\tilde{g}^2 \mathstrut}{\Lambda^d} \, V(\varphi)$ vs.\ $\tilde{g} \, \varphi / \Lambda^{\frac{d-2}{2}}$ for various values of $m_0^2 / \Lambda^2$ which demonstrate the different qualitative behaviors for the potential as described in the main text.}

\label{fig:1}

\end{figure}

In \autoref{fig:1}, we plot $V(\varphi)$ vs.\ $\varphi$ for various values of $m_0^2 / \Lambda^2$, recalling that all values of $m_0^2 / \Lambda^2$ naively seem consistently when considering the tree-level amplitudes alone. The qualitative behavior of $V(\varphi)$ may be summarized as follows:
\begin{itemize}
\item If $m_0^2 / \Lambda^2 < 0$, then $V(\varphi)$ has a local minimum (i.e.\ a metastable vacuum) at ${\varphi = \varphi_c}$ with $0 < \varphi_c < \varphi_m$, a local maximum at $\varphi = 0$ (the tachyonic extremum of the non-interacting theory), and a global minimum at $\varphi = -\infty$. In this case, the theory exhibits a metastable version of tachyon condensation and is classically metastable with a positive effective mass-squared given by~${V''(\varphi_c) = m_0^2 (m_0^2/\Lambda^2 - 1) > 0}$. 

\item  If $m_0^2 / \Lambda^2 = 0$, then $V(\varphi)$ is flat at $\varphi = \varphi_c = 0$ with a global minimum at $\varphi = -\infty$. In this case, the theory is classically unstable.

\item If $ 0 < m_0^2 / \Lambda^2 < 1$, then $V(\varphi)$ has a local minimum (i.e.\ a metastable vacuum) at ${\varphi = 0}$, a local maximum at $\varphi = \varphi_c < 0$, and a global minimum at $\varphi = -\infty$. In this case, the theory is classically metastable with a positive effective mass-squared given by~${V''(0) = m_0^2 > 0}$.

\item If $m_0^2 / \Lambda^2 \geq  1$, then $V(\varphi)$ has a global minimum (i.e.\ a stable vacuum) at ${\varphi = 0}$. In this case, the theory is classically stable with a positive effective mass-squared given by~${V''(0) = m_0^2 > 0}$.

\end{itemize}
Thus, for a wide range of parameters, the theory is classically stable or metastable. In each case except for $m_0^2 = 0$, the effective theory has a classical vaccuum with a positive effective mass-squared, even when the original mass-squared $m_0^2$ was negative. To ensure full stability, we must demand $m_0^2 / \Lambda^2 \geq  1$ which fixes $m_0^2 > 0$. We recall that $\Lambda^2 = q^{-1} \mu^2$ was necessarily positive but essentially a free parameter otherwise. It is also possible that higher-derivative correction, $\cO(q^{-1})$ corrections, or loop corrections stabilize the theory (as in the case of tachyon condensation in string theory).

Although we have only considered the simplified scalar theory without a global symmetry group, we can extend our results to the adjoint scalar theory with global flavor symmetry group $\mathrm{U}(N_F)$ by considering the subset of field configurations proportional to the $N_F \times N_F$ identity matrix $\phi = \mathds{1} \varphi$. In any case, we have demonstrated that (a simplified version of) the adjoint scalar theory has a classically stable vacuum and is thus a viable field theory. Of course, our results hold strictly at tree-level, and loop-level corrections may modify our conclusions.


\bibliography{BCR}

\end{document}